\documentclass[12pt]{article}

\usepackage{color}
\usepackage[english]{babel}
\usepackage{amsmath, amsfonts, amssymb}
\usepackage{mathrsfs}
\usepackage{bbm}
\usepackage{graphicx}
\usepackage[colorlinks=true,linkcolor=blue]{hyperref}
\usepackage{cleveref}
\usepackage{physics}
\usepackage{cite}
\usepackage{fontawesome5}
\usepackage[dvipsnames]{xcolor}
\usepackage{color}
\usepackage{upgreek}
\usepackage{caption}
\usepackage{feynmp}
\usepackage{tikz}
\usepackage{slashed}
\usepackage{subcaption}
\newcommand{\github}[1]{%
   \href{#1}{\faGithubSquare}%
}
\def\ba{\begin{eqnarray}}
\def\ea{\end{eqnarray}}
\def\be{\begin{equation}}
\def\ee{\end{equation}}

\def\d{\partial}

\renewcommand{\l}{\lambda}

\renewcommand{\o}{\omega}

\renewcommand{\a}{\alpha}
\renewcommand{\b}{\beta}

\newcommand{\vk}{\varkappa}

\newcommand{\vf}{\varphi}

\newcommand{\diff}{\mathrm{d}}
\newcommand{\e}{{\rm e}}

\renewcommand{\Im}{\mathop{\rm Im}\nolimits}

\renewcommand{\P}{\mathcal{P}}
\newcommand{\R}{\mathcal{R}}
\newcommand{\bz}{\mathbf{z}}
\newcommand{\ve}{\varepsilon}

\newcommand{\Rin}{\R_\text{in}}
\newcommand{\Rout}{\R_\text{out}}
\newcommand{\Rref}{\R_\text{ref}}
\newcommand{\Es}{E_{\rm ts}}

\newcommand{\GammaTST}{\Gamma_\text{TST}}
\newcommand{\tDyn}{t_{\text{dyn.}}}
\newcommand{\tTh}{t_{\text{th.}}}
\newcommand{\tDec}{t_{\text{dec.}}}
\newcommand{\tPlat}{t_{\text{plat.}}}
\newcommand{\tDrift}{t_{\text{drift}}}
\newcommand{\Nrel}{N_{\text{rel.}}}
\newcommand{\Erel}{E_{\text{rel.}}}
\newcommand{\brel}{\beta_{\text{rel.}}}
\newcommand{\Rpert}{R_{\text{pert.}}}
\newcommand{\Rnp}{R_{\text{non-pert.}}}

\newcommand\bseq{\begin{subequations}}
\newcommand\eseq{\end{subequations}}

\newcommand{\dint}{\int\displaylimits}

\newcommand{\thetato}[2]{\theta_{\text{#1}\to\text{#2}}(o_\bz)}
\newcommand{\thetatau}[2]{\theta_{\text{#1}\to\text{#2}}(\tau_\bz)}

\usepackage{etoolbox}
\preto\subequations{\ifhmode\unskip\fi}

\allowdisplaybreaks[4]

\definecolor{JHcolor}{rgb}{1.0, 0.5, 0.15}

\begin{document}
\title{
\vspace{-1cm}
Dynamics of nucleation in thermal phase transitions 
}

\author{
Oliver Gould$^{1}$\thanks{oliver.gould@nottingham.ac.uk}~,~
Joonas Hirvonen$^{1}$\thanks{joonas.hirvonen@nottingham.ac.uk}~,~
Andrey Shkerin$^{2}$\thanks{ashkerin@perimeterinstitute.ca}~,~
Sergey Sibiryakov$^{2,3}$\thanks{ssibiryakov@perimeterinstitute.ca}
\\[2mm]
{\small\it $^1$School of Physics and Astronomy, University of Nottingham, Nottingham NG7 2RD, U.K.}\\
{\small\it $^2$Perimeter Institute for Theoretical Physics, 31 Caroline St N, Waterloo, ON N2L 2Y5, Canada}\\
{\small\it $^3$Department of Physics and Astronomy, McMaster University,}\\
{\small\it 1280 Main St W, Hamilton, ON L8S 4M1, Canada}\\[1.5mm]
}
\date{}
\maketitle

\begin{abstract}

We study dynamical effects during nucleation in thermal first-order phase transitions in field theory.
Focusing on the classical regime of the decay of a metastable state, we present the general formula for the thermal decay rate
including
the dynamical prefactor and give a recipe for its systematic evaluation.
We describe the physical mechanism which reduces the actual thermal decay rate with respect to the statistical rate obtained in equilibrium theory.
We also discuss the thermality conditions ensuring the existence of a steady-state thermal rate, in which case 
our formula is exact up to exponentially small corrections.
We show that it reproduces the known results for the nucleation rate in stochastic mechanics and field theory, and allows us to unify and go beyond them. We illustrate this in real-time numerical simulations of simple field theory models.
We observe significant non-perturbative contributions which can dominate the dynamical prefactor in weakly-coupled field theories at moderate exponential suppression of the decay rate. We explore the connection of these non-perturbative effects to oscillons. Notably, our
numerical
method requires exponentially less computing time than direct simulations of decays and is thus applicable to systems with arbitrarily strong exponential suppression.
Finally, we discuss small or poorly thermalized systems when the thermality conditions are violated and the steady-state rate does not exist.
\end{abstract}

\newpage
{\hypersetup{hidelinks}
\tableofcontents
}

\section{Introduction}
\label{sec:intro}

Thermal first-order phase transitions are ubiquitous in physics.
They are actively studied in physical chemistry, soft condensed matter, high-energy physics and cosmology. 
They play an important role in our everyday life and in the life of the entire Universe \cite{Mazumdar:2018dfl}.
First-order phase transitions proceed via the nucleation and growth of bubbles, whose theoretical study was pioneered by Gibbs nearly 150 years ago \cite{gibbs1928collected}.
Major cornerstones of nucleation theory are the works of Becker and D\"{o}ring \cite{Becker}, which suggested the discrete model of the growth of bubbles of a new phase inside an old, metastable phase; Wigner \cite{wigner1938transition}, where the transition state method was applied to the calculation of rates of chemical reactions; Zeldovich \cite{zeldovich1942theory}, who studied bubble nucleation by treating the bubble radius as a particle undergoing Brownian motion. Cahn and Hilliard \cite{cahn1958free} and Langer \cite{Langer:1969bc} applied nucleation theory to systems with many degrees of freedom, and the application to quantum field theory was initiated by Coleman \cite{Coleman:1977py}, Affleck \cite{Affleck:1980ac} and Linde \cite{Linde:1980tt,Linde:1981zj}.

The physical mechanism of the decay of metastable phase (false vacuum) to the stable one (true vacuum) depends on the temperature $T$ of the system. At small $T$, 
the decay happens via tunneling through the potential barrier separating the vacua \cite{Coleman:1977py,Callan:1977pt}. Thermal fluctuations assist the tunneling process \cite{Linde:1980tt,Linde:1981zj,Weinberg:2012pjx}.
At sufficiently large $T$, the decay happens due to classical thermal fluctuations~\cite{Affleck:1980ac,Gould:2021ccf}. 
In the semiclassical regime, where the decay is exponentially suppressed, the system crosses the barrier in the vicinity of the transition state (TS) \cite{wigner1938transition}, which is a saddle point of the Hamiltonian of the system. In different physical contexts,
the TS is referred to as the critical droplet, critical bubble, or sphaleron.
The classical regime is realized at 
$T\gg T_\text{q} \sim \o_-$,
where $-\o_-^2$ is the curvature of the barrier along the unstable direction at the TS \cite{Affleck:1980ac}.\footnote{
Throughout the paper, we use the natural units $c=\hbar=k_B=1$.
In field theories with polynomial potentials and no hierarchical parameters, $\o_-$ is of the same order as the mass of the field at the metastable minimum.
}
In this case, an effective classical description exists which describes the nucleation process.
The decays remain suppressed as long as $T\ll E_\text{ts}$ where $E_\text{ts}$ is the height of the energy barrier at the TS. 
In this paper we consider thermal false vacuum decay in the limits $T_\text{q} \ll T\ll E_\text{ts}$. 

\begin{figure}[t]
\begin{center}
\includegraphics[scale=0.35]{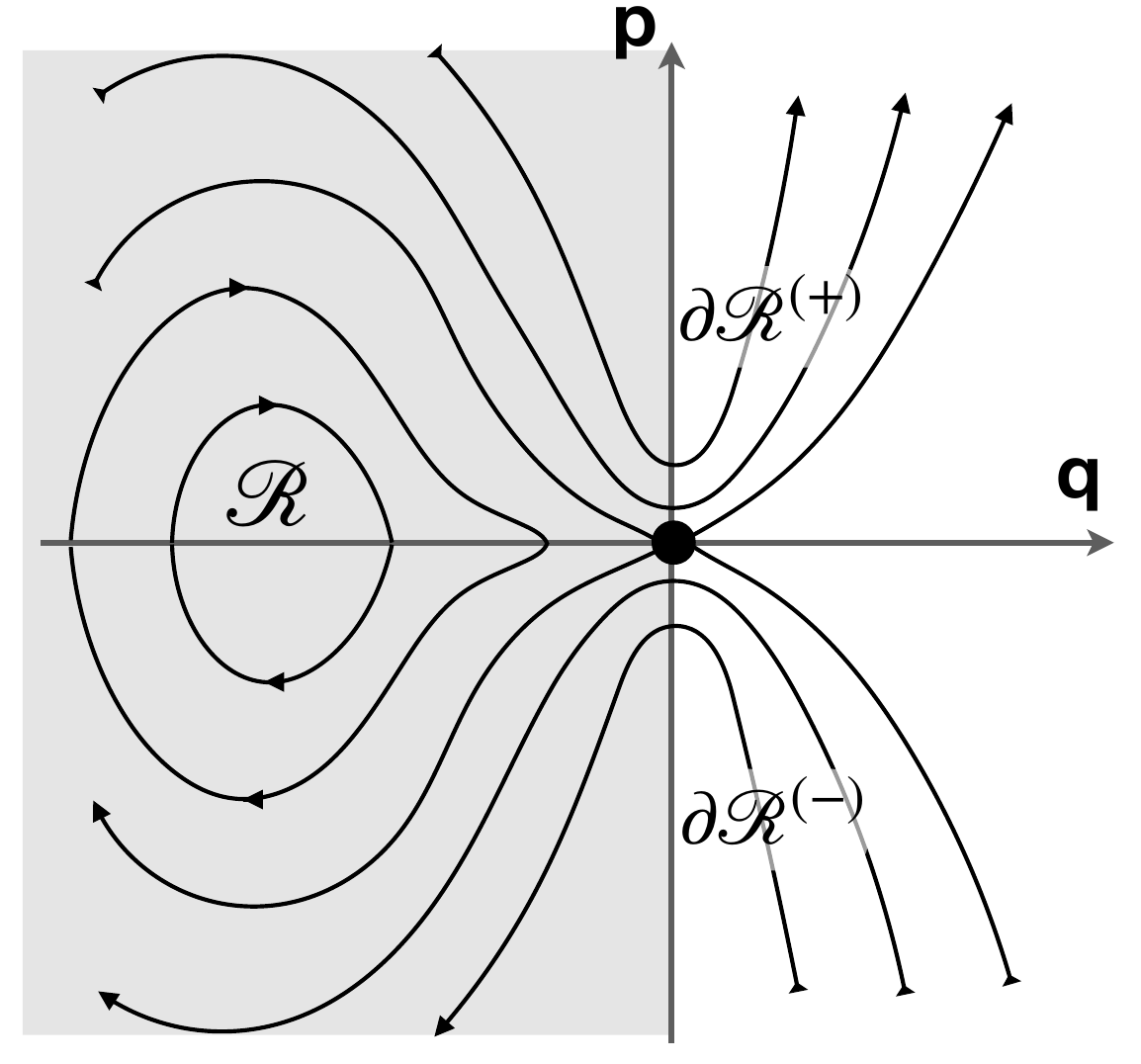}
\end{center}
\caption{Phase-space trajectories and the metastable region $\R$ (shaded gray) for the calculation of the decay rate. The flux out of $\R$ crosses the surface $\d\R^{(+)}$, selected by the theta-function in Eq.~(\ref{Gamma_eq_gen}). Trajectories from the outer region enter $\R$ through $\d\R^{(-)}$. The black dot indicates the transition state (TS). }
\label{fig:flux}
\end{figure}

The transition state theory (TST) approach to calculating observables associated with the phase transition, such as the rate of nucleation of the bubbles of true vacuum, relies on the assumption of local thermodynamic equilibrium during the nucleation process. 
The TST rate is defined as follows.
Denote by $\R$ the region encompassing the metastable minimum, whose boundary $\d\R$ contains the TS, see Fig.~\ref{fig:flux}. 
Assume that the system is in full equilibrium, i.e.\ that both $\R$ and its complement including the true vacuum region are thermally populated.\footnote{For simplicity, we assume here that the true vacuum exists and has finite energy. For a field theory, this requires putting the system in a finite box. Equation (\ref{Gamma_eq_gen}) is, however, insensitive to this assumption.
}
Define the TST rate as the
phase space probability flux through $\d\R$,\footnote{
We understand the rate as the decay probability per unit time. In extensive systems, such as field theory in finite volume, the rate will be proportional to the volume.
}
\be \label{Gamma_eq_gen}
\GammaTST = \frac{1}{Z_\R}\dint_{\d\R} \diff S\, \textbf{n}\cdot \dot{\textbf{z}} \; \theta( \textbf{n}\cdot\dot{\textbf{z}} ) \, \e^{-H(\textbf{z})/T} \;.
\ee
Here $\textbf{z} \equiv (\textbf{q},\textbf{p})$ is the phase space coordinate of the system with the Hamiltonian $H(\textbf{z})$, dot means derivative with respect to time, and 
\be
Z_{\R}=\dint_{\R} \diff {\bf z}\, \e^{-H(\textbf{z})/T} \;
\ee
is the normalization constant that can be interpreted as the partition function of the false vacuum.  
The vector $\textbf{n}$ is normal to $\d\R$ and points outwards. Due to the $\theta$-function, the integral runs only over the part of the surface $\d\R^{(+)}$ which is crossed by phase-space trajectories leaving the region $\R$, as shown in Fig.~\ref{fig:flux}. For future reference, we denote the part of the boundary crossed by trajectories entering $\R$ by $\d\R^{(-)}$.

Evaluating the integral in the saddle-point approximation, one arrives at the standard relation between $\GammaTST$ and the imaginary part of the false vacuum free energy~\cite{Affleck:1980ac},
\be
\label{GammaE}
\GammaTST\overset{\text{1-loop}}{=} \frac{\o_-}{\pi T} \cdot\Im F\;.
\ee
The imaginary part of the free energy can be calculated using an analytic continuation in the Euclidean time formalism where its exponential suppression is due to the large Euclidean action of the TS. 
Note, however, that the original integral in Eq.~(\ref{Gamma_eq_gen}) is given in the real time, and the analytic continuation is not necessary to define the TST rate.

The results (\ref{Gamma_eq_gen}), (\ref{GammaE}) are obtained within the purely statistical framework which is known to be incomplete, since it disregards dynamics of the nucleation process~\cite{Hanggi:1990zz,kalikmanov2012nucleation,Gould:2021ccf}. 
In a real decaying system the true vacuum region lying outside $\R$ \textit{is not} thermally populated; in fact, the vacuum decay is precisely the process leading to the global equilibration assumed in the TST expression (\ref{Gamma_eq_gen}).
This means that some phase-space trajectories leaving $\R$ through $\d\R^{(+)}$ may not contribute into the true decay rate, since they come into $\R$ from outside.
Besides, trajectories can cross the dividing surface $\d\R$ multiple times before escaping the metastable region, and only the final crossing-out must be counted. These {\it re-crossings} lead to suppression of the decay rate compared to the TST result, with the ratio being referred to as the {\it dynamical prefactor}.

At a fundamental level, the absence of thermal equilibrium in real decaying systems implies a qualitative change of the nucleation picture. The TS is not merely sampled from the Boltzmann ensemble, but emerges as a result of non-linear evolution starting deeply inside $\R$. This gives rise to strong temporal correlations preceding the nucleation which can potentially lead to new signatures; see e.g.\ \cite{Pirvu:2023plk, Pirvu:2024nbe}.

Methods for the calculation of the dynamical prefactor have developed in parallel in physical chemistry \cite{Hanggi:1990zz} and in thermal field theory \cite{langer1973hydrodynamic, Moore:1998swa, Moore:2000jw, Ekstedt:2022tqk, Hirvonen:2024rfg}. 
However, the connection between different approaches remains obscure. This led to several 
apparently incompatible expressions for the dynamical prefactor in field theory literature~\cite{Langer:1969bc, Affleck:1980ac, Linde:1981zj, Arnold:1987mh, Hirvonen:2024rfg}, 
leaving phenomenologists without any clear reason to prefer one expression over the others; see e.g.~\cite{Croon:2020cgk}. 
Importantly, the dynamical prefactor is not merely a small correction to the TST rate (\ref{Gamma_eq_gen}).
It has been known for a long time to significantly suppress the decay rate in systems with strong dissipation \cite{Langer:1969bc}. On the other hand, it has been commonly assumed to be close to unity for conservative systems with multiple degrees of freedom. This assumption has recently been disproved by numerical simulations of classical thermal decays in $(1+1)$-dimensional scalar field models which found a significantly lower decay rate than predicted by TST~\cite{Pirvu:2024nbe}.
A discrepancy was also observed between $(3+1)$-dimensional lattice simulations~\cite{Gould:2024chm} and the one-loop perturbative TST result~(\ref{GammaE}), calling for a systematic improvement of the perturbative expression for the decay rate, including the dynamical prefactor \cite{Ekstedt:2022tqk}; see, however \cite{Dutka:2025ghb}, which found a good agreement between theory and simulations at large dissipation.

In this paper we revisit the dynamical theory of nucleation and provide a unifying framework encompassing both mechanics and field theory. Our approach is similar to the reactive flux method in physical chemistry (see \cite{berne1985molecular,Hanggi:1990zz} and references therein), but is more general since it does not assume the existence of thermal equilibrium between the metastable and stable phases. In our development we establish the connection between several existing approaches and address the following puzzling aspects of nucleation:

\subsection{Puzzles of thermal nucleation}
\label{ssec:intro_puzzles}

\textit{\underline{Puzzle 1}: What is thermal nucleation rate? ---} 
As argued above, a real decaying system is not in global thermal equilibrium. In particular, the states in the vicinity of TS are not thermally populated, thereby violating the assumption used in the derivation of Eq.~(\ref{Gamma_eq_gen}). But if thermal equilibrium is violated, what does it mean to talk about a thermal rate?

A natural guess is to consider an ensemble of systems with only the metastable region 
$\R$ thermally populated. The probability distribution function of such an ensemble is,
\be \label{Rho_initial}
\rho(\textbf{z};t=0) = \frac{1}{Z_{\R}} \e^{-H(\textbf{z})/T} \theta(\textbf{z}\in \R) \;.
\ee
One would like to identify the rate with the fractional number of systems decaying per unit time, which is equal to the probability flux out of $\R$. 
The distribution (\ref{Rho_initial}), however, is not stationary and the rate defined in this way will depend on time \cite{Hirvonen:2025hqn}.  
One expects that if one waits long enough, the distribution function will get attracted to a steady state, with constant probability flux out of $\R$. We adopt this stationary flux as the definition of the thermal decay rate.  
Several
questions arise: Under what conditions does the steady state exist? How long does one need to wait until the system approaches it? What is the general expression for the decay rate in the steady state? Do different expressions for the nucleation rate in the literature agree?

\textit{\underline{Puzzle 2}: What is the right TS? ---} 
The TST rate (\ref{Gamma_eq_gen}) relies on the existence of a TS and depends exponentially sensitively on its energy. However, in systems with many degrees of freedom, such as field theory, the TS is not uniquely defined. 
Various quantum and thermal fluctuations contribute to the effective energy functional and shift its saddle point~\cite{Langer:1974cpa}. Including different fluctuations yields different solutions for the TS, which may not even exist before certain fluctuations are integrated out. 
In perturbative high-temperature quantum field theories, the decision about which fluctuations to include is fixed by the perturbative power counting~\cite{Gould:2021ccf, Lofgren:2023sep}, but no such guidance is available away from the weak-coupling limit.
Further, the TS is defined in equilibrium, yet many different dynamical equations lead to the same equilibrium distribution. This raises the question: can the specific form of the real-time dynamics modify the TS?

On the other hand, it is understood in the literature that the actual thermal decay rate must be independent of the TS, and more broadly, of the choice of the surface $\d\R$ bounding the metastable region~\cite{Langer:1969bc, chandler1978statistical, Moore:1998swa, Moore:2000jw, Hirvonen:2024rfg}. The dynamical prefactor is key for this property: it needs to compensate the surface-dependence of the statistical part of the rate.\footnote{Note that in strongly-coupled quantum field theories where the dynamical prefactor is unavailable, the estimates of the decay rate unavoidably depend on the TS~\cite{BarrosoMancha:2022mbj, Rummukainen:2026etj}.}
The problem of a precise TS determination then becomes irrelevant, giving way to another question: Is the steady-state rate defined above independent of the surface $\d\R$, irrespectively of whether it passes through a TS or not?

\textit{\underline{Puzzle 3}: Field theory vs.~low-dimensional mechanics.---} 
One can often get intuition about processes in field theory by looking at their analogs in systems with one or few degrees of freedom. 
This works well for quantum decays:
the path integral expression for the tunneling rate in quantum field theory is the natural generalization of that for one-dimensional quantum mechanics \cite{Coleman:1977py, Callan:1977pt, Coleman:1978ae}.
However, for classical decays, the situation is strikingly different. Indeed, consider a thermal ensemble of particles trapped in the metastable well with the energy barrier $V_0$. The initial flux of particles out of the well is thermal and given by Eq.~(\ref{Gamma_eq_gen}). 
However, in the absence of stochastic forces acting on particles, the flux drops to zero after a short time:
all particles with $E>V_0$ have escaped the well and those with $E<V_0$ will never escape. Thus, beyond the initial transient, the classical decay rate in a system with one degree of freedom vanishes. This differs drastically from field theory where decays persist within conservative evolution. 

It is clear where the difference comes from in this example: the one-particle conservative system cannot have an energy higher than the barrier energy and remain for long in the metastable state; 
thermal transitions are possible only in the presence of thermal noise.
On the other hand, in a field theory the total energy of the system is infinite, and there is always enough energy to nucleate.
Instead, the process of nucleation involves a local redistribution of energy.
Is it possible to have a unified framework to describe transitions in both particle mechanics and field theory?

\textit{\underline{Puzzle 4}: How do oscillons affect the decay rate? ---} Scalar field theories with anharmonic potentials have localized, long-lived nonlinear solutions called oscillons~\cite{Amin:2010jq, Zhang:2020bec, Levkov:2022egq, Levkov:2023ncb}.
Oscillons are formed in a thermal ensemble and survive for a long time if damping and stochastic noise are weak. 
Furthermore, numerical simulations show that the TS (critical bubble) evolves from a precursor oscillon, rather than from a spontaneous assembly of small field fluctuations~\cite{Pirvu:2023plk,Pirvu:2024nbe}. This is consistent with the observation that having a large-amplitude coherent field configuration, such as oscillon, in the initial state catalyzes formation of the critical bubble~\cite{Gleiser:2007ts,Gleiser:1991rf,Gleiser:1993pt}. 
One can be tempted to conclude that theories with oscillons should exhibit higher thermal nucleation rates than the naive TST prediction.

This conclusion, however, cannot be correct since it contradicts the general result that the dynamical prefactor does not exceed unity~\cite{Hanggi:1990zz}.
How do we reconcile our intuition about oscillons with this result? What is their actual effect on the nucleation rate?

\bigskip

The paper is organized as follows. In Sec.~\ref{ssec:prelim} we describe our setup and present the resolution of the puzzles listed above. We also give a general formula for the nucleation rate which accounts for the dynamical effects. In Secs.~\ref{ssec:surface-flux}, \ref{ssec:trajectories} we derive this formula using several methods and establish connections between them. In Sec.~\ref{sec:app} 
we illustrate the general approach by reproducing classical Langer's result \cite{Langer:1969bc} for the nucleation rate in the presence of moderate to strong dissipation, as well as the result of Mel'nikov and Meshkov \cite{Melnikov:1986} for one-dimensional mechanical systems with weak noise. Section~\ref{sec:num} is dedicated to the numerical investigation of various contributions to the dynamical prefactor in field theory. 
In Sec.~\ref{sec:finite} we study the approach of the nucleation rate to the stationary-flux regime and elaborate on the existence of the latter.
We present our conclusions and outlook in Sec.~\ref{sec:disc}.
Appendices contain technical details of the analysis.

The codes and testing routines used in this work are available on 
GitHub.\footnote{\url{https://github.com/Olborium/FVD_One} and \url{https://github.com/joonashir/ClasTermNucl11}.}

\section{The nucleation rate}
\label{sec:derivation}

\subsection{Setup and the rate formula}
\label{ssec:prelim}

We consider a general classical system with coordinates $q_i$ and momenta $p_i$, $i=1,\ldots,N$, which can be combined into a single phase-space coordinate $z_I$, $I=1,\ldots, 2N$, as in Eq.~(\ref{Gamma_eq_gen}). The Hamiltonian of the system is assumed to be at most quadratic in momenta,\footnote{This restriction is a pure matter of convenience which allows us to keep the discussion concrete. Our general nucleation rate formula, Eq.~(\ref{GammaFull}) below, does not rely on it.}
\be
\label{Hamiltonian}
H=\frac{1}{2}\sum_{ij} \big(p_i+g_i({\bf q})\big)G^{ij}({\bf q})\big(p_j+g_j({\bf q})\big) +V({\bf q})\;,
\ee
where $G^{ij}({\bf q})$ is a positive-definite metric. This form covers a large class of physically relevant situations, including field theories 
in the classical regime, in cases when degrees of freedom producing quantum behavior can be integrated out to yield an effective description which is local in time; see for instance Refs.~\cite{Bodeker:1996wb, Aarts:1997kp}.
In 
field theories the index $i$, apart from enumerating the types of fields, labels also the spatial coordinates, so that its range $N$ is infinite. A finite-dimensional phase space can still be recovered by putting the fields on a discrete spatial lattice and in a finite spatial volume, provided one can arrange appropriate continuum and infinite-volume limits. Note that if $g_i({\bf q})\neq 0$, the Hamiltonian contains terms linear in momenta. Such terms can arise, e.g., in the presence of external magnetic fields and break the time reversal symmetry ${\bf p}\mapsto -{\bf p}$.

If the system is in contact with a heat bath, its evolution is affected by dissipation and thermal noise. 
We model them by including a linear damping force and additive white noise in the momentum equation of motion,\footnote{We will comment on the generalization of our analysis to other types of dissipation and noise in
Sec.~\ref{sec:disc}.}
\be
\label{eoms}
\dot q_i=\frac{\d H}{\d p_i}~,~~~~~\dot p_i=-\frac{\d H}{\d q_i}-\sum_{j} \eta_{ij} \dot q_j +\xi_i\;,
\ee
where $\eta_{ij}$ is a constant symmetric positive-definite matrix of dissipation coefficients. By the classical fluctuation-dissipation theorem, the noise $\xi_i(t)$ has the correlation function
\be
\label{wnoise}
\langle \xi_i(t)\xi_j(t')\rangle=2\eta_{ij} T\,\delta(t-t')\;,
\ee
where $T$ is the temperature of the bath.
The phase-space distribution function $\rho({\bf z})$ of an ensemble of such systems  
obeys the
Fokker--Planck equation, 
\be
\label{FP}
\frac{\d \rho}{\d t}=-\sum_I\d_{I}  {\cal J}_I \;, ~~~~~~
{\cal J}_I=-\sum_K M_{IK} \left(\frac{\d H}{\d z_K}+T\frac{\d}{\d
    z_K}\right) \rho\;,
\ee
where $M_{IK}$ is the block matrix,
\be
M=\begin{pmatrix}
  0&-\mathbbm{1}\\
  \mathbbm{1}&\eta 
  \end{pmatrix} \;.
\ee

\begin{figure}[t]
\begin{center}
\includegraphics[scale=0.4]{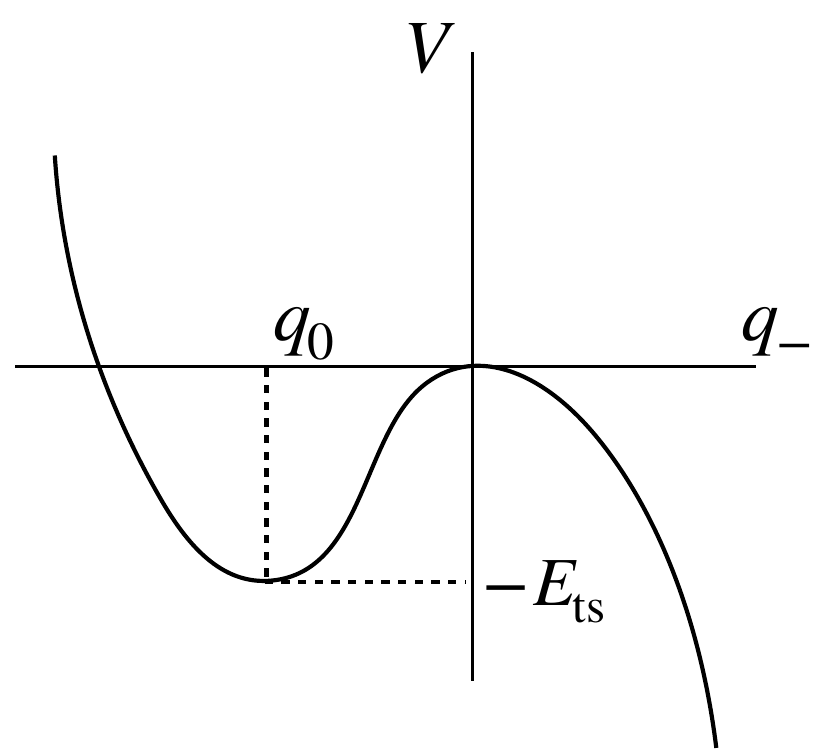}
\end{center}
\caption{\label{fig:potential} 
Potential with a metastable minimum at $q_-=q_0$, and TS at $q_{-}=0$. 
}
\end{figure}

The potential $V({\bf q})$ in Eq.~(\ref{Hamiltonian}) is assumed to have a local minimum --- the false vacuum --- at some point ${\bf q}={\bf q}_0$ (see Fig.~\ref{fig:potential} for illustration in the case of one-dimensional mechanics). The false vacuum is separated from regions of lower energy by a potential barrier with a saddle point --- the TS --- at ${\bf q}_{\rm ts}$. Assuming the TS is unique, we can set ${\bf q}_{\rm ts}=0$ and cast the kinetic term at it into the canonical form, $G^{ij}(0)=\delta^{ij}$, $g_i(0)=0$. The energy of the TS with respect to the false vacuum will be denoted by $E_{\rm ts}$. The matrix of second derivatives of $V({\bf q})$ at the TS contains a single negative eigenvalue $-\omega_-^2$, $\omega_->0$. The corresponding unstable direction in configuration space will be labeled by $q_-$. It represents a transition coordinate between the basin of attraction of the metastable vacuum at 
$q_-<0$ and the decay products at $q_->0$, as shown in Fig.~\ref{fig:potential}.

The assumption
that the TS is unique
does not hold in general. In fact, it is commonly violated in field theory where there is a continuous family of critical bubbles differing by spatial translations. Since translations do not cost any energy, they are associated with zero-frequency collective perturbations (zero modes). Further zero modes may exist if the TS breaks other internal or external symmetries. There are well-developed methods for the treatment of zero modes within the TST of false vacuum decay \cite{Weinberg:2012pjx}. We will first discuss systems without zero modes and will comment on how they get incorporated into our approach at the end of the subsection.    

As discussed in the Introduction, we consider an ensemble of systems with the initial phase-space distribution (\ref{Rho_initial}) confined to the neighborhood of the false vacuum $\R$. We let it evolve towards a steady state,
at which point the number of systems within the region $\R$ decreases at a constant rate. This steady-state decay rate $\Gamma$ is our main object of interest.
We do not need to specify the region $\R$ precisely: as we are going to see, the final result does not depend on $\R$ as long as it encompasses most of the steady-state distribution. A possible choice is to include into $\R$ all phase-space points with $q_-<0$, in which case the boundary $\d\R$ coincides with the surface $q_-=0$ passing through the TS (cf. Fig.~\ref{fig:flux}).

The existence of a well-defined thermal nucleation rate depends on a suitable hierarchy of time scales ({\it Puzzle 1} from Sec.~\ref{ssec:intro_puzzles}). The first time scale
is the {\it dynamical time} $\tDyn$ that characterizes the microscopic evolution of the degrees of freedom relevant for the nucleation. For example, in one-dimensional stochastic mechanics at weak to moderate dissipation ($\eta\lesssim \omega_-$), the dynamical time is set by the period of oscillations around the false vacuum, $\tDyn\sim 2\pi/\omega_0\sim2\pi/\omega_-$, where in the last equality we have assumed that the potential does not contain any hierarchical parameters. 
For strong damping, $\eta\gg\omega_-$, the dissipation slows down the particle motion and the dynamical time becomes $\tDyn\sim \eta^{-1}$.
In field theory, the relevant degrees of freedom are the Fourier modes of the decaying field with wavelengths comparable to the size of the critical bubble. The latter is typically $\sim m^{-1}$, where $m$ is the field mass in the false vacuum. The dynamical time is set by the oscillation period of such modes, so, in the absence of strong damping, we have  
$\tDyn\sim2\pi/m$.

The second time scale is the {\it thermalization time} $\tTh$ that controls how quickly the nucleating degrees of freedom exchange energy with the {\it thermostat}, which may be part of the Hamiltonian system or external to it. In the latter case, the thermostat
can be implemented explicitly through the stochastic terms in the equation of motion (see Eq.~(\ref{eoms})), in which case $\tTh$ is controlled by the dissipation coefficients. Taking again as an example the 1d stochastic mechanics, we have $\tTh\sim \eta^{-1}$ at weak and moderate damping and $\tTh\sim \eta/\omega_0^2$ at strong damping. Note that in both cases 
\be
\label{HierarchyOfScales1}
\tDyn\lesssim \tTh\;.
\ee
This hierarchy holds also in more complicated systems. For complex enough Hamiltonian dynamics, the role of the thermostat is played by
the many degrees of freedom that do not directly take part in the nucleation process. In field theory, these are short-wavelength modes of the decaying field, as well as all other fields in the system. In weakly coupled theories the thermalization time is typically much longer than the dynamical time, $\tDyn\ll \tTh$ (see e.g.\ the discussion in \cite{Pirvu:2024nbe}).  

The third time scale is the {\it decay time} itself, defined as $\tDec\sim \Gamma^{-1}$. If this is much longer than the thermalization time,
\be\label{HierarchyOfScales}
\tTh\ll\tDec \;,
\ee
the nucleating degrees of freedom have ample opportunities to randomize before the system leaves the region $\R$. 
Their phase-space distribution will thus be maintained over time.
Moreover, this distribution is expected to be independent of the initial conditions which get forgotten after time $\tTh$. 

Of course, for this last property to hold, the thermostat must contain enough energy to thermally populate the nucleating degrees of freedom, without any appreciable change in temperature. To derive the relevant criterion, we note that when the nucleating degrees of freedom absorb energy $\Es$, the thermostat temperature gets lowered by $\Delta T=-\Es/{\cal C}$, where ${\cal C}$ is the thermostat heat capacity. Not to affect the exponential suppression of the decay rate, $\Gamma_{\rm TST}\propto \exp(-\Es/T)$, this must obey $\Es \Delta T/T^2\ll 1$, yielding the condition
\be
\label{LargeBath}
{\cal C}\gg \frac{\Es^2}{T^2}\;.
\ee
This condition can be alternatively written as $\delta E_{\rm thermo.}\gg \Es$, 
where $\delta E_{\rm thermo.}=\sqrt{\cal C}T$ is the fluctuation of the thermostat energy in the canonical ensemble (i.e.\ assuming that the thermostat is itself embedded in an infinite thermal bath) \cite{landau1980statistical}.
For a thermostat with $N_{\rm thermo.}\gg 1$ degrees of freedom, we typically have ${\cal C}\propto N_{\rm thermo.}$, so the condition (\ref{LargeBath}) is stronger than a naive requirement $E_{\rm thermo.}\gg \Es$, where $E_{\rm thermo.}\sim N_{\rm thermo.}T$ is the total thermostat energy. Note that the stochastic dynamics (\ref{eoms}) with $\eta_{ij}\neq 0$ corresponds to a thermostat with infinite heat capacity.

The conditions (\ref{HierarchyOfScales}), (\ref{LargeBath}) appear to be sufficient for the existence of a well-defined thermal nucleation rate. They are naturally satisfied in many physical systems where the thermal bath is macroscopically large and $\tDec$ is exponentially long. For example, in cosmological first-order phase transitions, the thermal bath is provided by the primordial plasma in a Hubble patch and $\tDec$ is of order the inverse Hubble rate,
whereas $\tTh$ is much shorter, being determined by local perturbative interactions in the plasma.

On the other hand, the conditions (\ref{HierarchyOfScales}), (\ref{LargeBath}) can be violated in small systems, or if the nucleating degrees of freedom do not thermalize efficiently on the time scales of decay.
This can be relevant for laboratory experiments with cold-atom systems and Bose--Einstein condensates~\cite{Fialko:2014xba, Fialko:2016ggg, Billam:2021nbc,Song:2021pyy, Tian:2022dzv, Zenesini:2023afv, Jenkins:2023eez, Jenkins:2023npg, Darbha:2024srr, Zhu:2024dvz,Cominotti:2025qia}. The regime $\tTh\gtrsim \tDec$ also occurs in
direct numerical simulations of the decays~\cite{Pirvu:2023plk,Pirvu:2024nbe}.
In this case, an approximate notion of the steady-state rate may exist, depending on the model and the initial ensemble distribution. We will return to this question in Sec.~\ref{sec:finite}. From now on, we will assume that the conditions (\ref{HierarchyOfScales}), (\ref{LargeBath}) are satisfied, unless stated otherwise.

The thermality conditions (\ref{HierarchyOfScales}), (\ref{LargeBath}) do not mean that the nucleation rate coincides with the TST prediction (\ref{Gamma_eq_gen}). Indeed, the dynamical evolution of the TS takes a time of order $\tDyn$. If $\tDyn\ll \tTh$, then the relevant degrees of freedom do not have enough time to thermalize during nucleation. In other words, the steady-state distribution $\rho({\bf z})$ deviates from thermal in the vicinity of TS, and so does the probability flux. We are going to see that the correct expression for the steady-state rate is
\be
\label{GammaFull}
\Gamma=\GammaTST \cdot (1-R)\;,
\ee
where $\GammaTST$ is the TST rate (\ref{Gamma_eq_gen}) and
the {\it re-crossing probability} $R$ accounts for dynamical effects. It consists of two contributions,
\be \label{R}
R = R^{(+)}+R^{(-)} \;,
\ee
having the following meaning: $R^{(+)}$ is
the probability that a phase-space trajectory crossing the surface $\d\R$ 
from the false to true vacuum turns 
around and ends up back to the false vacuum. 
Similarly, $R^{(-)}$ is
the probability for a trajectory crossing $\d\R$
from the true to false vacuum to re-cross back and end up in the true vacuum. 
The expression (\ref{GammaFull}) is exact, modulo exponentially small corrections of order $\mathcal{O}(\e^{-E_{\rm ts}/T})$. 
Note that since the probabilities $R^{(\pm)}$ are non-negative, the thermal rate (\ref{GammaFull}) cannot exceed the TST rate. 

The probabilities $R^{(\pm)}$ can be practically evaluated as follows. Consider, for simplicity, the choice of the dividing surface $\d\R$ given by the equation $q_-=0$. 
Define statistical ensembles of systems with the initial data restricted to this surface and distribution functions 
\be\label{Rho_init2}
\rho^{(\pm)}(\textbf{q},\textbf{p};t=0) = \mathcal{N}\delta(q_-)\theta(\pm \dot q_-)|\dot q_-| \e^{-H(\textbf{q},\textbf{p})/T} \;,
\ee
where $\mathcal{N}$ is the normalization and $\dot q_-$ is understood as function of coordinates and momenta. 
Let these ensembles evolve for a time longer than $\tDyn$, but much shorter than $\tDec$. The re-crossing probabilities are then given by 
\be\label{Ppmdef1}
R^{(\pm)} = \lim_{\tPlat\ll t\ll \tDec}\int \diff \textbf{q}\diff \textbf{p}\,  \rho^{(\pm)}(\textbf{q},\textbf{p};t) \theta(\mp q_-) \;.
\ee
Here $\tPlat$ is the time, at which the integrals on the r.h.s. stabilize to a plateau. Depending on the system, it varies between $\tDyn$ and $\tTh$. 
The formulas (\ref{Ppmdef1}) are suitable for analytical (see Sec.~\ref{sec:app}) and numerical (see Sec.~\ref{sec:num}) evaluation of the dynamical prefactor. 
Importantly, the numerical evaluation requires evolving the ensembles until $\tPlat$, which takes exponentially less computing time than a direct simulation of decays from the metastable vacuum.
Thus, our method can be used numerically at strong exponential suppression when direct simulations are impossible.

\begin{figure}[t]
\begin{center}
\includegraphics[scale=0.35]{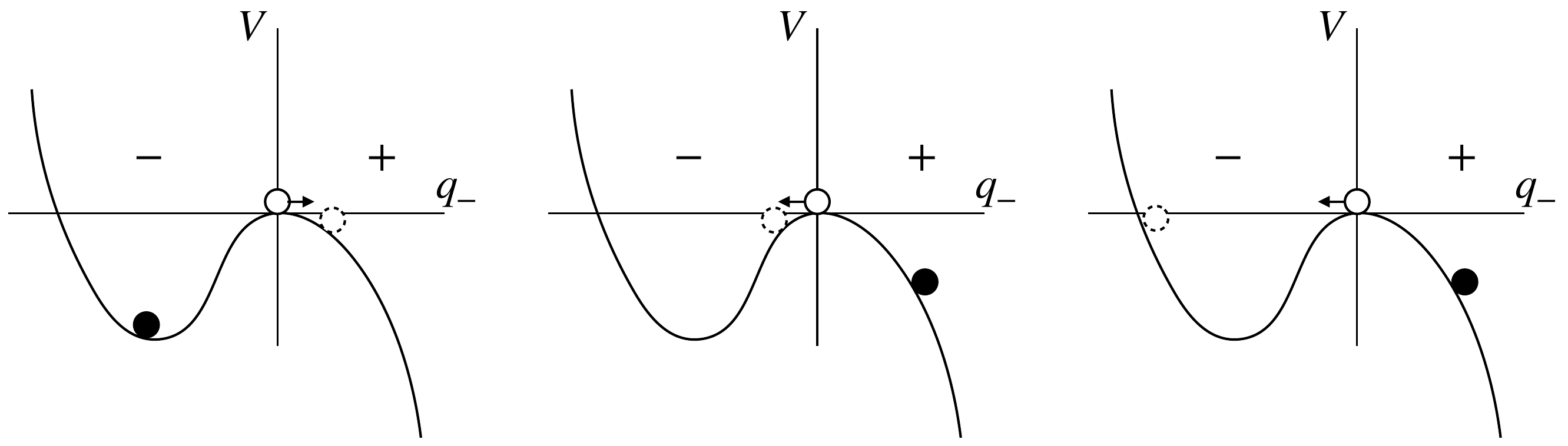} \\
\qquad\qquad\qquad $(a)$ \qquad\qquad\qquad\qquad \qquad \qquad  $(b)$\qquad\qquad\qquad \qquad \qquad\qquad $(c)$\qquad\qquad\qquad\qquad
\end{center}
\caption{\label{fig:particles} 
Different types of re-crossing motion. The particle starts from the top of the barrier   
(white solid circle) towards the true (\textit{a}) or false (\textit{b,c}) vacuum. Due to interactions with the heat bath, the particle turns around (white dashed circle), re-crosses the barrier and ends up in the region opposite its initial direction of motion (black circle). The re-crossing can happen promptly, before the particle leaves the vicinity of the barrier (\textit{a,b}), or after oscillating in the false vacuum region (\textit{c}). 
}
\end{figure}

Let us give a heuristic proof of Eq.~(\ref{GammaFull}). Assume that both the region $\R$ and its complement $\bar{\R}$ are in thermal equilibrium. Then the probability flux $I^{(+)}$ from $\R$ coincides with the TST rate.
However, not all of the
trajectories that contribute to this flux originate from the false
vacuum: some of them come from $\bar{\R}$ and turn
around, as illustrated in Fig.~\ref{fig:particles}b,c. The contribution of these trajectories into $I^{(+)}$
is equal to $I^{(-)}R^{(-)}$, where $I^{(-)}$ is the flux from $\bar{\R}$ to $\R$; due to detailed balance, this flux is the same as
$I^{(+)}$. To get the physical decay rate, we need to subtract this
contribution. In addition, we must subtract the contribution
$I^{(+)}R^{(+)}$ of trajectories that cross the barrier from $\R$ to $\bar{\R}$, but re-cross back into $\R$, as shown in Fig.~\ref{fig:particles}a.  
In this way we obtain Eq.~(\ref{GammaFull}). 

This argument is not completely satisfactory. First, the definition of the fluxes $I^{(\pm)}$ 
requires clarification, given that the full probability current ${\cal J}_I$ defined in (\ref{FP}) vanishes in thermal equilibrium.
An alternative current ${\bf J}=\dot{\bf z}\rho({\bf z})$ is not uniquely defined by a phase-space point in the stochastic case because of the random force, so an averaging procedure must be invoked to calculate a flux from it. Second, a single phase-space trajectory can cross the surface $\d\R$ back and forth multiple times. The above argument is unclear on how such multiple re-crossings must be counted. Third, the argument assumes thermal equilibrium between the metastable and stable phases, which, as emphasized in the Introduction, is an unphysical situation for a first-order phase transition. In the next subsections we present dynamical derivations of Eq.~(\ref{GammaFull}) free from these drawbacks and obtain expressions describing the approach of the rate to the steady-state value.

Before we proceed, a few comments are in order:

{\bf 1)} The re-crossing probabilities $R^{(\pm)}$ depend on the choice of the dividing surface $\d\R$, but so does $\GammaTST$, rendering the full expression (\ref{GammaFull}) independent of $\d\R$ in a wide range. 
In more detail, consider two surfaces $\d\R$ and $\d\R'$ containing the false vacuum and use them to evaluate the steady-state rates $\Gamma$ and $\Gamma'$ according to (\ref{GammaFull}). The two rates are the same as long as any phase-space trajectory escaping from the false to the true vacuum crosses both surfaces (possibly multiple times, see Sec.~\ref{ssec:trajectories}). 
Note that the surface does not need to pass through the TS. This answers the {\it Puzzle\,2} from Sec.~\ref{ssec:intro_puzzles}.
Of course, different choices of $\d\R$ can be more or less
convenient. For $\d\R$ passing
through the TS, the whole exponential suppression is
contained in $\GammaTST$.
If, on the other hand, we move $\d\R$ far from the TS, the equilibrium flux (\ref{Gamma_eq_gen}) increases exponentially, whereas the dynamical prefactor $(1-R)$ becomes exponentially small. Its numerical evaluation would then be impractical. Further, the perturbative, saddle-point approximation requires that $\d\R$ is chosen so that it is perturbatively close to the TS.

{\bf 2)} The approach leading to Eqs.~(\ref{GammaFull})--(\ref{Ppmdef1}) is general and applies both to stochastic mechanics and (Hamiltonian or stochastic) field theory, thereby addressing the {\it Puzzle\,3}. The only modification required in the latter case is connected to the translational zero modes. It is well-known how to deal with them in the statistical part of the rate \cite{Weinberg:2012pjx}. One introduces collective coordinates ${\bf X}_c$ that describe the spatial position of the critical bubble and integrates over them explicitly, producing a volume factor ${\cal V}$. The TST rate thus takes the form $\GammaTST={\cal V} \cdot \gamma_{\rm TST}$, where $\gamma_{\rm TST}$ is given by the expression (\ref{Gamma_eq_gen}), with the integral running over all perturbations of the critical bubble, except the collective coordinates ${\bf X}_c$, which are frozen at some fixed values.\footnote{Note that only ${\bf X}_c$ are frozen, but not their conjugate momenta, which are still included in the integral. Note also that the Jacobian from changing variables to collective coordinates appears in $\gamma_{\rm TST}$.} 
The same procedure applies to the dynamical prefactor. Since the re-crossing probability does not depend on the position of the critical bubble, one can arbitrarily fix ${\bf X}_c$ in the initial surface distributions (\ref{Rho_init2}), 
by inserting into $\rho^{(\pm)}({\bf q},{\bf p};t=0)$ an additional delta-function $\delta({\bf X}_c)$. The rest of the procedure and the formulas (\ref{Ppmdef1}) for the re-crossing probabilities remain the same.

{\bf 3)} There are different types of re-crossing trajectories that contribute to the dynamical factor $R$. We illustrate them in Fig.~\ref{fig:particles} using one-particle stochastic mechanics as an example.
The first type, shown in Figs.~\ref{fig:particles}a and \ref{fig:particles}b, is \textit{prompt re-crossings} whereby the trajectory turns around due to the kicks the particle (the amplitude of the negative mode of the critical bubble in field theory) receives from the stochastic force (other excitations of the system).
If there is no strong coupling, this is a perturbative contribution $R_\text{pert.}$ to the re-crossing probability.
We will see in Sec.~\ref{ssec:Langer} that, at leading order, it reproduces the classical Langer's dynamical prefactor \cite{Langer:1969bc} for systems with moderate to strong damping. 
Another type is \textit{sloshing re-crossings}. These are
trajectories that go back to the false vacuum, perform one or several oscillations around it, but then rebound and escape; see Fig.~\ref{fig:particles}c.
They constitute a non-perturbative contribution $R_\text{non-pert.}$.\footnote{In principle, there can also be sloshing trajectories visiting the vicinity of the true vacuum and rebounding back into the false vacuum. Clearly, this does not happen in field theory as long as the phase transition is sufficiently strongly first order, but can occur in mechanics with a few degrees of freedom. 
}
As we discuss in Sec.~\ref{ssec:Kramers}, they are responsible for the well-known suppression of the rate in 1d stochastic mechanics at weak noise \cite{kramers1940brownian}.

Sloshing re-crossings are generically present not only in mechanics, but also in field theory where they can give the dominant contribution to $R$ at moderate ratio of the temperature to the TS energy $T/\Es$ and weak damping.
They are particularly 
enhanced in theories with oscillons.
This is because the oscillon formed from the decay of the critical bubble survives for many field oscillations around the false vacuum, and during this time the system has a higher chance to decay. By Eq.~(\ref{GammaFull}), increasing $R$ decreases the ratio $\Gamma/\GammaTST$. This answers the {\it Puzzle $4$}: oscillons suppress the decay rate relative to TST by increasing the re-crossing probability.
In Sec.~\ref{sec:num} we will investigate in detail the sloshing re-crossings in field theory and their enhancement due to oscillons.

\subsection{Derivation using probability flux}
\label{ssec:surface-flux}

The rate formula has been traditionally derived in physical chemistry using linear-response theory and the fluctuation-dissipation theorem \cite{chandler1978statistical}. Here we give an alternative derivation that does not rely on the existence of thermal equilibrium between the metastable and stable phases, but considers instead the evolution of the initial distribution (\ref{Rho_initial}). We will restrict to the case of conservative dynamics, i.e.\ we set the dissipation coefficients $\eta_{ij}$ and noise $\xi_i$ in Eq.~(\ref{eoms}) to zero.
This is not a loss of generality because we can always extend the system to include the thermostat as its part. In particular, the Markovian dissipation and noise of Eq.~(\ref{eoms}) can be reproduced by coupling the system to an infinite set of harmonic oscillators \cite{zwanzig1973,CALDEIRA1983374,grabert1988,pollak1989theory,weiss2021quantum}. 
Different realizations of the white noise then correspond to different initial conditions for the oscillators, weighted with the thermal distribution.  
Note that for the Hamiltonian dynamics, we can replace the probability current ${\cal J}_I$ introduced in (\ref{FP}) by a simpler expression,
\be
 \mathbf{J}(\bz;t) =\dot{\bz}(\bz)\rho(\bz;t) \;, \label{J(t)}
\ee
because the divergence of the second term in ${\cal J}_I$ vanishes due to the anti-symmetry of the matrix $M_{IK}$. In Eq.~(\ref{J(t)}) we have assumed that the Hamiltonian is time independent, so that  
$\dot{\bz}(\bz)=\{\bz,H(\bz)\}$, where $\{\,,\,\}$ stands for the Poisson bracket,
is only a function of $\bz$ and not of $t$.

Under Hamiltonian evolution, the probability distribution changes according to
\be
    \rho(\bz;t) = \frac{1}{Z_\mathcal{R}}\e^{- H(\bz)/T}\theta\big(\bz(-t)\in\R\big) \;, \label{rho(t)} 
\ee
where 
$\bz(t)$ is the phase-space trajectory satisfying the initial condition $\bz(0)=\bz$ and we have used the reversibility of the Hamiltonian flow. 
Given this, we can evaluate the probability flux out of $\R$ as
\begin{align} \label{eq:Flux1}
    I_\R(t) &= -\frac{\d}{\d t} \dint_{\R}\diff {\bf z}\, \rho(\bz;t) 
    =\frac{1}{Z_{\R}}\int \diff {\bf z}\, \theta({\bf z}\in \R) \e^{- H(\bz)/T} \,
    \dot{\bf z}({\bf z}')\cdot\nabla' \theta({\bf z}'\in \R)
    \,,
\end{align}
where ${\bf z}'={\bf z}(-t)$ and $\nabla'$ is the gradient operator on the phase space ${\bf z}'$. We now change the integration variables from ${\bf z}$ to ${\bf z}'$. By the Liouville theorem, the Jacobian of this replacement is unity. Using further the conservation of energy along the Hamiltonian flow, $H({\bf z})=H({\bf z}')$, we obtain
\begin{align} \label{eq:Flux2}
    I_\R(t) &= \frac{1}{Z_{\R}}\int \diff {\bf z}'\, \theta\big({\bf z}'(t)\in \R\big) \e^{- H(\bz')/T} \,
    \dot{\bz}({\bz}')\cdot\nabla' \theta({\bf z}'\in \R)
    \,,
\end{align}
where we have used $\mathbf{z}=\mathbf{z}'(t)$. 
From now on we will omit prime on the integration variable. 

Next we observe that the gradient of the $\theta$-function localizes the integral to the boundary of the region $\R$,
\be
\label{gradtheta}
\nabla\theta(\bz \in \R)=-{\bf n}\,\delta(\bz \in\d\R) \;,
\ee
with ${\bf n}$ being the unit vector normal to the surface $\d\R$ and pointing outwards. Substituting into (\ref{eq:Flux2}), we obtain
\begin{align} \label{eq:Flux3}
    I_\R(t) &= -\frac{1}{Z_{\R}}\dint_{\d\R}\diff S\, {\bf n}\cdot \dot\bz(\bz) \,\e^{- H(\bz)/T} \,
\theta\big({\bf z}(t)\in \R\big)
    \,.
\end{align}
The flux is discontinuous across $t=0$ and satisfies
\bseq
\begin{align}
    &I_\R(t\to 0_-) = -\frac{1}{Z_{\R}}\dint_{\d\R^{(+)}}\diff S\, {\bf n}\cdot {\bf J}_{\rm TST}(\bz)
=-\GammaTST\,,\\
   &I_\R(t\to 0_+) = -\frac{1}{Z_{\R}}\dint_{\d\R^{(-)}}\diff S\, {\bf n}\cdot {\bf J}_{\rm TST}(\bz)
=\GammaTST\,,
\end{align}
\eseq
where we have introduced the auxiliary time independent ``TST current"
\be
\label{Jtherm}
{\bf J}_{\rm TST}(\bz)\equiv \frac{1}{Z_{\R}} \,\dot\bz(\bz) \,\e^{-H(\bz)/T}\;.
\ee
We recall that $\d\R^{(+)}$ and $\d\R^{(-)}$ are the parts of the boundary where this current is directed outwards ($\mathbf{n}\cdot\dot{\bz}(\bz)>0$) or inwards ($\mathbf{n}\cdot\dot{\bz}(\bz)<0$), respectively (see Fig.~\ref{fig:flux}).

Let us define the expectation values of a quantity $X(\bz;t)$ with respect to the \textit{surface flux ensembles},
\bseq
\begin{align}
\label{outaverage}
&\left\langle X(\bz;t) \right\rangle_{\d \R^{(+)}} \equiv\frac{\dint_{\d\R^{(+)} } \diff S\, {\bf n}\cdot{\bf J}_{\rm TST}(\bz)\,X(\bz;t)}{\dint_{ \d\R^{(+)} } \diff S\, {\bf n}\cdot {\bf J}_{\rm TST}(\bz)}\;,
\\
\label{inaverage}
&\left\langle X(\bz;t) \right\rangle_{\d \R^{(-)}} \equiv
\frac{\dint_{\d\R^{(-)} } \diff S\, \big(-{\bf n}\cdot{\bf J}_{\rm TST}(\bz))\big)\,X(\bz;t)}{\dint_{ \d\R^{(-)} } \diff S\, \big(-{\bf n}\cdot {\bf J}_{\rm TST}(\bz)\big) }\;,
\end{align}
\eseq
where the minus sign in the second equation ensures that the averaging weight is positive semi-definite. 
In both cases the denominator is simply $\GammaTST$. Let us further define 
\bseq
\label{R_R}
\begin{align}\label{R_R_plus}
    &R^{(+)}_{\mathcal{R}}(t) \equiv \left\langle \theta\big(\bz(t)\in\R\big) \right\rangle_{\d \R^{(+)}}\,,
    \\
    \label{R_R_minus}
    &R^{(-)}_{\mathcal{R}}(t) \equiv 
    \left\langle \theta\big(\bz(t)\notin\R\big) \right\rangle_{\d \R^{(-)}}
    =1-\left\langle \theta\big(\bz(t)\in\R\big) \right\rangle_{\d \R^{(-)}}\,.
\end{align}
\eseq
These quantities have an intuitive interpretation. Equation~(\ref{R_R_plus}) is the probability that a trajectory which starts from $\d\R$ in the outward direction at time $0_+$ has returned to $\R$ by time $t$. Whereas Eq.~(\ref{R_R_minus}) is the probability that a trajectory entering into $R$ at time $0_+$ has escaped from $\R$ by time $t$. Note that in both cases the trajectory can cross $\d\R$ multiple times in the time interval from $0_+$ to $t$. 

With these notations, Eq.~(\ref{eq:Flux3}) can be written as
\begin{align}
    \frac{I_\R(t)}{\GammaTST} &= 1 - R_{\mathcal{R}}^{(+)}(t) - R_{\mathcal{R}}^{(-)}(t)
    \,.\label{eq:FluxRatio}
\end{align}
Dividing out the probability $p_\R(t)$ of remaining inside $\R$, we can then \textit{define} the time-dependent decay rate as
\begin{align} \label{eq:RateFromFlux}
    \Gamma_\R(t) &\equiv \frac{I_\R(t)}{p_\R(t)} \,.
\end{align}
For sufficiently short times, $t\ll\tDec$, the normalization is unchanged,
\begin{align}
    p_\R(t) = 1 - \dint_{0}^t \diff t' I_\R(t') \approx 1\,,
\end{align}
up to exponentially small corrections, and thus can be dropped.
We arrive at the expression
\begin{align} \label{eq:RateFromFluxWithoutp}
    \Gamma_\R(t) = \GammaTST\left(1 - R_{\mathcal{R}}^{(+)}(t) - R_{\mathcal{R}}^{(-)}(t) \right)\,,
\end{align}
which has the same form as Eq.~(\ref{GammaFull}), except that the terms are time dependent. 
Note that, from their definitions above, both $R_{\R}^{(\pm)}(t)$ are positive, and hence lead to a reduction of the rate $\Gamma_{\mathcal{R}}(t)$ compared to $\GammaTST$. The time dependence is not necessarily monotonic though.

Assume that the system is thermal in the sense that it satisfies Eqs.~(\ref{HierarchyOfScales}), (\ref{LargeBath}). We then expect that on timescales, $\tDyn\ll t\ll \tDec$,
there should be a plateau in the time dependence of $\Gamma_\R(t)$, where the flux is stationary. We identify this plateau value with the physical thermal decay rate,
$\Gamma_\R(t)\approx\Gamma$. Denoting by $\tPlat$ the time when the plateau is reached and defining
\be \label{R_asympt}
R^{(\pm)}\equiv\lim_{\tPlat\ll t \ll \tDec} R_{\mathcal{R}}^{(\pm)}(t) \;,
\ee
we arrive at Eq.~(\ref{GammaFull}).
The upper bound in the limit in Eq.~(\ref{R_asympt}) is needed to distinguish re-crossings from the trajectories that get mixed in the metastable region and represent genuine decays of the false vacuum. The limit is well defined if the stationary-flux regime exists, a question we return to in Sec.~\ref{sec:upperBoundForDoubleScalingLimit}.

In applications, it is often convenient to define the region $\R$ in configuration space. Select the transition coordinate $q_-$ as described in Sec.~\ref{ssec:prelim}. 
Then $\textbf{n}\cdot \textbf{J}_{\rm th}\propto \dot q_-\e^{-H(\textbf{q},\textbf{p})/T}$ and the parts $\d\R^{(\pm)}$ of the dividing surface are selected by the theta-function $\theta(\pm \dot{q}_-)$. Thus the surface flux averages become averages with respect to the distributions (\ref{Rho_init2}). Equivalently, we can write
\be
R_{\R}^{(\pm)}(t)=\int \diff{\bf q}\diff{\bf p}\,\rho^{(\pm)}({\bf q},{\bf p};t) \,\theta(\mp q_-)\;.
\ee
Taking the large-time limit brings us to Eq.~(\ref{Ppmdef1}). 

The analysis of this section was based on starting from fully thermalized initial conditions in $\R$, Eq.~(\ref{Rho_initial}). Alternative choices of initial conditions will lead to different decay rates at early times, $t <\tPlat$. In the context of computer simulations of decays from metastability, a number of alternative initial conditions have been considered in the literature. For instance, Refs.~\cite{Valls:1990, Borsanyi:2000ua, Batini:2023zpi} use Gaussian initial conditions for perturbations around the metastable well.
In the context of stochastic dynamics, Refs.~\cite{Valls:1990,soskin2001noise}
start from the metastable minimum itself. In both these cases, the time-dependent decay rate starts from zero (or close to zero) and grows.
The steady-state rate $\Gamma$, on the other hand, is expected to be independent of this choice as long as the nucleating degrees of freedom can efficiently thermalize (conditions~(\ref{HierarchyOfScales}),~(\ref{LargeBath})).

\subsection{Derivation using trajectories}
\label{ssec:trajectories}

In this subsection, we give an alternative derivation of the rate formula (\ref{GammaFull}) by considering the phase-space trajectories contributing to false vacuum decay. This will allow us to establish a connection to the approach of Refs.~\cite{Moore:1998swa, Moore:2000jw, Moore:2001vf} (see also~\cite{bolhuis2002transition}) and to show that the result for the steady-state rate (\ref{GammaFull}) does not depend on the choice of the dividing surface $\d\R$. 
As explained in Sec.~\ref{ssec:surface-flux}, we can assume Hamiltonian dynamics
without loss of generality.

As argued in the Introduction and Sec.~\ref{ssec:prelim}, the equilibrium flux through the surface $\d\R^{(+)}$ used in the TST rate (\ref{Gamma_eq_gen}) contains trajectories that only recently entered into $\R$.
In the physical situation where e.g.\ a system is cooled slowly until the nucleation happens, such trajectories cannot arise: they correspond to initial conditions that were never part of the metastable phase. These trajectories must be excluded from the rate, as well as trajectories that, escaping the metastable phase through $\d\R^{(+)}$, return back to it through $\d\R^{(-)}$.   

\begin{figure}[t]
    \centering
\includegraphics[width=0.45\textwidth]{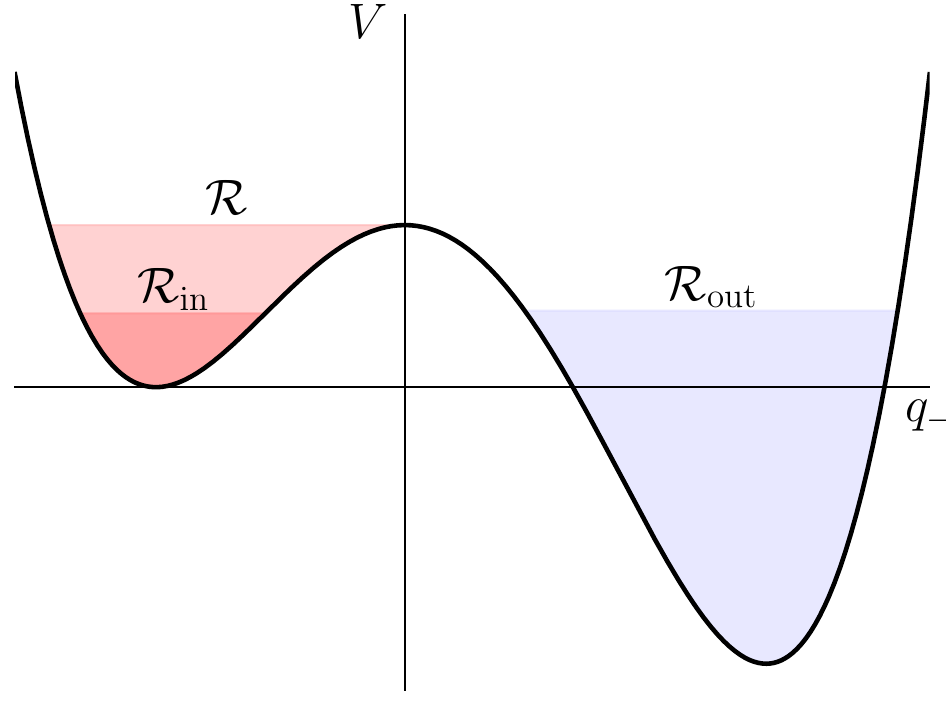}
\caption{Three regions of phase space, here shown on a one-dimensional potential: $\R_\text{in}\subset \R$, deep in the metastable phase; $\R$ the full metastable phase; and $\R_\text{out} \subset \bar{\R}$, where $\bar{\R}$ is the complement of $\R$, far outside the metastable phase. The nucleation rate is the flux through $\d\R$ carried by trajectories going from $\R_\text{in}$ to $\R_\text{out}$.}
\label{fig:potential_regions}  
\end{figure}

Following Moore, Rummukainen and Tranberg (MRT) \cite{Moore:2000jw, Moore:2001vf}, we define two regions in phase space: $\R_\text{in}$ and $\R_\text{out}$. For this purpose, let us assume that the true vacuum has finite energy, so that the metastable and stable phases can be in thermal equilibrium; for field theory, this requires putting the system in a finite box.\footnote{This assumption is not essential. It is just convenient for a general definition of $\R_\text{in}$, $\R_\text{out}$, but is not used in the rest of the argument. In specific systems, one can often find alternative definitions of $\R_\text{in}$, $\R_\text{out}$ that avoid this assumption.} Then $\R_\text{in}$ covers the equilibrium probability peak around the metastable well, and $\R_\text{out}$ covers the bulk of the probability outside the metastable phase, see Fig.~\ref{fig:potential_regions}. 
The boundaries of these regions, $\d\R_\text{in}$ and $\d\R_\text{out}$, are chosen to be sufficiently far from $\d\R$ that configurations on $\d\R$ are exponentially less likely than those on either $\d\R_\text{in}$ or $\d\R_\text{out}$.
The equilibrium probabilities of being in each region satisfy ${p_{\rm eq}(\R\setminus\R_\text{in}) \ll p_{\rm eq}(\R_\text{in})\ll p_{\rm eq}(\R_\text{out})}$.
Under time evolution, on average it takes an exponentially long time to escape either $\R_\text{in}$ or $\R_\text{out}$.

We wish to classify the trajectories going through $\d\R^{(+)}$. Starting from a point $\bz\in\d\R^{(+)}$ at time $t=0$ and evolving backward in time, the trajectory $\bz(t)$ will reach  either $\R_\text{in}$ or $\R_\text{out}$. Let us denote by $t_1$ the time when this first happens in the backward evolution; in other words, $t_1$ is the negative time with the smallest absolute value, such that $\bz(t_1)\in \R_\text{in} \cup \R_\text{out}$. Similarly, for the forward evolution, we define the smallest positive time $t_2$, such that $\bz(t_2)\in \R_\text{in} \cup \R_\text{out}$. 
Both $t_1$ and $t_2$ are of order the microscopic dynamical time $\tDyn$. Due to the deterministic nature of the Hamiltonian evolution, these times depend only on the phase space point $\bz$. We then define the {\it orbit}
\begin{align} \label{eq:orbits_static}
    o_\bz \equiv \left\{ \bz(t):\, \bz(0)=\bz;\, t\in [t_1(\bz), t_2(\bz)]\right\}\,,
\end{align}
and the {\it indicator function} which equals $1$ if the orbit goes from $\R_{\rm in}$ to $\R_{\rm out}$ and vanishes otherwise:
\begin{align} \label{eq:indicator_static}
    \thetato{in}{out} &=\theta\big(\bz(t_1)\in \R_\text{in}\big)\times \theta\big(\bz(t_2)\in \R_\text{out}\big)\,.
\end{align}
We can define $\thetato{out}{out}$, etc. analogously.
In the study of chemical reactions, $\thetato{in}{out}$ is called the characteristic function for the reaction~\cite{miller1974classical}.

\begin{figure}[t]
\begin{center}
\includegraphics[scale=0.42]{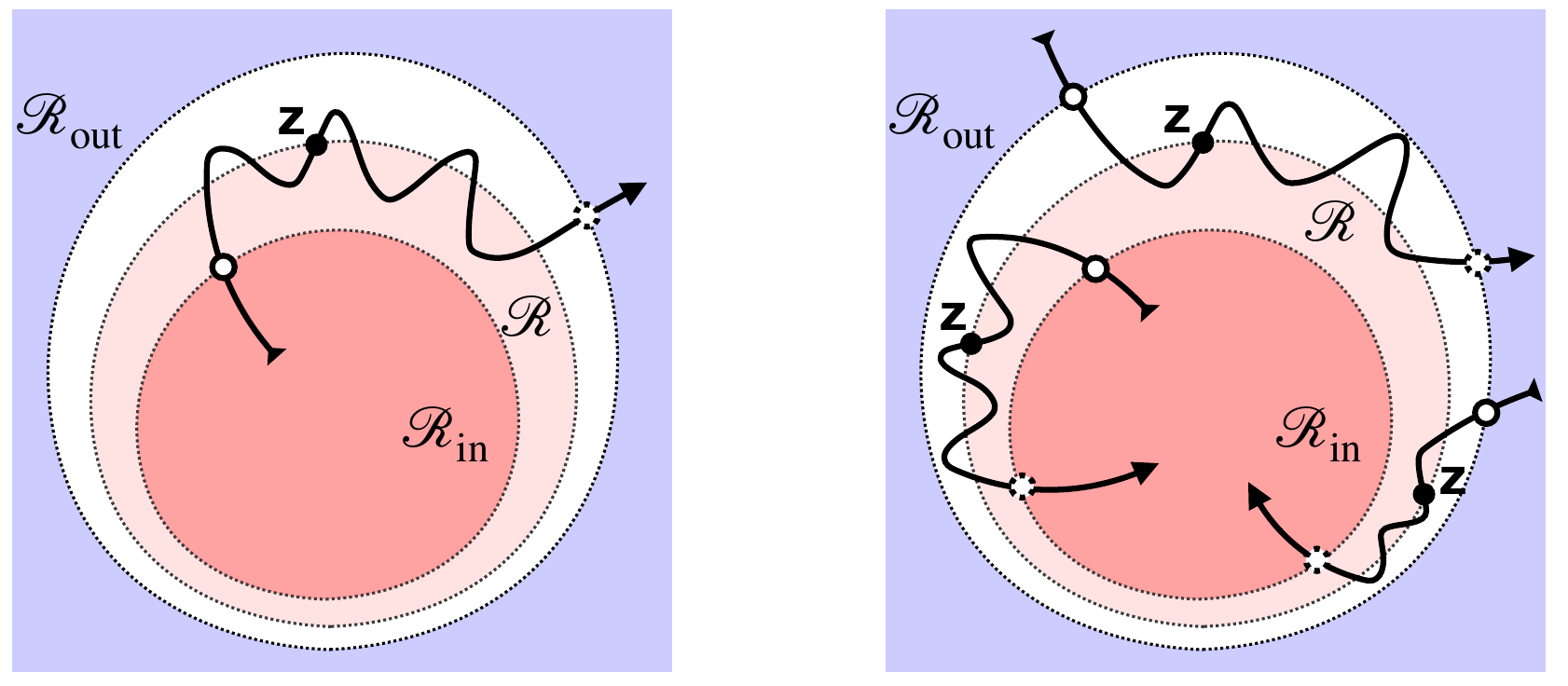}\\
$(a)$ \qquad\qquad\qquad\qquad\qquad\qquad $(b)$ 
\end{center}
\caption{
Different types of trajectories crossing the boundary of the metastable region $\R$ in the outward direction at a point $\bz$ marked by a black dot. The part of the trajectory between the solid and dashed white circles is the orbit $o_\bz$. The indicator function $\thetato{in}{out}$ equals unity for the orbit in panel (a) and vanishes for all orbits in panel (b). Only the trajectory in panel (a) contributes to the MRT rate (\ref{GammaMRT0}). 
}
\label{fig:trajectories}  
\end{figure}

The meanings of these orbits and indicator functions are illustrated in Fig.~\ref{fig:trajectories}. All trajectories shown in the figure cross $\d\R^{(+)}$ multiple times and thus contribute into the TST rate (\ref{Gamma_eq_gen}). However, only the trajectory in Fig.~\ref{fig:trajectories}a interpolates from the metastable to the stable region and thus describes an actual nucleation. Inserting the indicator function (\ref{eq:indicator_static}) into the integral in Eq.~(\ref{Gamma_eq_gen}) will keep the contributions of such trajectories and eliminate the contributions of the other ones shown in Fig.~\ref{fig:trajectories}b. This is not yet sufficient to get the correct rate, because a single trajectory of Fig.~\ref{fig:trajectories}a will contribute into the integral multiple times due to its multiple crossings of the surface $\d\R^{(+)}$. Due to the Liouville theorem, each crossing contributes the same amount of flux. Thus, to avoid double-counting, one divides the flux density at a point $\bz \in \d\R^{(+)}$ by $N_{\rm cross.}(o_\bz)$, the number of times the orbit $o_\bz$ crosses the surface $\d\R^{(+)}$. In this way we arrive at the expression~\cite{Moore:2000jw, Moore:2001vf},
\begin{equation}\label{GammaMRT0}
    \Gamma_\text{MRT} = \dint_{\d\R^{(+)}}\diff S\, \mathbf{n}\cdot {\bf J}_{\rm TST}(\bz) \, \frac{\thetato{in}{out}}{N_{\rm cross.}(o_\bz)} 
    =\Gamma_{\rm TST}\left\langle\frac{\thetato{in}{out}}{N_{\rm cross.}(o_\bz)}\right\rangle_{\d\R^{(+)}}
    \,,
\end{equation}
where ${\bf J}_{\rm TST}(\bz)$ is defined in (\ref{Jtherm}) and in the second equality we used the definition of the surface flux average (\ref{outaverage}). Since the quantity inside the average varies between $0$ and $1$, we conclude 
that $\Gamma_{\rm MRT}\leq \Gamma_{\rm TST}$.

An equivalent expression for $\Gamma_{\rm MRT}$ can be obtained by noting that the number of times an in-out orbit crosses the surface $\d\R^{(+)}$ is bigger by one than the number of times it crosses the surface $\d\R^{(-)}$. Thus, spurious contributions of the orbit into the flux through $\d\R^{(+)}$ are compensated by its negative flux through $\d\R^{(-)}$. This leads to the expression,
\begin{align}
    \Gamma_\text{MRT} = \Gamma_{\rm TST} \left(\big\langle\thetato{in}{out} \big\rangle_{\d\R^{(+)}}
    -\big\langle\thetato{in}{out} \big\rangle_{\d\R^{(-)}} \right)\;.\label{GammaMRT}
\end{align}
In the first term of Eq.~(\ref{GammaMRT}), we substitute the identity 
\be
\label{thetaid}
\thetato{in}{out} =1-
\thetato{in}{in} -
\thetato{out}{in} 
-\thetato{out}{out} 
\ee
and use $\left\langle \thetato{out}{out}\right\rangle_{\d\R^{(+)}}=\left\langle \thetato{out}{out}\right\rangle_{\d\R^{(-)}}$ which holds because an out-out trajectory crosses the surfaces $\d\R^{(+)}$ and $\d\R^{(-)}$ an equal number of times (see Fig.~\ref{fig:trajectories}b). In this way we arrive at
\begin{equation}\label{MRT-to-TST}
  \Gamma_\text{MRT}= \GammaTST\left( 1 - R_\text{MRT}^{(+)}  - R_\text{MRT}^{(-)}\right)\,,
\end{equation}
where we have introduced the notations:
\bseq
\label{eq:returns_MRT}
\begin{align} 
\label{eq:returns_MRT+}
    R_\text{MRT}^{(+)} &= \big\langle \thetato{in}{in} \big\rangle_{\d\R^{(+)}} 
   +\big\langle \thetato{out}{in} \big\rangle_{\d\R^{(+)}} =    
  \big\langle \thetato{any}{in} \big\rangle_{\d\R^{(+)}} \,,\\
\label{eq:returns_MRT-}
    R_\text{MRT}^{(-)} &=  
    \big\langle \thetato{out}{out} \big\rangle_{\d\R^{(-)}} 
   +\big\langle \thetato{in}{out} \big\rangle_{\d\R^{(-)}}=
    \big\langle \thetato{any}{out}\big\rangle_{\d\R^{(-)}} \,.
\end{align}
\eseq
The quantities $R^{(\pm)}_{\rm MRT}$ are manifestly positive semi-definite and time independent. 

The physical meaning of $R^{(+)}_{\rm MRT}$ ($R^{(-)}_{\rm MRT}$) is the probability that an orbit crossing $\d\R$ in the outward (inward) direction, no matter where it came from, turns around and gets back to $\R_{\rm in}$ ($\R_{\rm out}$). 
Since an orbit, by definition, is contained in the vicinity of $\d\R$ between $\R_{\rm in}$ and $\R_{\rm out}$, the re-crossings contributing to $R^{(\pm)}_{\rm MRT}$ happen before the trajectory leaves this vicinity. We called these prompt re-crossings in Sec.~\ref{ssec:prelim}. As long as the surface $\d\R$ passes through (or close to) the TS, the probabilities $R^{(\pm)}_{\rm MRT}$ can be found by perturbative expansion of trajectories in the neighborhood of the TS. The MRT rate (\ref{GammaMRT0}) is therefore an all-order perturbative expression. 

\begin{figure}[t]
\begin{center}
\includegraphics[scale=0.45]{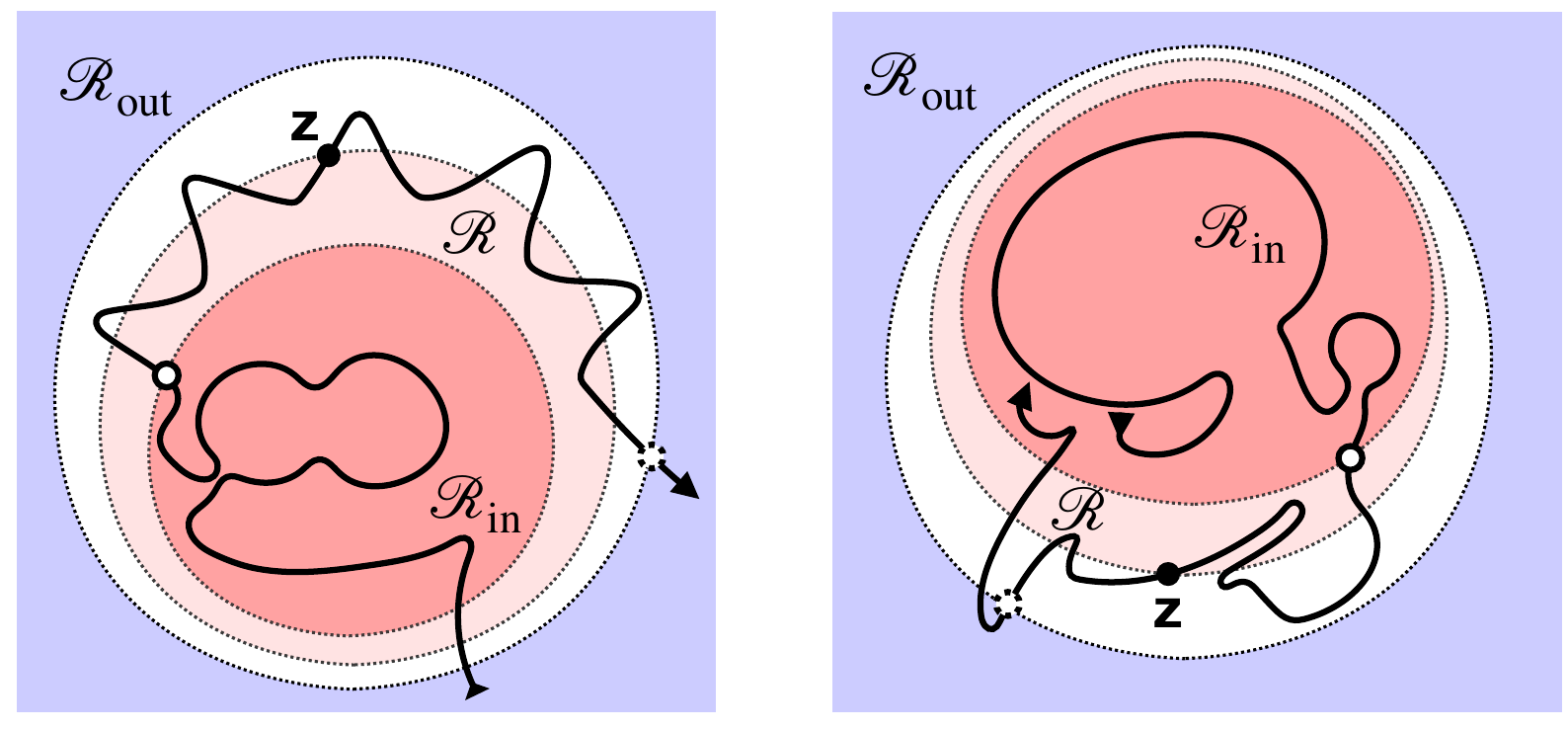}\\
$(a)$ \qquad\qquad\qquad\qquad\qquad\qquad $(b)$
\end{center}
\caption{
Sloshing re-crossing trajectories that contribute into the MRT rate (\ref{GammaMRT0}) but must be excluded from the steady-state rate $\Gamma$.
}
\label{fig:trajectories1}  
\end{figure}

Does $\Gamma_\text{MRT}$ equal the steady-state rate $\Gamma$? In general, Eq.~(\ref{GammaMRT0}) contains the contribution of trajectories, like those shown in Fig.~\ref{fig:trajectories1}. These trajectories leave the vicinity of $\d\R$, make a large excursions into $\R_{\rm in}$ or $\R_{\rm out}$ on a microscopic time scale, and cross $\d\R$ again. These are the sloshing re-crossings of Sec.~\ref{ssec:prelim}. The simplest example where they occur is one-dimensional stochastic mechanics at weak damping,\footnote{Recall that this is equivalent to a Hamiltonian system with the particle coupled to a bath of harmonic oscillators \cite{grabert1988, pollak1989theory}.} cf.~Fig.~\ref{fig:particles}. We will see below that field theory also admits sloshing solutions around the false vacuum (Fig.~\ref{fig:trajectories1}a); on the other hand, sloshing around the true vacuum (Fig.~\ref{fig:trajectories1}b) does not occur in field theory since, upon nucleation, the bubble of the new phase expands indefinitely.\footnote{Until collision with other bubbles, which is beyond the scope of our study.} Sloshing trajectories are non-perturbative in the sense that they are not accessible to the analysis focused on a small vicinity of the TS. Their contribution must be removed to get the true steady-state rate.

To achieve this, we modify the indicator function. Instead of focusing on the behavior of the orbit in the vicinity of $\d\R$, we take into account the whole portion of the trajectory extended from a point $\bz\in \d\R$ into positive and negative times until the sloshings die out. We denote this portion by $\tau_\bz$. It can include several oscillations around the vacuum, so its duration is in general longer than $\tDyn$. 
In any case, the sloshings are guaranteed to die out no later than the thermalization time $\tTh$, after which the trajectory becomes completely mixed in $\R_{\rm in}$ or $\R_{\rm out}$. We identify the disappearance time of sloshing re-crossings with the plateau time $\tPlat$ introduced in the previous subsections. This naturally leads us to the following definitions:
\bseq
\begin{align} 
\label{eq:sloshing_portion}
    \tau_\bz &\equiv \left\{ \bz(t):\, \bz(0)=\bz;\, t\in [-\tPlat, \tPlat]\right\}\,,\\
\label{eq:Sloshing_indicator}
    \thetatau{in}{out} &\equiv\theta\big(\bz(-\tPlat)\in \R_\text{in}\big)\times \theta\big(\bz(\tPlat\big)\in \R_\text{out})\,.
\end{align}
\eseq
Similarly to Eqs.~(\ref{eq:orbits_static}), (\ref{eq:indicator_static}), they depend only on the phase-space point $\bz\in\d\R$, for fixed~$\tPlat$. 

Accounting for multiple crossings of $\d\R^{(+)}$ by a single trajectory, we write for the rate,
\begin{equation}\label{GammaSteady}
    \Gamma= \dint_{\d\R^{(+)}}\diff S\, \mathbf{n}\cdot {\bf J}_{\rm TST}(\bz) \, \frac{\thetatau{in}{out}}{N_{\rm cross.}(\tau_\bz)} = \GammaTST \left\langle\frac{\thetatau{in}{out}}{N_{\rm cross.}(\tau_\bz)}\right\rangle_{\d\R^{(+)}}\,.
\end{equation}
Performing the same steps that lead from Eq.~(\ref{GammaMRT0}) to (\ref{MRT-to-TST}), we obtain, 
\begin{equation}\label{Gamma-to-TST}
  \Gamma = \GammaTST\left( 1 - R^{(+)}  - R^{(-)}\right)\,
\end{equation}
with
\be
\label{eq:returns_traj+-}
    R^{(+)} = \big\langle \thetatau{any}{in} \big\rangle_{\d\R^{(+)}} \,,\qquad
    R^{(-)} =  
    \big\langle \thetatau{any}{out}\big\rangle_{\d\R^{(-)}} \,.
\ee
Since any trajectory through $\d\R$ ends up either in $\R_{\rm in}$ or $\R_{\rm out}$ on the time scale $\tPlat$, the averages (\ref{eq:returns_traj+-}) coincide with the late-time limit of Eqs.~(\ref{R_R}) and we recover Eq.~(\ref{GammaFull}). 

Note that, unlike Sec.~\ref{ssec:surface-flux}, we have obtained Eq.~(\ref{GammaFull}) without considering the time evolution of the distribution function. Instead, we directly focused on trajectories that interpolate between $\R_{\rm in}$ and $\R_{\rm out}$ in the steady state. This confirms that the steady-state rate $\Gamma$ is independent of the choice of the initial distribution in the region $\R$, as long as the system has enough time to thermalize. Further, the trajectory approach makes it clear that $\Gamma$ is independent of the specific choice of the boundary $\d\R$. 
First, the indicator function $\thetatau{in}{out}$ depends only on the endpoints of the trajectory, so that its value on any given trajectory is unchanged as $\d\R$ is moved between $\d\R_\text{in}$ and $\d\R_\text{out}$. Thus, the set of trajectories which contribute to the rate remains the same for all $\d\R$ in this range. 
Second, the probability density is conserved along phase space trajectories, so that it is immaterial where along a given trajectory the flux is measured. 
This resolves \textit{Puzzle 2} from Sec.~\ref{ssec:intro_puzzles}.

\section{Analytical examples}
\label{sec:app}

The rate formula (\ref{GammaFull}) is accurate up to, perhaps, exponentially small corrections, suppressed relatively by ${\cal O}(\e^{-\Es/T})$. The formula provides a convenient basis for various expansions. We illustrate this point here by reproducing two classical results: Langer's perturbative leading-order dynamical prefactor for systems with moderate to strong dissipation \cite{Langer:1969bc}, and Mel'nikov--Meshkov's non-perturbative formula for the dynamical prefactor in one-dimensional stochastic mechanics with weak noise \cite{Melnikov:1986}. The re-crossing probability $R$ in these two examples is dominated, respectively, by prompt and sloshing re-crossings.

\subsection{Langer's rate from prompt re-crossings}
\label{ssec:Langer}

We consider a system with the Hamiltonian (\ref{Hamiltonian}) obeying the stochastic equations of motion~(\ref{eoms}). For simplicity, we assume the canonical kinetic term and diagonal dissipation matrix, 
\be
\label{Langer_simpl}
G^{ij}({\bf q})=\delta^{ij}~,~~~~g_i({\bf q})=0~,~~~~\eta_{ij}=\eta\,\delta_{ij}\;.
\ee
In this case, Langer's result \cite{Langer:1969bc} for the false vacuum decay rate can be written as \be
\label{GammaLanger}
\Gamma_\text{L}=\left(\sqrt{1+\frac{\eta^2}{4\omega_-^2}}
  -\frac{\eta}{2\omega_-}\right)\GammaTST\;,
\ee
where $\GammaTST$ is given by the leading-order saddle-point expression (\ref{GammaE}). The original derivation uses the flux-over-population method \cite{Hanggi:1990zz} which requires finding a stationary probability distribution function in the vicinity of the TS. In this approach, the corrections to the dynamical prefactor in (\ref{GammaLanger})
get out of control
at low damping, $\eta<\o_- T/\Es$ \cite{Pirvu:2024nbe}, apparently invalidating the saddle-point expansion~\cite{Ekstedt:2022tqk}.
 Our derivation based on Eq.~(\ref{GammaFull}) will ascertain that Eq.~(\ref{GammaLanger}) correctly captures the perturbative re-crossing probability $R_{\rm pert.}$ at the leading order in $T/\Es$ expansion even for weak damping. Langer's formula is thus valid, irrespective of the strength of dissipation, iff the non-perturbative contribution to the re-crossing probability $R_{\rm non-pert.}$ can be neglected. A derivation of Eq.~(\ref{GammaLanger}) using the reactive flux formalism, which is similar to ours, was given in Ref.~\cite{tannor1994}. 

We want to find the re-crossing probabilities $R^{(\pm)}$ using Eq.~(\ref{Ppmdef1}). To this end, we consider evolution of the flux ensembles (\ref{Rho_init2}) under the Fokker--Plank (FP) equation (\ref{FP}). At leading order in $T/\Es$, the thermal fluctuations of the phase-space variables around the TS are small, so we can restrict the Hamiltonian to quadratic terms. Then the solution of the FP equation factorizes,
\be
\label{FPfactor}
\rho^{(\pm)}({\bf q},{\bf p};t)=\rho^{(\pm)}_1(q_-,p_-;t)\,\rho_{\rm eq.}(\bz_{>0})\;,
\ee
where $p_-$ is the canonical momentum conjugate to $q_-$ and $\rho_{\rm eq.}(\bz_{>0})$ stands for the time independent thermal equilibrium distribution of the positive normal modes around the TS.\footnote{If zero modes are present, the r.h.s. of (\ref{FPfactor}) will contain one more factor $\rho_2({\bf X}_c,{\bf P}_c;t)$, where ${\bf X}_c$, ${\bf P}_c$ are the collective coordinates and momenta along the flat directions. The distribution $\rho_2({\bf X}_c,{\bf P}_c;t)$ is the same as for a cloud of particles with Maxwellian velocities spreading from the origin ${\bf X}_c$. The integral of this distribution over ${\bf X}_c$ and ${\bf P}_c$ is constant and equals unity, so it does not affect the probabilities (\ref{Ppmdef1}).
}
The re-crossing probabilities (\ref{Ppmdef1}) are determined by the first factor describing the distribution in the phase-space of the negative mode. Note that, within our approximation, the potential along the negative mode, $V\approx -\o_-^2q_-^2/2$, is invariant under reflection, implying that the two re-crossing probabilities are equal, $R^{(-)}=R^{(+)}$.
We focus on $R^{(+)}$. 

From now on, we restrict to the phase-space of the negative mode and denote $(q_-,p_-)$ simply by $(q,p)$. We also omit the sub- and superscripts on the distribution function. The FP equation reads
\be
\label{FPred}
\frac{\d\rho}{\d t}=\eta T \frac{\d^2\rho}{\d p^2} +\eta\frac{\d}{\d
  p}(p\rho)-\omega_-^2 q\frac{\d\rho}{\d p} -p\frac{\d\rho}{\d q}\;.
\ee
We need to solve it with the initial condition (see Eq.~(\ref{Rho_init2})),
\be\label{Pinit}
\rho(q,p;t=0) = \delta(q) \theta(p)\frac{p}{T}\e^{-p^2/2T} \;.
\ee
Note that (\ref{Pinit}) is normalized. 
The solution can be expressed using the Fourier transform 
\be 
\label{PpFour}
\rho(q,p;t)=\int \frac{\diff x \diff y}{(2\pi)^2}\, \tilde\rho(x,y;t)\e^{iqx+ipy} \;,
\ee
where $\tilde \rho$ is found in Appendix~\ref{AppB}.
We do not need the full solution here. Substituting (\ref{PpFour}) into the re-crossing probability (\ref{Ppmdef1}), we obtain
\be
\label{Ppmexpr}
R^{(+)}=\lim_{t\to+\infty}\dint_{-\infty}^{+\infty} \frac{\diff x}{2\pi i}
\frac{\tilde\rho(x,0;t) }{x-i\epsilon}\;,
\ee
where $\epsilon\to 0_+$.
We see that the probability depends on $\tilde \rho$ only at $y=0$.

Simplifying the solution from Appendix~\ref{AppB} at late times, we have
\be
\label{PspecAsymp}
\tilde \rho(x,0;t)\simeq \mathcal{F}_0\bigg(x\frac{\e^{-\l_-t}}{\l_+-\l_-}\bigg)
\exp\bigg[-x^2\frac{T\eta\l_+}{2\omega_-^2(\l_+-\l_-)^2}\e^{-2\l_-t}\bigg]\;,
\qquad t\gg \frac{1}{\l_+}\;,
\ee
where 
\be
\label{lambdapm}
\lambda_{\pm}=\frac{\eta}{2}\pm\sqrt{\omega_-^2+\frac{\eta^2}{4}}\;,\qquad
 \mathcal{F}_0(z)=\frac{1}{T}\dint_0^\infty \diff p\,p\,\e^{-p^2/2T -ipz}\;.
\ee
Substitution into (\ref{Ppmexpr}) then yields
\begin{equation}
R^{(+)}
  = \frac{\eta\l_+}{\omega_-^2} \dint_0^\infty \diff \zeta
  \,\zeta\,\e^{-\eta\l_+\zeta^2/2\omega_-^2}
  \dint_{-\infty}^{+\infty}\frac{\diff z}{2\pi i(z-i \epsilon)}\e^{-z^2/2-i\zeta z}
  \;.
    \label{Ppmexpr1}
\end{equation}
Integrating by parts in $\zeta$ and evaluating the remaining Gaussian integrals, we arrive at 
\be\label{Ppmexpr3}
R^{(+)}=\frac{1}{2}\left(1-\frac{1}{\sqrt{1+\eta\l_+/\omega_-^2}}\right)
=\frac{1}{2}\left(1-\sqrt{1+\frac{\eta^2}{4\omega_-^2}}+\frac{\eta}{2\omega_-}\right)
\;,
\ee
where in the second equality we used the explicit form
 of $\l_{+}$. Finally, substituting this expression into
Eq.~(\ref{GammaFull}) and taking into account that
$R^{(-)}=R^{(+)}$, we obtain
Eq.~(\ref{GammaLanger}). 

Note that the distribution function in this subsection has been restricted to a small vicinity of the TS and the re-crossing probabilities attain the limiting value right after a single dynamical time, see Eq.~(\ref{PspecAsymp}). This implies that the result (\ref{Ppmexpr3}) accounts for prompt re-crossings.

\subsection{Sloshing re-crossings in 1d mechanics}
\label{ssec:Kramers}

The perturbative re-crossing probability (\ref{Ppmexpr3}) vanishes in the limit $\eta\to 0$, which would imply that the steady-state rate becomes equal to the TST rate. On the other hand, in one-dimensional stochastic mechanics in this limit, the steady-state decay rate of a metastable state in fact tends to zero, implying that the re-crossing probability must go to $1$. This mismatch is accounted for by the non-perturbative contribution of the sloshing re-crossings. In this subsection we show that Eq.~(\ref{GammaFull}) reproduces the so-called Kramers turnover formula for mechanics with weak damping originally derived using the flux-over-population method by Mel'nikov and Meshkov \cite{Melnikov:1986} and refined by Pollak, Grabert and H\"anggi \cite{pollak1989theory}. This is a rare example where the sloshing re-crossings can be studied analytically. 

We consider the potential shown in Fig.~\ref{fig:potential} and assume $\eta\ll
\omega_-$. 
The kinetic term is assumed to be canonically normalized. The coordinate will be denoted simply by $q$. 
Since the noise is
weak, it produces only a small perturbation of the deterministic
motion. Then, a particle starting from the barrier $q=0$ in the positive direction will
not come back, so $R^{(+)}=0$. On the other hand, a particle starting
from $q=0$ backward will bounce from the rising potential on the left
and return to the vicinity of the barrier. If it returns with positive
energy, it will cross the barrier and escape. If it returns with
negative energy, it will make another oscillation, upon which it may
or may not escape, and so on.
The total turn-around probability $R^{(-)}$ is the sum of probabilities to
escape at every oscillation.

To assess the problem quantitatively, we decompose the
particle trajectory,
$q(t)=\bar q(t) +\hat q(t)$,
where $\bar q(t)$ and $\hat q(t)$ satisfy the deterministic equation and the linear stochastic equation for perturbations, respectively,
\begin{align}
\ddot{\bar q}+\eta\dot{\bar q}+\frac{\diff V}{\diff \bar q}=0\;,\qquad
\ddot{\hat q}+\eta\dot{\hat q}+\frac{\diff^2 V}{\diff \bar q^2}\hat q=\xi\;. 
\label{qhateq}
\end{align}
Without the noise, the second equation has two linearly independent
solutions, $\hat q_1(t)$ and $\hat q_2(t)$. 
We choose $\hat q_1(t)=\dot{\bar q}(t)$ and normalize the second solution, so that 
their Wronskian is $ \hat q_1\dot{\hat q}_2-\dot{\hat q}_1\hat
q_2=\e^{-\eta t}$.
Using these conventions, the stochastic part of the trajectory with 
the initial conditions $\hat q(0)=\dot{\hat q}(0)=0$ reads as follows,
\be
\label{hatqsol}
\hat q(t)=\dint _0^t \diff t'\,\xi(t')\,\e^{\eta t'} \big(\hat q_1(t') 
\hat q_2(t)-\hat q_2(t')\hat q_1(t)\big)\;.
\ee
Clearly, it is proportional to the noise $\xi(t)$ which vanishes in the limit $\eta\to 0$. We will return to the precise conditions under which this perturbation is small at  
the end of the section. 

The friction and random force acting on the particle
change 
its energy, 
which we also decompose into the deterministic and stochastic
parts,
\be
\label{Edecomp}
E(t)=\bar E(t)+\hat E(t)\;,\qquad \bar E(t)=\frac{\dot{\bar q}^2}{2}+V(\bar q)\;,
\qquad
\hat E(t)=\dot{\bar q}\dot{\hat q}+\frac{\diff V(\bar q)}{\diff\bar q}\,\hat q\;.
\ee
The deterministic part decreases due to friction,
\be
\label{Edeterm}
\bar E(t)=E_0-\dint_0^t
\diff t'\,\eta\, \dot{\bar q}^2(t')\;.
\ee
In the stochastic part, we substitute the expression for the
perturbation of the trajectory (\ref{hatqsol}). Simplifying the result by means of the relations
\bseq
\label{linerelat}
\begin{align}
  \label{linerelat1}
&\dot{\bar q}\dot{\hat q}_1+\frac{\diff V}{\diff \bar q}\hat q_1
=\dot{\bar q}\ddot{\bar q}+\frac{\diff V}{\diff \bar q}\dot{\bar
                       q}=\frac{\diff\bar E}{\diff t}=-\eta\dot{\bar q}^2\;,\\
  \label{linerelat2}
 & \dot{\bar q}\dot{\hat q}_2+\frac{\diff V}{\diff \bar q}\hat q_2=
  \dot{\bar q}\dot{\hat q}_2-(\ddot{\bar q}+\eta\dot{\bar
  q})\hat q_2
  =\hat q_1\dot{\hat q}_2-\dot{\hat q}_1\hat q_2-\eta\hat
  q_1\hat q_2=\e^{-\eta t} - \eta\hat q_1\hat q_2\;,
\end{align}
\eseq
we obtain,
\be
\label{Estochast}
\hat E(t)=-\eta\dot{\bar q}(t)\hat q(t)+\dint_0^t \diff t'\,\xi(t') \dot{\bar
  q}(t')\,\e^{\eta(t'-t)}\;.
\ee
Note that the first term is suppressed by a factor $\eta$, compared to the second one. 
Since $\hat E(t)$ is linearly related to the Gaussian noise $\xi(t)$, it is itself a Gaussian random variable.

Consider a collection of particles starting at $t=0$ from $q=0$ with
a given energy $E_0$. At $t>0$ the energy distribution shifts
and broadens. It is Gaussian, with mean $\bar E(t)$ and variance
$\langle \hat E^2(t)\rangle$:
\be
\label{PEE0}
{\cal P}(E;t,E_0)=\frac{1}{\sqrt{2\pi \langle \hat E^2(t)\rangle}}
\exp\bigg[-\frac{\big(E-\bar E(t)\big)^2}{2 \langle \hat E^2(t)\rangle}\bigg]\;,
\ee
where the dependence on $E_0$ on the r.h.s.\ is implicit inside $\bar E(t)$
and $\langle \hat E^2(t)\rangle$.
When the particles come back to the barrier after one oscillation
in the potential well,
the mean and variance of their energy
distribution take the form,
\be
\label{E1meanE1var}
\bar E_1=E_0-\ve~,~~~~\langle \hat E^2_1\rangle = 2T\ve+\ldots\;,
\ee
where
\be
\label{Eloss}
\ve=\eta \oint \dot{\bar q}^2(t) \,\diff t = \eta\oint p \,\diff q
\ee
is the energy loss in one oscillation.
In the last equality we have
assumed $E_0\ll \Es$ and expressed $\ve$ through
the action integral along the separatrix of the conservative system
(i.e.\ the trajectory with $E=0$).
Dots in the expression for the variance (\ref{E1meanE1var}) stand for
terms suppressed by powers of $\eta/\omega_-$.
Note that only the second term
from (\ref{Estochast}) contributes to the energy variance at leading order.

In the re-crossing problem the initial energy of the particles is not
fixed but is distributed according to Eq.~(\ref{Pinit}). To get the energy
distribution after one oscillation, we have to
take a convolution of (\ref{PEE0}) with the initial distribution,
\be
\label{PE1}
\P_1(E)=\dint_0^\infty \frac{\diff E_0}{T}\e^{-E_0/T}
\frac{1}{\sqrt{4\pi T\ve}}
\exp\bigg[-\frac{(E-E_0+\ve)^2}{4 T\ve}\bigg]\;.
\ee
The fraction of particles escaping after one oscillation is
$R_{1}^{(-)}=\dint_0^\infty \diff E \,\P_1(E)$.
On the other hand, the particles with negative energies remain in the
well and will come again to the barrier after a second oscillation
with the energy distribution
\be
\label{PE2}
\P_2(E)=\dint_{-\infty}^0 \frac{\diff E_1}{\sqrt{4\pi T\ve}}
\exp\bigg[-\frac{(E-E_1+\ve)^2}{4 T\ve}\bigg] \P_1(E_1)\;.
\ee
The probability to escape at the second oscillation is then
$R_{2}^{(-)}=\dint_0^\infty \diff E \,\P_2(E)$.
Continuing this reasoning, we write for the total escape probability,
\be
\label{Pescape}
R^{(-)}=\sum_{n=1}^\infty R_{n}^{(-)}=
\dint_0^\infty \diff E\,\sum_{n=1}^\infty \P_n(E)\;,
\ee
where the distributions $\P_n(E)$ are constructed inductively using
the convolution similar to Eq.~(\ref{PE2}).
Summation of this series is performed in Appendix~\ref{app:WH} with the result
\be
\label{Pmpfin}
R^{(-)}=1-\exp\bigg[\dint_{-\infty}^\infty
  \frac{\diff z}{\pi(z^2+1)}\ln\Big(1-\e^{-(z^2+1)\ve/4T}\Big)\bigg]\;.
  \ee
Substituting into Eq.~(\ref{GammaFull}) for the decay rate and using that $R^{(+)}=0$, we recover
the result of Ref.~\cite{Melnikov:1986}.

The limiting behavior of the expression (\ref{Pmpfin}) is $R^{(-)}\approx 1-\ve/T$ at $\ve/T\ll 1$. For a potential without large hierarchy of parameters, we can estimate $\ve\sim\eta \Es/\omega_-$, so the re-crossing probability indeed goes to $1$ at vanishing damping. On the other hand, $R^{(-)}$ vanishes exponentially for $\ve/T\sim \eta \Es/\omega_- T\gg 1$. 
This provides an idea of how many sloshings the particle makes before escaping. The escapes become exponentially improbable once the net energy loss exceeds $T$, which gives an estimate
\be
\label{nslosh}
n_{\rm slosh}\sim \frac{T}{\ve}\sim \frac{\omega_-T}{\eta \Es}\;.
\ee
Let us also estimate the total sloshing time $t_{\rm slosh}$. For this, we need the typical period of a single oscillation. Due to the smallness of the energy,
$|E|\sim T \ll \Es$,
the particle spends a long time in the vicinity of the barrier, so that its oscillation period gets logarithmically enhanced,
\be
\label{tosc}
t_{\rm osc}\sim \frac{1}{\omega_-}\ln \frac{\omega_-\Delta q}{\sqrt{|E|}}\sim \frac{1}{\omega_-}\ln \sqrt\frac{\Es}{T} \;.
\ee
Here $\Delta q\sim \sqrt{\Es}/\omega_-$ is the characteristic range, at which the potential changes significantly.
Multiplying this by (\ref{nslosh}), we obtain the sloshing time,
\be
\label{tslosh}
t_{\rm slosh}\sim \frac{1}{\eta}\cdot \frac{T}{\Es} \ln \sqrt\frac{\Es}{T}\;.
\ee
For an ensemble of particles that initially thermally populate the region $q<0$, the re-crossings will die out after $t_{\rm slosh}$ and the decay rate will approach its steady-state value. Thus $t_{\rm slosh}$
is identified with the plateau time $\tPlat$ defined in Sec.~\ref{ssec:prelim}. Note that, since $T/\Es\ll 1$, it is parametrically shorter than the thermalization time $\tTh\sim 1/\eta$.  

Finally, let us verify the consistency of our approximations. Since we have expanded the potential in Taylor series in the stochastic perturbation $\hat q(t)$, the latter must be smaller than $\Delta q$ within the time interval between two successive visits of the barrier by the particle. From Eq.~(\ref{hatqsol}) we have
\be
\label{qhatsquare}
\langle\hat q^2(t)\rangle=2\eta T\dint_0^t dt'\e^{2\eta t'}
\big[\hat q_1(t')\hat q_2(t)-\hat q_2(t')\hat q_1(t)\big]^2\;.
\ee
Denote the combination in the square brackets by $\hat q_{12}(t',t)$. As a function of $t'$, it satisfies the second equation in (\ref{qhateq}) with $\xi=0$ and obeys the conditions $\hat q_{12}(t,t)=0$, ${\d_{t'}\hat q_{12}(t',t)|_{t'=t}=-\e^{-\eta t}}$. 
Integrated backwards from $t'=t$,
the absolute value of $\hat q_{12}(t',t)$ stays roughly constant if 
$\frac{\diff^2 V}{\diff \bar q^2}>0$ and grows exponentially if $\frac{\diff^2 V}{\diff \bar q^2}<0$. The latter regime occurs when the unperturbed solution $\bar q(t')$ is in the vicinity of the barrier. Then the maximal value of $\hat q_{12}(t',t)$ over a single oscillation of $\bar q(t')$ can be estimated as 
\be
\label{Gbound}
|\hat q_{12}(t',t)|\lesssim \frac{\e^{-\eta t}}{\omega_-} \cdot\e^{\omega_- t_{\rm osc}}
\lesssim \frac{\e^{-\eta t}\Delta q}{\sqrt{|E|}}\;,
\ee
where we have used Eq.~(\ref{tosc}). Substituting into Eq.~(\ref{qhatsquare}), we obtain a bound,
\be
\label{qhatbound}
\langle \hat q^2(t)\rangle \lesssim 
(\Delta q)^2\,\frac{2T}{|E|}\cdot\frac{\eta}{\omega_-} \ln\sqrt{\frac{\Es}{T}}\;,\qquad t\lesssim t_{\rm osc}\;.
\ee
Since for a typical trajectory $|E|\sim T$, we conclude that the condition 
${(\eta/\omega_-)\ln\sqrt{\Es/T}\ll 1}$ is sufficient to ensure that the deviation of the trajectory from the mean remains small over a whole oscillation.

\section{Re-crossings in field theory}
\label{sec:num}

This section is dedicated to the study of the re-crossing probability $R$ in the stationary flux regime in weakly-coupled field theory. 
The investigation of how the system approaches this regime starting from the initial thermal distribution (\ref{Rho_initial}) in the false vacuum region is postponed to Sec.~\ref{sec:finite}.

\begin{figure}[t]
    \centering
    \includegraphics[width=0.85\linewidth]{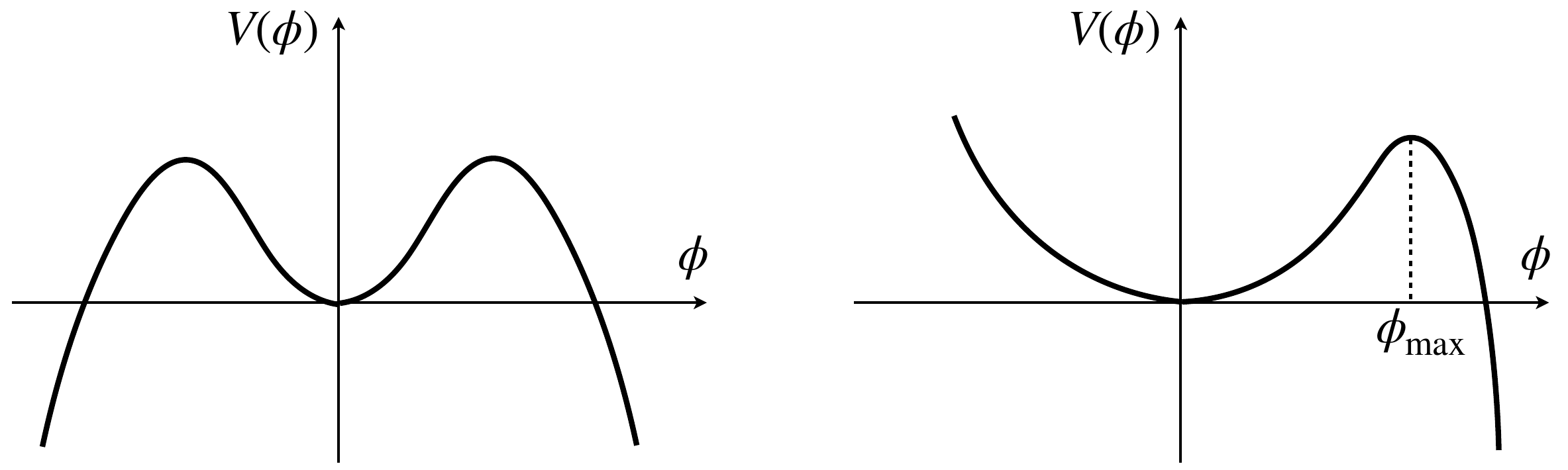}\\
    \qquad\qquad (a)  \qquad\qquad\qquad\qquad \qquad\qquad\qquad\qquad   (b) \qquad\qquad \qquad
    \caption{Scalar field potentials for the numerical investigation of re-crossing probability. (a): the quartic potential, (b): the Liouville potential.}
    \label{fig:potentials}
\end{figure}

Our goal is to analyze two types of contributions to $R$: prompt (perturbative) re-crossings and sloshing (non-perturbative) re-crossings.
The perturbative contribution $\Rpert$ can, in principle, be evaluated analytically~\cite{Ekstedt:2022tqk,ToAppear}, whereas 
the non-perturbative one, $\Rnp$, generally requires the use of numerical 
methods.\footnote{With a notable exception of 1d stochastic mechanics at weak noise considered in Sec.~\ref{ssec:Kramers}.}
We perform real-time classical lattice simulations of a single scalar field in $(1+1)$ dimensions, with the action
\be \label{S_gen}
    S=\int\diff t \, \diff x\left(\frac{(\d_\mu\phi)^2}{2} - V(\phi) \right) \, .
\ee
We consider two benchmark potentials: one with the quartic nonlinearity and another with the exponential (Liouville) nonlinearity,
\bseq 
\begin{align}
& V_1(\phi) = \frac{m^2\phi^2}{2} - \frac{\l \phi^4}{4} \;, \label{V4} \\
& V_2(\phi) = \frac{m^2\phi^2}{2} -\vk\,\e^{\phi} \;, \label{Vexp}
\end{align}
\eseq
where $\l,\vk>0$. The potentials are plotted in Fig.~\ref{fig:potentials}. The theory with the quartic potential is known to have oscillons~\cite{Pirvu:2024nbe}; on the other hand, the Liouville potential does not support oscillons at small enough $\vk$ (see below).

Based on the results of Sec.~\ref{sec:derivation}, we give a practical recipe of evaluating $R$ in numerical simulations, at a much lower cost than the direct decay simulations require. 
Similar numerical techniques have been used in physical chemistry within the reactive flux approach~\cite{berne1985molecular,Hanggi:1990zz}.
In the absence of external damping and noise, 
we will see that the non-perturbative contribution to $R$ can be surprisingly big and can significantly reduce the dynamical prefactor $(1-R)$ at 
$T/\Es\gtrsim 0.05$.
Moreover, in the model (\ref{V4}) the sloshing re-crossings dominate over prompt ones down to 
${T/\Es\sim 10^{-2}}$, due to the presence of long-lived oscillons.
Nevertheless, we will show that the non-perturbative re-crossing probability vanishes exponentially fast in the limit $T/\Es\to0$,
\be
\label{RnpNum}
R_{\text{non-pert.}}\propto \exp(-\alpha\,\Es/T)\;,\qquad \alpha\simeq 0.044\;.
\ee
Furthermore, we will see that $\Rpert=\mathcal{O}(T/\Es)$ if the dividing surface passes through the tree-level TS (critical bubble).
Thus, $\Gamma/\GammaTST\to 1$ as $T/\Es\to0$ in the conservative case. In the dissipative case we will see that in the limit of small $T/\Es$ the re-crossing probability approaches the perturbative prediction (\ref{Ppmexpr3}).

\subsection{Re-crossings with oscillons}
\label{ssec:non-pert}

We begin with the model with the quartic potential (\ref{V4}) and zero dissipation and noise.
Thermal vacuum decay in this model was studied in Refs.~\cite{Pirvu:2024nbe,Shkerin:2025hui,Hirvonen:2025hqn}.
The model with flipped sign of the interaction, $\l<0$, was used in Ref.~\cite{Boyanovsky:2003tc} for the study of thermalization (see also \cite{Destri:2004ck} for the study in $(3+1)$ dimensions).
The TS (critical bubble) profile and its energy are given by
\be \label{Sph_Esph}
\phi_{\rm ts}(x;x_0) = \sqrt{\frac{2}{\l}}\frac{m}{\cosh m(x-x_0)} \;, ~~~ \Es = \frac{4m^3}{3\l} \;,
\ee
where $x_0$ is the position of the center of the bubble.
The weak coupling expansion of the theory is controlled by the dimensionless parameter
\be \label{TT}
\hat{T} = \frac{\l T}{m^3} \;.
\ee
The theory is weakly coupled as long as $\hat{T} \ll 1$. From Eq.~(\ref{Sph_Esph}) we see that $\hat{T}= 4T/(3\Es)$, so this coincides with the domain of applicability of the saddle-point approximation $T/\Es\ll 1$. 

We discretize the model on a periodic spatial lattice with lattice spacing $a$ and size $L$
using 2nd-order accurate spatial derivatives.
We evolve the resulting multi\--dimensional Hamiltonian system numerically using the 4th order ope\-ra\-tor\--split\-ting pse\-udo\--spec\-tral scheme \cite{Pirvu:2024nbe}.
In most of our simulations we take $a=0.024/m$, $L=100/m$, corresponding to the number of points $N=4096$. The time-step is taken as $h\approx 0.8\,a$.\footnote{The high accuracy of the numerical scheme allows us to take a large time step $h\lesssim a$ \cite{Pirvu:2024nbe}.} We checked that our results do not depend significantly on $a$, $L$ and $h$.
Furthermore, we cross-checked some of the results with a second code using the 4th-order Forest--Ruth scheme~\cite{Forest:1989ez}.

To calculate the re-crossing probability, we must first select the dividing surface $\d\R$. It lies in the 
multi-dimensional (functional) space, and we choose it to pass through the field configuration 
$\phi_{\rm ts}(x;x_0)$. Due to the space translation symmetry, we can fix the TS location at $x_0=L/2$. Next, we
expand the field difference $\phi(x)-\phi_{\rm ts}(x;x_0)$ and momentum $\dot{\phi}(x)$ in the basis of linear perturbations around the TS. 
We choose the dividing surface as $q_-=0$, where 
$q_-$ is the amplitude of the negative mode of the TS. 
We further fix $X_c=0$, where $X_c$ is the amplitude of the zero mode associated to the spatial shifts of the TS. 
We pick the momentum of the negative and zero modes, as well as the amplitudes and momenta of all other modes, according to the surface-flux distribution 
(\ref{Rho_init2}). 
In particular, we pick the momentum of the zero mode corresponding to the center-of-mass velocity of the TS according to the Maxwell distribution~\cite{Pirvu:2023plk}; see Appendix~\ref{app:instate1} for details on the preparation of the initial ensemble.

Note that for the purpose of studying the leading-order perturbative and non-perturbative contributions to the re-crossing probability, it is sufficient to prepare the surface-flux ensemble in the quadratic approximation, by drawing the amplitudes of various modes from their respective Gaussian distributions.
The quadratic spectrum on $\d\R$ correctly reproduces the leading in $T/\Es$ behavior of $\Rpert$. 
This is because prompt re-crossings are due to interactions between the field modes during the system's evolution, and changing (perturbatively) the initial distribution does not change the leading non-zero contribution to this effect.

We consider the range of temperatures $10^{-3}\lesssim T/\Es\lesssim 10^{-1}$. For each temperature, we prepare a suite of $10^5$ trajectories\footnote{We run a large number of simulations since we want to analyze contributions to $R$ from different types of trajectories. For the purpose of just measuring $R$, much fewer simulations would suffice.} with the initial conditions drawn from the surface-flux distribution and evolve them for $t_\text{sim.}=200/m$,
which ensures that $\tDyn\ll t_\text{sim.} \ll \tDec $.\footnote{We estimate the decay time as $\tDec^{-1}=\Gamma=\GammaTST\cdot(1-R)$ with the one-loop TST rate computed in \cite{Pirvu:2024nbe}.}
If a decay is detected ($\max|\phi(x)|>10$), the simulation is stopped and the stopping time is identified as the re-crossing time.
If, during the evolution, the maximum of $|\phi(x)|$ over the lattice drops below $1$, we mark this trajectory as sloshing; all non-sloshing re-crossings are labeled as prompt. The fraction of decayed trajectories as a function of time gives us $R^{(-)}(t)$. We see numerically that it reaches a plateau during the simulation, i.e.\ $\tPlat<t_{\rm sim.}$. Measurement of $R^{(-)}(t_{\rm sim.})$ gives us the stationary-flux value $R^{(-)}$.

Similarly, one can find $R^{(+)}$ by evolving trajectories with the initial momentum of the negative mode directed towards the true vacuum and measuring the fraction of non-decayed trajectories. 
Our simulations showed no such trajectories, $R^{(+)}=0$.
This is consistent with the theoretical expectation~\cite{ToAppear}.

\begin{figure}[t]
\begin{center}
\includegraphics[scale=0.7]{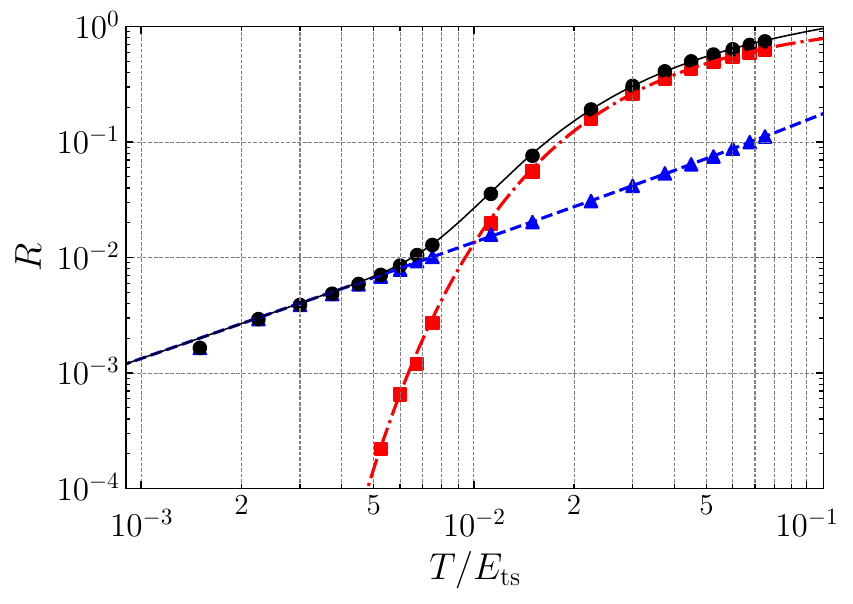}
\end{center}
\caption{The re-crossing probability $R$ measured in simulations in the model with the quartic potential (\ref{V4}) (\textit{black dots}). Perturbative (\textit{blue triangles}) and non-perturbative (\textit{red squares}) contributions to $R$ are also shown. The blue dashed line is the linear fit to the perturbative contribution, the red dash-dotted line is the exponential fit to the non-perturbative contribution, and the black solid line is the sum of the two.
}
\label{fig:returns_phi4}
\end{figure}

The re-crossing probability measured in simulations is shown in Fig.~\ref{fig:returns_phi4}.
The plot also shows separately the contributions from prompt and sloshing re-crossings. 
We observe that the perturbative contribution scales linearly with temperature, $R_\text{pert.}\propto T/\Es$. 
Fitting the data points gives the proportionality coefficient, $R_\text{pert.}\simeq 1.0\,\hat{T}$, in agreement with the result of Ref.~\cite{Hirvonen:2025hqn}.
The perturbative contribution dominates the re-crossing probability at small $T/\Es$. At $T/\Es\to 0$, the re-crossing probability vanishes and the steady-state decay rate coincides with the TST rate.

On the other hand, at moderate temperatures $ T/\Es\gtrsim 10^{-2}$, the re-crossing probability is dominated by the non-perturbative contribution.
Thanks to it, the total re-crossing probability becomes 
$R\approx 0.74$ at $T/\Es\simeq 0.1$. This explains the large discrepancy between the TST decay rate and the rate measured in direct simulations~\cite{Pirvu:2024nbe}.
Remarkably, even at $T/\Es\simeq 10^{-2}$
the effect from sloshing is still comparable to the prompt re-crossings.\footnote{We recall that the suppression typically considered in the context of cosmological phase transitions is $\Es/T\lesssim 170$\cite{Mazumdar:2018dfl}.}
We find that the sloshing contribution is well fitted by an exponential function (\ref{RnpNum}), which is shown by the red dash-dotted line in the plot. The exponential behavior is consistent with the non-perturbative nature of this contribution. 
Note, however, that the coefficient $\alpha$ in the exponent is much smaller than $1$, implying that  $R_\text{non-pert.}$ vastly exceeds the corrections of order ${\cal O}(\e^{-\Es/T})$ neglected in the derivation of the rate formula (\ref{GammaFull}).

\begin{figure}[t]
	\centering 
	\includegraphics[width=0.49\textwidth]{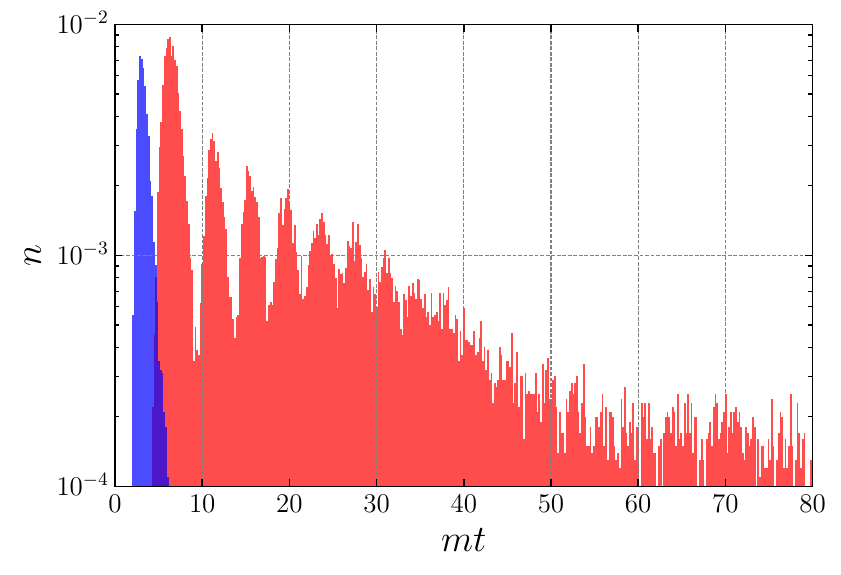}
	\includegraphics[width=0.49\textwidth]{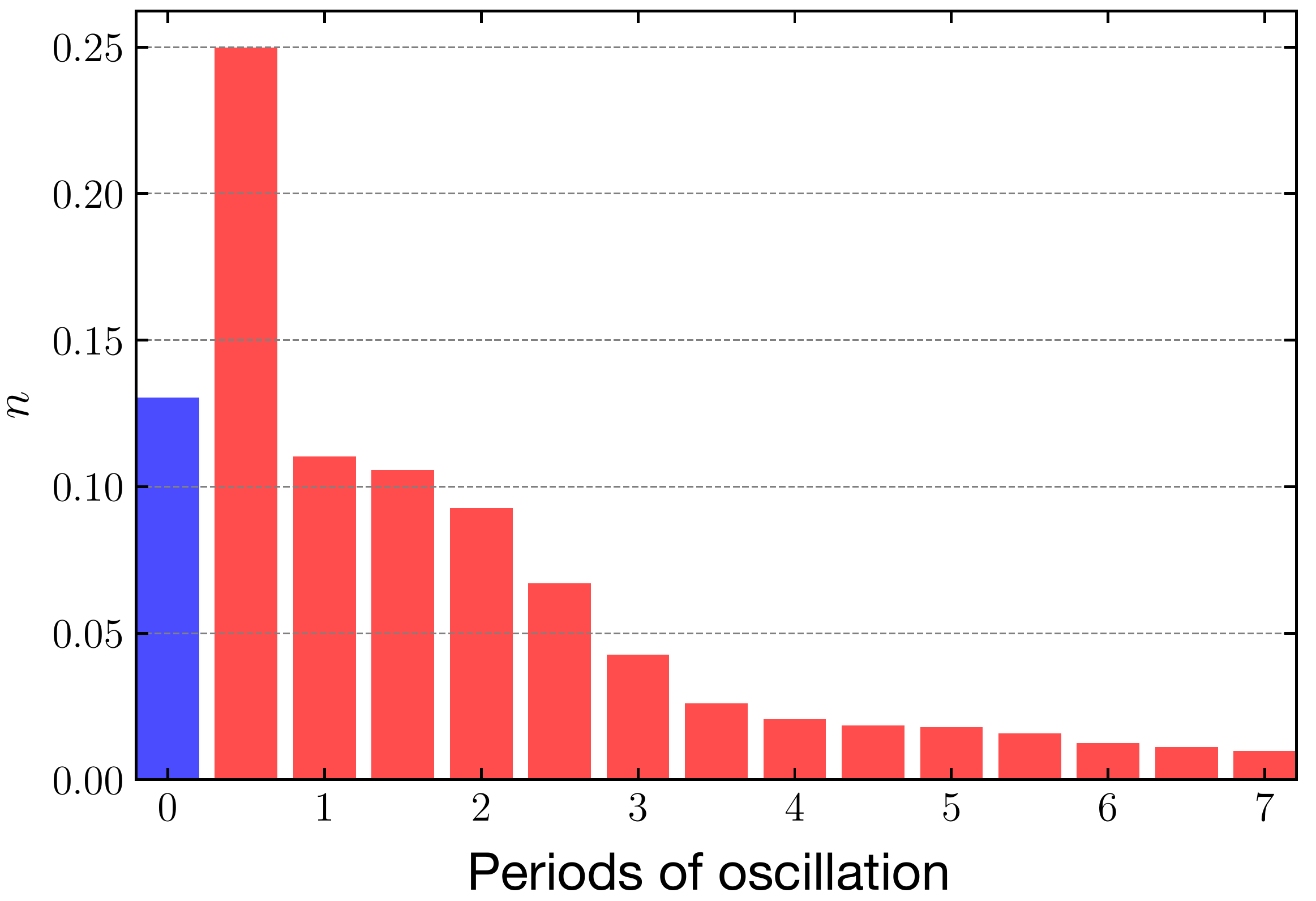}
	\caption{\textit{Left:} Number density of re-crossings as a function of time. Perturbative prompt re-crossings are shown in blue and non-perturbative sloshing re-crossings are shown in red. We take $\hat{T}=0.05$ corresponding to $T/\Es=0.0375$. The gap at small times is due to the decay trigger requiring $\max|\phi(x)|>10$. \textit{Right:} Distribution of re-crossings in number of oscillations across the false vacuum, at the same temperature.}
	\label{fig:density_phi4}
\end{figure}

Sloshing trajectories are clearly distinguishable from prompt re-crossings since they survive longer. To cross-check that we correctly identify the type of trajectories, we plot in the left panel of Fig.~\ref{fig:density_phi4}
the distributions of their re-crossing times. We take the temperature $\hat{T}=0.05$ as an example. We see that all prompt re-crossings occur within $t<5/m$, which is consistent with the expected barrier crossing time
$\sim 1/\o_-\cdot \log\sqrt{\Es/T}$, 
where $\o_-^2=3m^2$ for the quartic potential~\cite{Pirvu:2024nbe}.
On the other hand, sloshing trajectories start decaying later and have a broad time-distribution, with clear peaks corresponding to integer number of half-oscillations around the false vacuum.\footnote{The re-crossing times are correlated with half-periods, rather than full oscillation periods, because in the model with the potential (\ref{V4}) the field can decay both towards positive and negative $\phi$, thereby doubling the decay channels.}
The distribution of re-crossings in the number of oscillations is shown in the right panel of Fig.~\ref{fig:density_phi4}.

The surprisingly big non-perturbative contribution to the re-crossing probability can be explained by noticing that the quartic potential (\ref{V4}) supports long-lived nonlinear oscillatory solutions -- oscillons. 
Studies of the nucleation dynamics have revealed that in systems with long thermalization times, 
the TS (critical bubble) is always preceded by an oscillon~\cite{Pirvu:2023plk,Pirvu:2024nbe}.
Conversely, the oscillon is an intermediate product of the TS decay. Until the oscillon is dissipated, the system maintains an exponentially enhanced decay probability. 
This implies that oscillons enhance the re-crossing probability, hence, by Eq.~(\ref{GammaFull}), they {\it suppress} the dynamical prefactor in the decay rate. This clarifies the role of oscillons in dynamics of thermal vacuum decay and resolves the {\it Puzzle 4} raised in Introduction.

\subsection{Re-crossings without oscillons}\label{ssec:EliminatingOscillons}

\begin{figure}[t]
\begin{center}
\includegraphics[scale=0.7]{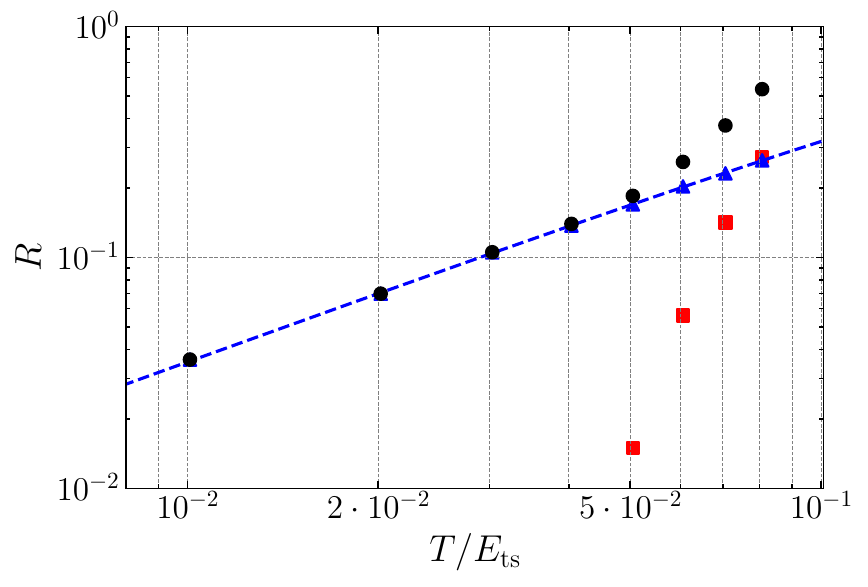}
\end{center}
\caption{The re-crossing probability for the model with the Liouville potential (\ref{Vexp}) featuring no oscillons: total probability ({\it black dots}), perturbative ({\it blue triangles}) and non-perturbative ({\it red squares}) contribution.
We take $\vk/m^2=0.05$. The blue dashed line is the linear fit to the perturbative contribution.}
\label{fig:returns_L}
\end{figure}

It is natural to wonder how critical are oscillons for the existence of sloshing re-crossings. To address this question, 
we repeat the analysis of Sec.~\ref{ssec:non-pert} for the model with the Liouville potential (\ref{Vexp}).
Assume a large hierarchy between the mass and the nonlinear coupling: $\ln(m^2/\vk)\gg 1$. Then the scalar potential is almost quadratic at $\phi<\phi_{\rm max}\approx \ln (m^2/\vk)$, while at $\phi>\phi_{\rm max}$ it abruptly falls down; see Fig.~\ref{fig:potentials}b.
This shape precludes the existence of oscillons in the thermal system.
The reason is that for the nearly quadratic potential the oscillonic frequency $\o_{\text{osc.}}$ tends to $m$; at the same time, in $(1+1)$ dimensions the oscillon size scales as $\ell_{\text{osc.}}\propto(m^2-\o_{\text{osc.}}^2)^{-1/2}$~\cite{Levkov:2022egq}, and the oscillon becomes too big to form.
Indeed, we have verified explicitly that our simulations do not contain any oscillons.
In simulations we take $\vk/m^2=0.05$ corresponding to $\ln(m^2/\vk)=3$. The numerical calculation of the TS energy at this value of $\vk$ gives $\Es\approx 24.8\,m$.

\begin{figure}[t]
	\centering 
	\includegraphics[width=0.49\textwidth]{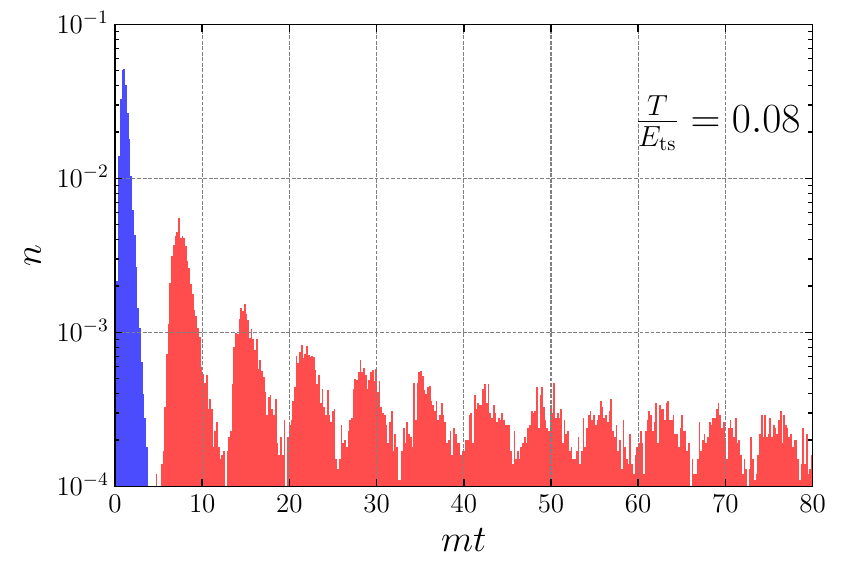}~~~~~~~~~~~~~
	\includegraphics[width=0.235\textwidth]{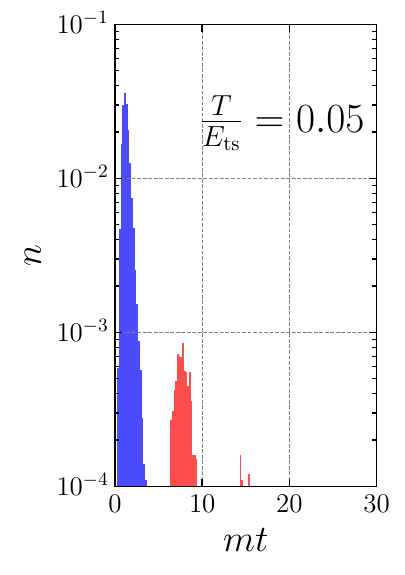}
	\caption{Number density of re-crossings as a function of time, in the theory with the Liouville potential (\ref{Vexp}), at two values of temperature: $T/\Es=0.08$ (right) and $T/\Es=0.05$ (left). 
    Prompt re-crossings are shown in blue and sloshing re-crossings are shown in red.}
	\label{fig:density_L}
\end{figure}

For each value of $T$ we prepare a suite of $10^5$ configurations using the mode decomposition around the TS, as explained in Sec.~\ref{ssec:non-pert} and Appendix~\ref{app:instate1}, evolve them for $t_\text{sim.}=200/m$, ensuring that $\tPlat\lesssim t_\text{sim.}\ll \tDec$, and measure the re-crossing probability. The result is shown in Fig.~\ref{fig:returns_L} in the temperature range 
$10^{-2}\lesssim T/\Es\lesssim 10^{-1}$. Comparing with Fig.~\ref{fig:returns_phi4}, we observe that the contribution from 
sloshing trajectories dies out at low temperature much faster than for the theory with the quartic potential, and at $T/\Es\sim 10^{-2}$ it is completely negligible. 
The re-crossings are then entirely given by the perturbative contribution well fitted by the linear function
$R_\text{pert.}\simeq 3.5\, T/\Es$.
On the other hand, at moderate temperatures, $T/\Es\sim 10^{-1}$, the non-perturbative contribution is still comparable to the perturbative one, and at $T/\Es>0.08$ it becomes dominant.

To further compare the models with and without oscillons, we plot in Fig.~\ref{fig:density_L} the time-distribution of re-crossings for the Liouville potential at two values of temperature. At $T/\Es=0.08$ (left plot), the non-perturbative contribution into the re-crossing probability is significant, manifesting itself by a long tail in the distribution of the re-crossing times. On the other hand, at a somewhat lower temperature $T/\Es=0.05$ (right plot) the sloshing re-crossings are strongly suppressed and completely disappear after the first oscillation. This should be contrasted with Fig.~\ref{fig:density_phi4} where, in the model with oscillons, sloshings dominate the re-crossing probability at temperature as low as $T/\Es=0.0375$.

To sum up, we find that even in theories without oscillons, sloshings persist and contribute to $R$ on par with the prompt re-crossings at moderately high temperature $T/\Es\sim 10^{-1}$. The absolute size of $R_\text{non-pert.}$ is, however, smaller and decreases faster with temperature than in theories with oscillons.

\subsection{Re-crossings with dissipation}
\label{ssec:diss_phi4}

Next, we study re-crossings in dissipative systems governed by the Langevin equation (\ref{eoms}) with thermal, additive white noise (\ref{wnoise}).
In Sec.~\ref{ssec:Langer} we derived the leading perturbative
expression for the re-crossing probability (\ref{Ppmexpr3}). Our goal here is to see
numerically 
how $R$ is modified by higher-order perturbative and non-perturbative contributions.

We take again the model with the quartic potential (\ref{V4}) and promote the Klein--Gordon equation of motion to the Langevin equation
\be \label{LangevinEq}
\ddot{\phi} + \eta \dot{\phi} - \phi'' + m^2\phi - \l \phi^3 = \xi \;,
\ee
where $\eta$ is the dissipation coefficient and $\xi=\xi(t,x)$ is a stochastic white noise satisfying
\be 
\langle \xi(t,x) \rangle = 0 \;, ~~~ \langle\xi(t,x)\xi(t',x')\rangle = 2\eta T\,\delta(t-t')\delta(x-x') \;.
\ee 
We solve Eq.~(\ref{LangevinEq}) numerically using the 3rd order stochastic, spectral, operator-splitting scheme described in Refs.~\cite{Pirvu:2024nbe,Shkerin:2025hui}. We use the lattice parameters listed in Sec.~\ref{ssec:non-pert} and the time step $h=2.5\cdot 10^{-3}/m$. For each value of $T$ and $\eta$ we prepare a suite of $2\cdot 10^4$ configurations\footnote{This suite is smaller than in Secs.~\ref{ssec:non-pert} and \ref{ssec:EliminatingOscillons}, since the stochastic systems are more computationally costly to evolve. It is sufficient for accurate measurement of the re-crossing probabilities $R^{(\pm)}$.}
as explained in Appendix~\ref{app:instate1}, evolve them for $t_\text{sim.}=200/m\gtrsim \tPlat$ and find the fraction of decayed trajectories. 

\begin{figure}[t]
	\centering 
	\includegraphics[width=0.49\textwidth]{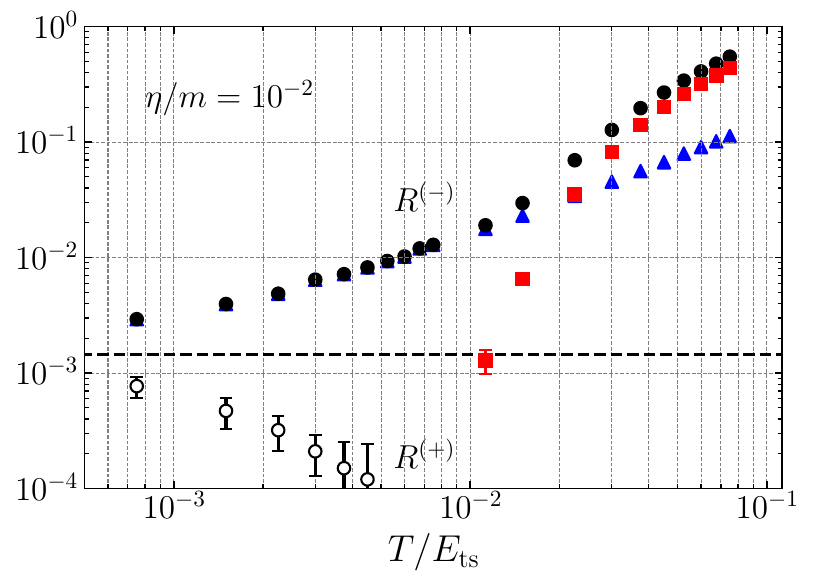}
	\includegraphics[width=0.49\textwidth]{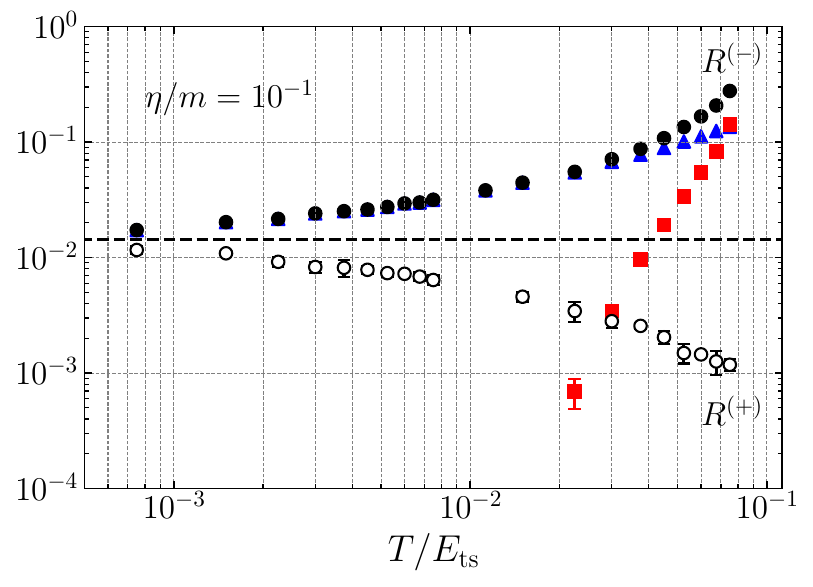}
	\caption{Re-crossing probabilities $R^{(\pm)}$ in the dissipative case in the theory with the potential (\ref{V4}), as functions of temperature and at two values of the dissipation coefficient:
    $\eta/m=10^{-2}$ (left) and $\eta/m=10^{-1}$ (right). 
    The black dashed line is the leading-order perturbative prediction (\ref{Ppmexpr3}). Filled and empty dots show $R^{(-)}$ and $R^{(+)}$, respectively. Blue triangles (red squares) stand for perturbative (non-perturbative) contribution into $R^{(-)}$.
    The error bars reflect the statistical uncertainty of the measurement.}
	\label{fig:returns_diss}
\end{figure}

Figure~\ref{fig:returns_diss} shows the probabilities $R^{(\pm)}$ in the range $10^{-3}\lesssim T/\Es\lesssim 10^{-1}$ and for two values of the dissipation coefficient, $\eta/m=10^{-2}$ and $10^{-1}$. 
We observe that the stochastic noise leads to non-zero forward re-crossing probability $R^{(+)}$; it is due entirely to prompt re-crossings.
In the limit $T/\Es\to 0$ at fixed $\eta$, the probabilities $R^{(\pm)}$ are equal and tend to the theory prediction (\ref{Ppmexpr3}).
At nonzero temperature, the probabilities deviate from this value, with $R^{(-)}$ growing and $R^{(+)}$ decreasing at higher $T/\Es$. At $T/\Es\gtrsim \eta/2m$, the forward probability $R^{(+)}$ becomes negligible, whereas the perturbative contribution into $R^{(-)}$ essentially coincides with its value for Hamiltonian dynamics (cf. Fig.~\ref{fig:returns_phi4}). One the other hand, the non-perturbative contribution is suppressed. For $\eta/m=10^{-2}$, it still dominates the re-crossings at moderately high temperature, but for $\eta/m=10^{-1}$ it is almost as small as for the Liouville potential (cf.~Fig.~\ref{fig:returns_L}).
This points again at the role of oscillons in sloshing re-crossings, since large dissipation damps the oscillons \cite{Pirvu:2024nbe}. 

Though we do not present explicitly the temperature dependence of $R^{(\pm)}$ for $\eta/m\gtrsim 1$, its form is clear from the previous discussion: The non-perturbative contribution is negligible, whereas the perturbative part is given by Eq.~(\ref{Ppmexpr3}) and is order-one, up to perturbative corrections of order ${\cal O}(T/\Es)$. 

\begin{figure}[t]
\begin{center}
\includegraphics[scale=0.75]{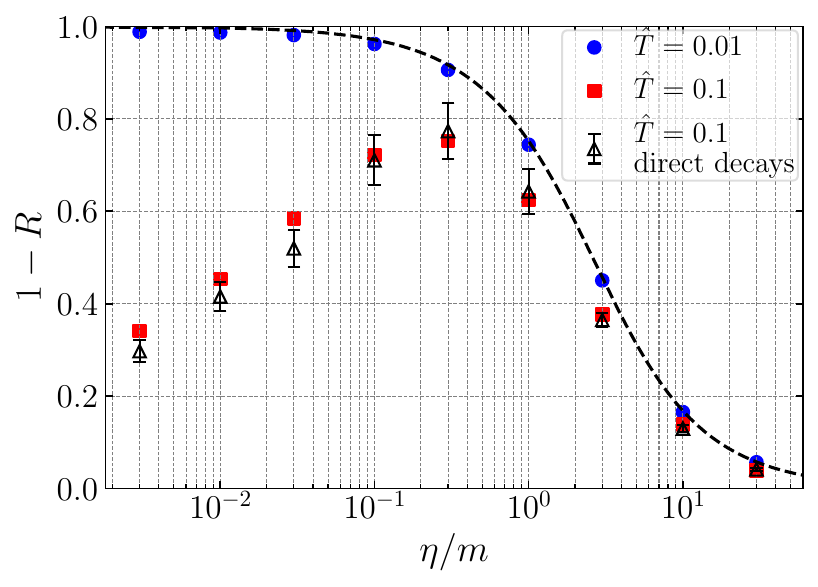}
\end{center}
\caption{Dynamical prefactor in the dissipative case, in the theory with the potential (\ref{V4}), as a function of the dissipation coefficient,
for temperature $\hat{T}=0.01$ ({\it blue dots}) and $\hat{T}=0.1$ ({\it red squares}).
The black dashed line shows the leading-order analytic prediction~(\ref{Ppmexpr3}).  
The black triangles show $\Gamma/\GammaTST$ at $\hat{T}=0.1$, with $\Gamma$ measured from the direct simulations of decays \cite{Shkerin:2025hui}. 
}
\label{fig:returns_etas}
\end{figure}

It is instructive to measure the dependence of the dynamical prefactor $(1-R)$ on the strength of the dissipation at fixed temperature. This dependence is compared in Fig.~\ref{fig:returns_etas} with the leading-order analytic result following from Eq.~(\ref{Ppmexpr3}). 
At the low temperature ${\hat{T}=0.01}$, the numerical data perfectly agrees with the analytic result, implying that the perturbative and non-perturbative corrections are negligible. 
On the other hand, at $\hat{T}=0.1$ the dynamical prefactor deviates from the prediction of Eq.~(\ref{Ppmexpr3}). 
Rather, the non-monotonic shape of $(1-R)$ resembles the behavior of the dynamical prefactor in 
1d stochastic systems \cite{kramers1940brownian, Hanggi:1990zz}.
This is not accidental: in both cases, this behavior is caused by sloshing re-crossings that are active at small dissipation, $\eta/m\lesssim 10^{-1}$.
At larger dissipation, sloshings disappear and the residual $\sim 20\%$ discrepancy between the numerical data and the analytic expression is due to 2-loop perturbative corrections.

At $\hat{T}=0.1$, when the exponential suppression is not too large, the nucleation rate can be measured in direct numerical simulations of decays. 
Black triangles in Fig.~\ref{fig:returns_etas} show the ratio of the rate measured in Ref.~\cite{Shkerin:2025hui} to the 1-loop TST rate. This ratio is in very good agreement with the dynamical prefactor $(1-R)$ determined from re-crossings.

\section{Finite-time behavior of the rate}
\label{sec:finite}

\subsection{Approaching the stationary regime}
\label{ssec:NumTest}

Here we investigate the time-dependence of the rate beyond the regime of stationary flux. We will combine ideas from Secs.~\ref{ssec:surface-flux} and \ref{ssec:trajectories} to single out the effects of the sloshing re-crossings. 
In doing so, we will construct a rate definition which interpolates
monotonously between the MRT rate (\ref{GammaMRT0}) at $t=0$ and the thermal rate (\ref{GammaFull}) at $t\gtrsim \tPlat$.

Perturbative re-crossings correspond to orbits, $o_\bz$ in Eq.~\eqref{eq:orbits_static}, crossing $\d\R$ multiple times. To eliminate perturbative re-crossings, we 
define a new phase space region $\Rref$ as a refinement of $\R$ based on the orbits.
The original boundary $\d\R$ is deformed so that each orbit only crosses $\d\Rref$ at most once.
This ensures that every orbit contributing to the flux through $\d\Rref^{(+)}$ is an in-out trajectory.
For such a refined region, the MRT rate coincides with the TST rate:
\begin{equation}\label{GammaMRT0ref}
    \Gamma_\text{MRT} = \Gamma_{\rm TST}\left\langle\frac{\thetato{in}{out}}{N_{\rm cross.}(o_\bz)}\right\rangle_{\d\Rref^{(+)}} \overset{~~\R\to\Rref}{=} \Gamma_{\rm TST}\,.
\end{equation}
This follows directly from Eq.~\eqref{GammaMRT0} applied to $\Rref$. 
In addition, $R_{\Rref}^{(+)}(t)$ in Eq.~\eqref{R_R_plus} is zero at $t>0$, assuming that there are no sloshing returns from the true vacuum. This follows from the same properties of $\d\Rref^{(+)}$ as the MRT = TST result. 
Consequently, the time dependence in Eq.~\eqref{eq:RateFromFluxWithoutp} simplifies to
\begin{align}\label{eq:Gamma_Rref}
    \Gamma_{\Rref}(t) = \Gamma_\text{MRT} \left(1 - R_{\Rref}^{(-)}(t) \right) = \Gamma_\text{MRT} \left\langle \theta\big(\bz(t)\in\Rref\big) \right\rangle_{\d \Rref^{(-)}}\,.
\end{align}

Given an initial region $\R$, it is possible to construct a corresponding refined region $\Rref$ in different ways.
Here, we propose the following algorithm:
\begin{itemize}
\item[\textit{(i)}] All points inside $\Rin$ are included into $\Rref$, by default; conversely, all points in $\Rout$ are excluded.
\item[\textit{(ii)}] For a point $\bz$ in the layer between $\d\Rin$ and $\d\Rout$, evolve the system forward and backward in time and decide on what type of orbit ${o}_{\bz}$ it lives.
\item[\textit{(iii)}] If the orbit is of the in-in type, the point $\bz$ is included into $\Rref$; if the orbit is of the out-out type, it is excluded.
\item[\textit{(iv)}] If the orbit is of the in-out type, then the point $\bz$ is included into 
$\Rref$ if its forward evolution crosses $\d\R$; otherwise, it is excluded.
\item[\textit{(v)}] If the orbit is of the out-in type, then the point $\bz$ is included into 
$\Rref$ if its forward evolution {\it does not} cross $\d\R$; otherwise, it is excluded.
\end{itemize}
It is straightforward to make sure that any point from the boundary of $\Rref$ defined this way belongs to an orbit that crosses $\d\Rref$ only once.

To perform our simulations, we take the spatially discretized version of the theory (\ref{S_gen}) with the quartic potential (\ref{V4}) and solve the resulting multi-dimensional Hamiltonian system numerically using 4th-order Forest--Ruth scheme~\cite{Forest:1989ez} with second-order accurate spatial derivatives; see Ref.~\cite{Hirvonen:2025hqn} for more details and tests of the code.
The parameters of the lattice are $a=0.1$, $h=0.02$ and $L=50$, in units of the inverse field mass. 
We take the moderate temperature $\hat{T}=0.1$, as defined in Eq.~(\ref{TT}), in order to be able to simulate the direct decays from the metastable state.

We define the metastable region $\R$ (not to be confused with $\Rref$) using a fully non-perturbative procedure based on a criticality condition described in Appendix~\ref{app:instate2}. 
The non-perturbative definition adopted here extends the perturbative definition of the boundary in Sec.~\ref{ssec:non-pert} to non-perturbative regions of the phase space akin to Ref.~\cite{Hirvonen:2025hqn}.
It is expected to be close to the MRT procedure \cite{Moore:2000jw, Moore:2001vf}, at least with the studied parameter points.
We also define $\Rin=\lbrace\max|\phi(x)|<1\rbrace$, $\Rout=\lbrace\max|\phi(x)|>3\sqrt{2}\rbrace$, satisfying $\Rin \subset \R\subset \bar{\R}_{\rm out}$; see Sec.~\ref{ssec:trajectories}.

\begin{figure}[t]
\begin{center}
\includegraphics[scale=0.75]{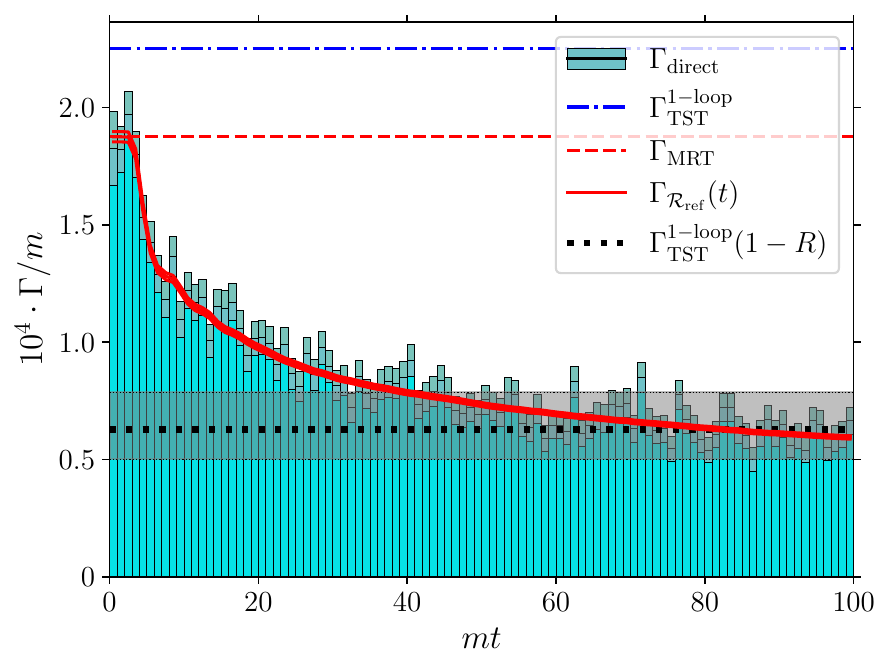}
\end{center}
\caption{The early-time behavior of the rate (\ref{eq:Gamma_Rref}) in the theory (\ref{S_gen}) with the quartic potential (\ref{V4}), at $\hat{T}=0.1$ and $Lm=50$. The rate is measured in direct simulations (blue bars) and calculated using the surface-flux ensemble (\ref{eq:Gamma_Rref}) (red solid line). The tips of the bars of different color and the width of the line show the statistical uncertainty. The dash-dotted blue line is the analytical 1-loop TST prediction; the dashed red line is the MRT rate (\ref{GammaMRT0}) determined from $\Gamma_{\rm MRT}=\Gamma_{\rm \Rref}(0)$; 
the dotted black line denotes the 
steady-state
rate (\ref{GammaFull}) computed using the analytical 1-loop TST rate and the numerical re-crossing probability measured at $t_\text{sim.}=100/m$.
The band accounts for the change in the measured value of $R$ when the measurement time is varied in the range $50/m<t_\text{sim.}<200/m$.
}
\label{fig:comp}
\end{figure}

Next, we define the initial state of the system for direct decays from the metastable minimum.
We prepare it using Hamiltonian Monte-Carlo sampling consisting of two steps.
First, we sample $\R$ in thermal equilibrium; see \cite{Hirvonen:2025hqn} for the description of the thermalization algorithm.
Note that here $\R$ is defined based on the configuration space alone, and the conjugate momenta for the configurations are sampled independently from their Gaussian distribution.
Sampled configurations $\phi(x)$ in the in-region, $\phi(x)\in \Rin$ are always accepted.
Second, if a sampled configuration
is not in the in-region, $\phi(x)\in \R \backslash \Rin$,
we evolve it, along its independently sampled conjugate momentum, backward in time and check if this evolution brings it into $\Rin$.
If this happens, the configuration is accepted;
otherwise it is rejected.
This way the prompt re-crossings are eliminated, but the sloshing re-crossings remain. 
Only the last exit of $\R$ is taken to contribute to the rate, to match exiting $\Rref$.

A surface-flux ensemble is more difficult to sample directly, because it lives on a lower-dimensional dividing surface in phase space. Further, we do not have a local definition of the surface $\d\Rref$.
Our procedure is first to construct the surface-flux ensemble on $\d\R$, which is detailed in Appendix~\ref{app:instate2}, and then to use this to construct that of $\d\Rref$.
The relation is given by the following:
An orbit crossing $\d\R$ $N_{\rm cross.}(o_\bz)$ times is overrepresented by this factor. The overcounting is corrected by multiplying the weight by $1/N_{\rm cross.}(o_\bz)$, which is the same factor as in the MRT rate, Eq.~\eqref{GammaMRT0}. Secondly, the time $t=0$ is fixed to the last crossing of $\d\R$ along the orbit, coinciding with the \textit{only} crossing of $\d\Rref$. Finally, the crossings corresponding to in-in and out-out trajectories are excluded from the boundary flux ensemble of $\d\Rref$.

We evolve a suite of $10^5$ trajectories for $t_{\text{sim.}}=100/m$ and compare the decay rate measured in direct simulations with the one 
evaluated
using the surface-flux ensemble.
The result is presented in Fig.~\ref{fig:comp}. It shows perfect agreement of the two 
measurements.
Note the difference between $\Gamma_{\Rref}(0)\equiv\Gamma_{\rm MRT}$ and the analytical prediction of the 1-loop TST rate \cite{Pirvu:2024nbe}. The difference is due to perturbative corrections to $\GammaTST$ and the perturbative dynamical prefactor (prompt re-crossings); see Eq.~(\ref{MRT-to-TST}).
After a short initial plateau for $mt\lesssim 3$ before sloshing re-crossings kick in,
the rate rapidly drops from the initial value due to the sloshing re-crossings.
Over the time of several $\tDyn$, the rate relaxes towards the steady-state value $\Gamma$, Eq.~(\ref{GammaFull}). The re-crossing probability in $\Gamma$ is found from simulations described in Sec.~\ref{sec:num}, using the same lattice parameters as for calculating $\Gamma_{\Rref}(t)$. Note the uncertainty band for $\Gamma$, reflecting the residual time-dependence of $R$ at $t\sim t_\text{sim.}$. This time-dependence comes mainly from the fact that the thermalization time condition (\ref{HierarchyOfScales}) is not satisfied 
in the system under study at $\hat{T}=0.1$, as shown in Ref.~\cite{Pirvu:2024nbe}.
We see, nevertheless, that at $t\gg \tDyn$ the rate still enters the quasi-stationary regime and the thermal formula (\ref{GammaFull}) is approximately valid.  
We will discuss the applicability of Eq.~(\ref{GammaFull}) in systems with slow thermalization in the next subsection.

\begin{figure}[t]
\begin{center}
\includegraphics[scale=0.7]{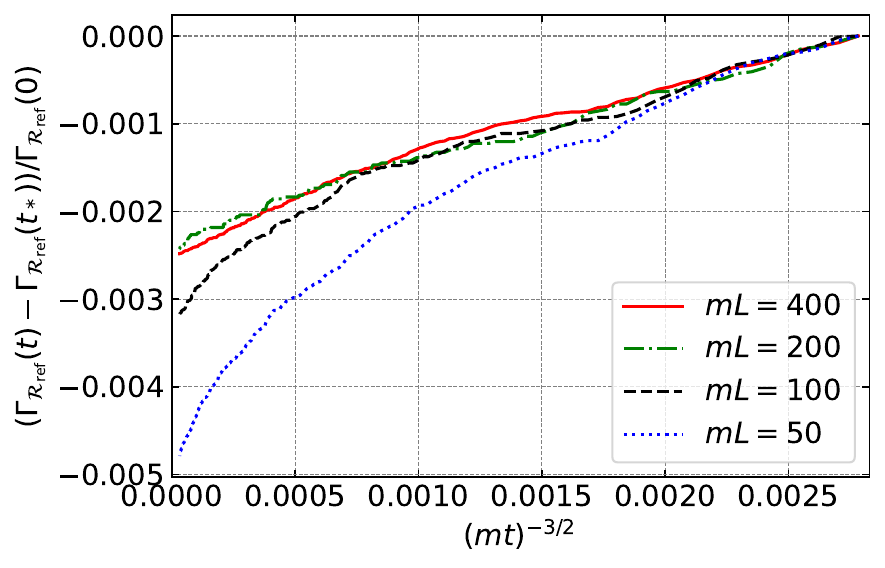}
\end{center}
\caption{Late-time behavior of the rate (\ref{eq:Gamma_Rref}), for different lattice size $L$. We take $\hat{T}=0.0125$ and $t_*=50/m$. The wiggles on the lines correspond to statistical fluctuations.
}
\label{fig:extrap}
\end{figure}

Let us see in more detail how the stationary-flux regime is approached. To this end, we calculate $\Gamma_{\Rref}(t)$ at a much lower temperature $\hat{T}=0.0125$, where the thermality condition (\ref{HierarchyOfScales}) is satisfied. 
The result is presented in Fig.~\ref{fig:extrap}, which shows $\Gamma_{\Rref}(t)$ normalized to the initial rate $\Gamma_{\rm MRT}$, as a function of $(mt)^{-3/2}$.
In order to see the asymptotics of the approach to the thermal rate, we subtract $\Gamma_{\Rref}(t_*)/\Gamma_{\rm MRT}$, where we choose $t_*=50/m$ as a benchmark time at which 
most sloshing re-crossings have already occurred; cf.~Fig.~\ref{fig:density_phi4}.
We perform simulations at several lattice sizes, $50\leqslant mL \leqslant 400$ to illustrate the magnitude of finite-size effects.
At large enough $L$, the convergence of the rate to a constant value $\Gamma$ appears to be governed by a power-law, $\Gamma_{\Rref}(t)-\Gamma \propto t^{-\a}$.
Fitting the data at the largest lattice size $Lm=400$ gives $\alpha\in[1.4,1.7]$.
Furthermore, the relative difference between $\Gamma$ and $\Gamma_{\Rref}(t_*)$ is $\sim 10^{-3}$, meaning that, practically, $t_*\gtrsim \tPlat$ at this (and lower) temperature.

We also see that at smaller $L$, the asymptotics of $\Gamma_{\Rref}(t)$ at large $t$ changes. 
Indeed, if the simulation time exceeds the size of the lattice with periodic boundary conditions and if dissipation is not strong enough, then large field fluctuations radiating from the collapsing TS will have enough time to travel around the box, converge at the original location and enhance the re-crossing probability by constructive interference.
This effect is unphysical if 
the actual system one wants to simulate is large;
note, however, that the relative difference between $\Gamma_{\Rref}(t_*)$ and 
$\Gamma_{\Rref}(t)$ for $t$ in the range $t_*<t<10^3/m$ shown in Fig.~\ref{fig:extrap} is still small.
On the other hand, if the system under investigation is itself relatively small, then the simulation of re-crossings must be performed with the lattice size matching the size of the system, in order to pick up the contribution to the re-crossing probability resulting from the finite-size effects.

\subsection{Quasi-stationary rate without thermality}
\label{sec:upperBoundForDoubleScalingLimit}

In Sec.~\ref{ssec:prelim} we formulated the conditions (\ref{HierarchyOfScales}), (\ref{LargeBath}) which state that the degrees of freedom relevant for nucleation efficiently thermalize. These conditions ensure the existence of a steady state with stationary nucleation rate. It is instructive to explore what happens if one or both of these conditions are violated. We are going to see that this gives rise to a new time scale $\tDrift\ll\tDec$. If $\tPlat\ll \tDrift$, the ensemble of systems starting from the initial distribution (\ref{Rho_initial}) enters at $\tPlat<t<\tDrift$ into a regime with approximately constant decay rate. The latter is given by Eq.~(\ref{GammaFull}) with the upper time limit in the expressions (\ref{Ppmdef1}) for the re-crossing probabilities being changed from $\tDec$ to $\tDrift$. This regime is, however, not truly stationary, since the decay rate drifts away from the constant value on time scale $\tDrift$ due to the statistical biasing of the ensemble. The decay probabilities thus do not follow the exponential distribution. This drift of the decay rate has been observed in beams of isolated molecular clusters \cite{klots1985evaporative,hansen2001observation,andersen2002thermionic,hansen2021decay}; it was called {\it classical Zeno effect} in Ref.~\cite{Pirvu:2024nbe}.

\subsubsection{Small systems}
\label{sssec:small}

To understand the origin of the new time scale, we first consider a situation when the condition (\ref{HierarchyOfScales}) is satisfied, but the condition (\ref{LargeBath}) is not. Imagine an ensemble of identical Hamiltonian systems. Each system has $N\gg 1$ degrees of freedom which we model by $N$ harmonic oscillators. The coupling between degrees freedom within a single system is strong enough, so that the system  efficiently thermalizes. However, individual systems are isolated, so their energies are conserved. Since the dynamics of each system is ergodic, the nucleation rate in the 
microcanonical ensemble of such systems with given energy $E$ is time independent, though it can depend on the energy. So, we write it as $\Gamma_{\rm mc}(E)$, where the subscript ``mc" stands for ``microcanonical". 

Consider now the ensemble (\ref{Rho_initial}). 
At early time, it evolves as described in Sec.~\ref{ssec:surface-flux}, and the flux approaches a plateau. Only a small fraction of systems decay by $t\sim \tPlat$, so the energy distribution of the systems remains the same as in the thermal state, i.e.\ Gaussian with the mean $N T$ and variance $N T^2$,
\be \label{RhoFieldT}
{\cal P}(E;0) =\frac{1}{\sqrt{2\pi N}\,T}\, \e^{-\frac{(E-N T)^2}{2N T^2}} \;.
\ee
We assume that the typical energy of the system is much higher than the TS energy,
\be
\label{highE}
NT\gg \Es\;,
\ee
but the variance is small, so that the condition (\ref{LargeBath}) is violated, 
\be
\label{smallVar}
NT^2< \Es^2\;.
\ee
At later time, the energy distribution gets biased due to the fact that systems with higher relevant energy decay faster and thus get washed out from the ensemble: 
\be \label{RhoEEE}
{\cal P}(E;t) = {\cal P}(E;0)\,\e^{-\Gamma_{\rm mc}(E)\,t} \;.
\ee
Defining, as before, the decay rate as the probability flux out of the metastable region in phase space, we have
\begin{equation}\label{eq:timeChangingRateLateTimes}
    \Gamma(t) = \langle \Gamma_{\rm mc}\rangle_{\cal P}
    =\frac{\int\dd E\, {\cal P}(E;t)\,\Gamma_{\rm mc}(E)}{\int\dd E\, {\cal P}(E;t)}
    \,.
\end{equation}
Note that the rate is now time dependent due to the depletion of the ensemble.
To estimate how fast it changes, we take the time derivative of Eq.~(\ref{eq:timeChangingRateLateTimes}): 
\begin{equation}
\label{eq:Gvariance}
    \dot{\Gamma}(t) = -\langle \Gamma_{\rm mc}^2 \rangle_{\cal P}
    +\langle\Gamma_{\rm mc}\rangle_{\cal P}^2\equiv-\sigma_{\Gamma_{\rm mc}}^2\,,
\end{equation}
where in the last equality we have introduced the notation for the variance of the microcanonical rate over the energy distribution. Note that $\dot\Gamma(t)$ is negative, implying that the rate decreases. The characteristic time of this decrease is estimated as 
\begin{equation}\label{eq:TimeScaleRateChange}
\tDrift \simeq \left.\frac{\Gamma}{\sigma_{\Gamma_{\rm mc}}^2}\right\vert_{t=0}\;.
\end{equation}

To proceed, we further assume that the microcanonical decay rate has the exponential form,
\be\label{GammaEField}
\Gamma_{\rm mc}(E)= A_{\rm mc}(E) \, \e^{-\Es/T_{\rm mc}(E)} \;, 
\ee
where $T_{\rm mc}(E)\equiv E/N$ is the effective ``microcanonical temperature" (cf. \cite{andersen2001}), and the prefactor $A_{\rm mc}(E)$ is a slowly varying function whose precise form is unimportant for us. This assumption is reasonable as long as the energies of the systems are well above the TS energy. Substituting this form into Eqs.~(\ref{eq:timeChangingRateLateTimes}), (\ref{eq:Gvariance}) and taking the integrals in the saddle-point approximation we obtain,
\be \label{GammaVarField1}
\Gamma|_{t=0}=A_{\rm mc}(NT)\e^{-\Es/T}\e^{\beta/2}\;,~~~
\sigma^2_{\Gamma_{\rm mc}}\big|_{t=0} =A^2_{\rm mc}(NT)\e^{-2\Es/T}\big(\e^{2\beta}-\e^{\beta}\big)\;,~~~ 
\beta\equiv\frac{\Es^2}{NT^2}\;.
\ee
From the first expression we observe that the rate averaged over the canonical ensemble is enhanced with respect to the microcanonical rate evaluated at the same temperature by a factor $\e^{\b/2}$. This difference is unimportant if the condition (\ref{LargeBath}) is satisfied, so that $\beta\ll 1$. For $\beta>1$, however, the enhancement is significant. It constitutes the so-called finite-heat-bath correction to the rate \cite{klots1989thermal,andersen2001} and is due to fast decay of the members of the ensemble at the high-energy tail of the energy distribution (\ref{RhoEEE}).

The second equation in (\ref{GammaVarField1}) shows that for small $\beta$ the variance $\sigma^2_{\Gamma_{\rm mc}}$ is also small and $\tDrift$ from Eq.~(\ref{eq:TimeScaleRateChange}) is longer than $\tDec$. This implies absence of drift on the decay time scales, confirming our assertion that, for ergodic systems, the condition $\b\ll 1$ is sufficient for existence of a steady-state decay rate.
On the other hand, for large $\b$, the variance exceeds the square of the rate and the drift time gets parametrically shorter than the decay time,
\be \label{TauField}
\tDrift\sim \tDec\, \e^{-\b}\;,
\ee
where we still define $\tDec$ through the inverse rate, $\tDec=\Gamma^{-1}|_{t=0}$.
Only at time scales shorter than $\tDrift$ can we view the rate as being stationary.

Does this result agree with the exact formula for the time-dependent rate (\ref{eq:RateFromFluxWithoutp}) obtained in Sec.~\ref{ssec:surface-flux}\,? The answer is yes. In evaluating the re-crossing probabilities $R_{\R}^{(\pm)}(t)$, we have to start from the surface flux ensembles describing thermal excitations with temperature $T$ on top of the TS. Consider the ensemble of systems evolving towards the metastable vacuum (i.e.\ contributing to $R_{\R}^{(-)}(t)$). After possible prompt and sloshing re-crossings, the systems remaining in the metastable well will form an approximately thermal distribution. However, the temperature of this distribution is higher than $T$ due to the energy release from the decay of the TS back to the metastable vacuum,
\be
T'\simeq T+\frac{\Es}{N}\;,
\ee
where we have assumed that inside each system the released energy gets equally distributed over all degrees of freedom.
This temperature shift, though small, enhances the decay rate of the newly formed distribution compared to that of the distribution with temperature $T$, 
\be
\Gamma_{T'}=\Gamma_T\,\e^\b\;.
\ee
The systems will leak from $\R$, i.e.\ they will re-cross the dividing surface, with the rate $\Gamma_{T'}$, leading to the drift of $R_{\R}^{(-)}(t)$ on the time scale $\Gamma_{T'}^{-1}$. This brings us back to Eq.~(\ref{TauField}). Previous discussion makes clear that Eq.~(\ref{TauField}) remains valid also for systems that cannot be modeled by a collection of harmonic oscillators: the only thing one needs to do is to replace $N$ in the definition of $\b$ (see Eq.~(\ref{GammaVarField1})) by the system's heat capacity ${\cal C}$.

One might entertain the possibility that at $t\gg \tDrift$ the rate would stabilize to a new steady-state value. However, this does not happen. To see this, we return to the expression (\ref{eq:timeChangingRateLateTimes}) and evaluate it at late time using the saddle-point approximation. It is more convenient to work with the probability flux $I(t)$ which stands in the numerator of Eq.~(\ref{eq:timeChangingRateLateTimes}), instead of the rate. We obtain,
\be
\label{verylaterate}
I(t)\simeq \frac{1}{t\sqrt{1+\b}}\exp\bigg\{-\frac{1}{2\b}\Big[
\ln\big(A_{\rm mc}(NT)\e^{-\Es/T}\,t\big)\Big]^2-1\bigg\} \;.
\ee
This expression is valid at $\tDrift\e^{\b/2}\ll t\ll \tDrift\e^{5\b/2}$ and in deriving it we have neglected relative corrections of order $\b T/\Es$ in the exponent. We observe the characteristic $1/t$ behavior of the flux, typical for emission from molecular clusters \cite{hansen2001observation,andersen2002thermionic,hansen2021decay}, slowly modulated by the logarithmic dependence in the exponent. This is very different from the exponential decay corresponding to a steady-state rate. 
One can check that if $\b\gg 1$, most of the original ensemble decays by the end of the regime (\ref{verylaterate}).

\subsubsection{Systems with slow thermalization}
\label{sssec:notherm}

Finally, we discuss systems where thermalization is not efficient enough to satisfy the condition (\ref{HierarchyOfScales}). An example of systems with slow thermalization is the Hamiltonian $(1+1)$-dimensional scalar field theory \cite{Boyanovsky:2003tc}. 
In the model with the potential (\ref{V4}), at $\hat T=0.1$, the thermalization time is $\tTh|_{\hat T=0.1}\sim 10^6/m$ \cite{Pirvu:2024nbe} and actually exceeds the nucleation time $\tDec|_{\hat T=0.1}\simeq 1.5\times 10^4/m$ in the simulation box with length $L=50/m$ considered in Sec.~\ref{ssec:NumTest}.\footnote{The hierarchy (\ref{HierarchyOfScales}) is recovered in the Hamiltonian system at low enough temperature, $\hat T\lesssim 0.06$, but the nucleation time is then too long to be simulated directly. The condition (\ref{HierarchyOfScales}) is also satisfied in the stochastic setting with $\eta\gtrsim 10^{-3}m$, irrespective of temperature. }

The dynamics of such system can be qualitatively understood as follows. Let us split all degrees of freedom (all Fourier modes of the field) into those relevant for nucleation and the rest providing the thermostat.
We denote the number of the relevant modes by 
$\Nrel$ and assume it to be large, $\Nrel\gg 1$. In the field theory example of Sec.~\ref{ssec:NumTest} this is the number of lattice modes with wavenumbers of order the size of the critical bubble, $\Nrel\simeq mL$. In what follows we refer to these modes as the {\it relevant subsystem}. Since thermalization is inefficient, we can neglect energy exchange between the relevant subsystem and the rest of degrees of freedom, so that the energy $\Erel$ of the relevant subsystem can be considered as constant. 
If we further assume that the dynamics within 
 the relevant subsystem is sufficiently ergodic, 
we recover the setup of Sec.~\ref{sssec:small}. Even if we start from an initially thermal population in the metastable region $\R$, the smallness of the relevant subsystem will introduce the drift of the decay rate on the time scale
\be
\label{eq:tdrift1}
\tDrift\sim \tDec\,\e^{-\brel}\;,\qquad \brel=\frac{\Es^2}{\Nrel T^2}\;,
\ee
which is 
significantly shorter than the decay time for $\brel>1$. If the time-dependent decay rate reaches a plateau before $\tDrift$, we can talk about the quasi-stationary regime in the interval $\tPlat\ll t\ll \tDrift$. Otherwise, no stationary rate exists.

Let us put in some numbers. In the example of Sec.~\ref{ssec:NumTest} we have $\Nrel\simeq 50$, $\Es/T=40/3$, implying $\brel\simeq 3.5$. This predicts $\tDrift\sim 450/m$, so we have a marginal separation between $\tPlat\sim 50/m$, where the sloshing re-crossings die out (see Fig.~\ref{fig:comp}), and the drift time. Still, the drift of the rate is quite appreciable, as illustrated by the band in Fig.~\ref{fig:comp} reflecting the variation of the rate in the time interval $50/m<t<200/m$. Note that Eq.~(\ref{eq:tdrift1}) predicts a significantly milder drift 
already for a twice bigger lattice: $mL=100$ implies $\brel\simeq 1.75$ and $\tDrift\simeq 1.3\times 10^3/m$, where we have used $\tDec|_{\hat T=0.1}\simeq 0.75\times 10^4/m$ for this lattice size.

To sum up, the slow thermalization effectively reduces the nucleation dynamics to the relevant subsystem. If the latter is too small and violates the condition (\ref{LargeBath}), the decays will deviate 
from the exponential law on long time scales. Such deviations --- called the classical Zeno effect --- have been observed in a variety of lattice simulations \cite{Pirvu:2023plk,Pirvu:2024nbe}. It can be relevant for laboratory experiments with cold atoms.

\section{Conclusions and outlook}
\label{sec:disc}

Nucleation is a dynamical, out-of-equilibrium process.
In this paper we have studied the dynamics and deviations from equilibrium which occur during nucleation in thermal systems.
Intriguingly, 
the physical decay rate $\Gamma$ is proportional to a purely equilibrium quantity, the TST rate,
\begin{align} \label{GammaFullRepeated}
    \Gamma = \GammaTST\cdot (1-R)\,,
\end{align}
repeated here from Eq.~(\ref{GammaFull}).
The TST rate counts all crossings of the dividing surface, but dynamically a fraction $R$ of these re-cross the dividing surface and are not real nucleation events.
The physical rate is always slower than the TST rate, because $R$ has the interpretation of a probability, so lies in $0\leq R \leq 1$.
Our derivation of this formula and the framework stemming from it offer resolutions to the four puzzles of nucleation theory introduced in Sec.~\ref{ssec:intro_puzzles}.

{\it Puzzle 1} asks what is the physical nucleation rate for a thermal system? We have argued that this should be identified with the rate in the stationary regime, during which the rate is approximately constant. Our derivations of this proceeded along two different routes: the first followed the time evolution of the probability flux out of an initially thermalized region of phase space, and the second selected nucleating trajectories using an appropriate indicator function. Both derivations agreed on the physical decay rate, yielding versions of Eq.~(\ref{GammaFullRepeated}) in Eqs.~\eqref{eq:RateFromFluxWithoutp}, \eqref{MRT-to-TST} and \eqref{Gamma-to-TST}. In so doing, we exposed the relationships between different expressions for the rate in the literature.

{\it Puzzle 2} raises the question of what is the correct dividing surface? In the stationary regime, the physical decay rate $\Gamma$ is in fact independent of the dividing surface within a wide range, and up to exponentially small corrections.
The specific choice is therefore a matter of convenience. This is however not the case for the rate outside the stationary regime, nor for $\GammaTST$ or $R$ separately.

Striking differences between nucleation in field theory and in low-dimensional mechanics provoked {\it Puzzle 3}, which asks can they be reconciled? In particular, in low-dimensional Hamiltonian mechanics, but ostensibly not in field theory, the nucleation rate from a thermalized metastable phase drops to zero in a microscopic time. As argued in Sec.~\ref{sec:finite}, the explanation of this apparent discrepancy is simply the different possibilities for thermalization.
For instance, there is also no nonzero stationary decay rate in Hamiltonian field theory models in sufficiently small volumes.
On the other hand, in both mechanics and field theory, either an external thermal bath or a sufficiently large and ergodic system yields a well-defined steady-state decay rate.
In Sec.~\ref{sec:upperBoundForDoubleScalingLimit} we showed that a quasi-stationary rate can exist even when the thermalization of the system is inefficient.
Future work is necessary to explore the necessary conditions for quasi-stationary nucleation in concrete systems.

Finally, previous lattice simulations~\cite{Pirvu:2023plk} found that oscillons are precursors to nucleation events, and so {\it Puzzle 4} asks how do oscillons affect the decay rate? The answer, demonstrated in Sec.~\ref{sec:num}, is that the presence of oscillons in a field theory enhances the re-crossing probability $R$, and hence reduces the decay rate compared to $\GammaTST$. This effect is most pronounced in models without noise and damping. In Sec.~\ref{ssec:EliminatingOscillons}, we showed that non-perturbative re-crossings do occur in models without oscillons, though with lower probability.

Our central expression, Eq.~(\ref{GammaFullRepeated}), is useful both perturbatively and non-perturbatively. It provides a target observable for perturbative derivations of the steady-state decay rate, as demonstrated in Sec.~\ref{sec:app} and to be extended in future work~\cite{ToAppear}. Moreover, it suggests a practical numerical algorithm to solve for the steady-state rate which does not require exponentially long run times. This is based on sampling the surface flux ensemble, which we applied to $(1+1)$ dimensional field theories with exponentially slow rates down to $\sim \e^{-700}$ in Secs.~\ref{sec:num} and \ref{ssec:NumTest}, and which could be applied to arbitrarily slow rates. Similar methods have been developed in physical chemistry~\cite{berne1985molecular, Hanggi:1990zz}. In field theory, the method extends that of Moore, Rummukainen and Tranberg~\cite{Moore:1998swa, Moore:2000jw, Moore:2001vf}, by correctly accounting for sloshing trajectories. In $(3+1)$ dimensions oscillons dissipate faster~\cite{Mukaida:2012qn}, so we expect the effect of sloshing trajectories will be smaller. However, there are some indications of oscillons playing a role in thermal nucleation in $(3+1)$ dimensions~\cite{Bian:2025twi}.

Our approach is applicable to a huge variety of classical decay processes, including the familiar example of bubble nucleation in liquid-gas transitions. In this case, it would be worthwhile to investigate whether non-perturbative sloshing trajectories can significantly increase $R$, extending previous studies including only perturbative, direct re-crossings in Langer's framework~\cite{langer1973hydrodynamic, turski1980dynamics, Csernai:1992tj}. This may improve upon the rather poor agreement found between theory and experiment~\cite{oxtoby1992homogeneous,Ganton1999}, assuming that the microphysical system can be modelled sufficiently accurately~\cite{langer1980kinetics, nellas2010exploring}. Non-perturbative sloshing trajectories become more important for small to moderate exponential suppression, so that they will be especially important for tabletop experiments that are inherently limited to observing not too slow transitions. This includes experiments in superfluid ${}^3$He~\cite{QUEST-DMC:2024crp}, and analogue systems~\cite{Fialko:2014xba, Fialko:2016ggg, Billam:2021nbc,Song:2021pyy, Tian:2022dzv, Zenesini:2023afv, Jenkins:2023eez, Jenkins:2023npg, Darbha:2024srr, Zhu:2024dvz,Cominotti:2025qia}.

While we have only directly addressed nucleation in classical particle dynamics and classical field theories, our results nevertheless apply to quantum particle dynamics and quantum field theories in the regime of high temperatures $T\gg T_\text{q} \sim \hbar|\o_-|$. This is particularly relevant for applications to early-universe cosmology, where microphysical models can be specified precisely as quantum field theories.
The form of the relevant classical effective description then depends on details of the underlying quantum field theory~\cite{Bodeker:1996wb, Aarts:1997kp, Arnold:1997gh, Blaizot:2001nr}, but at least in weakly coupled models can be derived explicitly.
In some cases,
the effective dynamical equations are integro-differential and the noise is non-Markovian~\cite{Boyanovsky:1996xx, Blaizot:2001nr, Calzetta:2001pp, Gautier:2012vh}.
This motivates generalising the present work to such descriptions, akin to Refs.~\cite{grote1980stable, hanggi1982thermally} for particle escapes.
With appropriate modifications, our methods can be applied to other thermal semiclassical processes, such as sphaleron processes in the electroweak broken phase~\cite{Klinkhamer:1984di, Arnold:1987mh, Rubakov:1996vz, Moore:1998swa, Annala:2025aci} and the thermal production of solitons~\cite{Grigoriev:1989ub, Bochkarev:1989tk}.

Generalising our dynamical framework for nucleation to field theories in the quantum regime $T\lesssim T_\text{q}$ is an important challenge. In vacuum tunnelling computations, such as Coleman's classic work~\cite{Coleman:1977py}, the whole computation can be analytically continued to Euclidean time, and in that sense there is no dynamics. We suspect that this is a special property of the vacuum, while for more general initial states the dynamics of nucleation should play a role.
Promising frameworks for generalising to such systems in the quantum regime include semiclassical path integral methods 
developed to describe false vacuum decay and similar non-perturbative processes in non-equilibrium  
conditions \cite{Rubakov:1992ec,Bezrukov:2003er,Levkov:2004ij,Demidov:2015bua,Demidov:2015nea,Shkerin:2021zbf}, and the formalism of time-evolving Wigner functionals, which has recently been applied to this problem~\cite{Hirvonen:2026zaq}.
While classical simulations do not apply in the quantum regime~\cite{Epelbaum:2014yja, Tranberg:2022noe}, there are a number of promising proposals for simulating such quantum systems~\cite{Lagnese:2021grb, Vodeb:2024tvo, Abel:2025pxa}, at least in low dimensions.

\section*{Acknowledgments}

We thank Andreas Ekstedt, Matthew Johnson and Sung-Sik Lee for insightful discussions.
This research was supported in part by Perimeter Institute for Theoretical Physics.
Research at Perimeter Institute is supported in part by the Government
of Canada through the Department of Innovation, Science and Economic
Development Canada and by the Province of Ontario through the Ministry
of Colleges and Universities.
This research
was enabled in part by support provided by Compute Ontario (www.computeontario.ca), Digital Research Alliance of Canada (alliancecan.ca), and by access to the University of Nottingham’s Ada HPC service.
Oliver Gould and Joonas Hirvonen were supported by the Royal Society Dorothy Hodgkin Fellowship with grant number DHF\textbackslash{}R1\textbackslash{}221001.
The work of Sergey Sibiryakov is supported by the
Natural Sciences and Engineering Research Council (NSERC) of Canada. 

\appendix

\section{Solution of the Fokker--Planck equation}
\label{AppB}

The coefficients of Eq.~(\ref{FPred}) are linear in $q$ and $p$, implying that the Fourier transform of the distribution $\tilde\rho$ defined in Eq.~(\ref{PpFour}) satisfies a first-order partial differential equation. This allows us to obtain a general solution for $\tilde\rho$. 

Let us define $u(x,y;t)\equiv \log\tilde\rho(x,y;t)$. It satisfies the equation
\be
\label{ueq}
\frac{\d u}{\d t}-\omega_-^2y\frac{\d
  u}{\d x}+(\eta y-x) \frac{\d u}{\d y}=-\eta T y^2\;.
\ee
Introduce a fictitious dynamical system depending on time $\tau$, with
equations of motion
\be
\label{ficteoms}
\dot t=1~,~~~~~\dot x=-\omega_-^2y~,~~~~~\dot y=\eta y-x\;,
\ee
such that Eq.~(\ref{ueq}) takes the form
\be
\label{ueq1}
\dot u=-\eta T y^2\;.
\ee
Equations (\ref{ficteoms}) are solved by
\be
\label{fictsol}
t=\tau+t_0~,~~~~x=C_1\l_-\e^{\l_+\tau}+C_2\l_+\e^{\l_-\tau}~,~~~~
y=C_1\e^{\l_+\tau}+C_2\e^{\l_-\tau}\;,
\ee
with $\lambda_\pm$ defined in Eq.~(\ref{lambdapm}), and $t_0$, $C_1$,
$C_2$ arbitrary integration constants. Setting $t_0=0$ by a trivial shift of the variable $\tau$, we can express
the integration constants through the coordinates:
\be
\label{intmot}
C_1=\frac{(x-\l_+ y)\e^{-\l_+ t}}{\l_--\l_+}~,~~~~~~~
C_2=\frac{(x-\l_- y)\e^{-\l_- t}}{\l_+-\l_-}\;.
\ee
These combinations represent two independent integrals of motion of the system
(\ref{ficteoms}). 
The solution of the homogeneous Eq.~(\ref{ueq1}) is an arbitrary function of
them,
\be
\label{uhom}
u_{\rm hom}(x,y;t)={\cal F}\Big( (x-\l_+ y)\e^{-\l_+ t}, (x-\l_- y)\e^{-\l_- t}\Big)
\ee
To find a particular solution of (\ref{ueq1}), we express $y$ as
function of $\tau$ from (\ref{fictsol}) and integrate. This gives,
\be
\label{upart}
u_{\rm part}(x,y;t)=-\frac{\eta T}{2\l_+}C_1^2\e^{2\l_+ \tau}-\frac{\eta
  T}{2\l_-}C_2^2\e^{2\l_- \tau}
-2TC_1C_2\e^{\eta \tau}=\frac{T}{2}\bigg(\frac{x^2}{\omega_-^2}-y^2\bigg)\;,
\ee
where in the second equality we used the expressions (\ref{intmot}). Note that the particular solution does not depend on time $t$.
The general solution of Eq.~(\ref{ueq}) is the sum of the 
homogeneous (\ref{uhom}) and 
particular (\ref{upart}) parts, $u_{\rm gen}=u_{\rm hom}+u_{\rm part}$. 

To get the solution relevant for the re-crossing problem, we subject the general solution to the initial condition (\ref{Pinit}). In the Fourier space, it reads
\be
\label{tilderhoinit}
\tilde\rho(x,y;t=0)={\cal F}_0(y)\;,
\ee
where the function ${\cal F}_0$ is defined in (\ref{lambdapm}).
This fixes the arbitrary function ${\cal F}$ in the general solution, with the result
\be \label{Pspec}
\begin{split}
    \tilde\rho(x,y;t)= \mathcal{F}_0(C_1+C_2)\,
\exp\bigg\{ \frac{T}{2}\left[ \eta \left( \frac{C_1^2}{\l_+} + \frac{C_2^2}{\l_-} \right) + 4 C_1C_2 +\frac{x^2}{\o_-^2}-y^2 \right] \bigg\} \;,
\end{split}
\ee
where $C_1$ and $C_2$ are functions of $x$, $y$, $t$ from Eq.~(\ref{intmot}).

At $y=0$ the solution simplifies, 
\be
\tilde\rho(x,y=0;t)=\mathcal{F}_0\left(x\frac{\e^{-\l_-t}\!-\!\e^{-\l_+t}}{\l_+-\l_-}\right)\,
\exp\bigg\{ x^2\frac{T}{2\omega_-^2}\!\left[1\!-\!\frac{\eta\l_+\e^{-2\l_-t}+\eta\l_-\e^{-2\l_+t}
+4\omega_-^2\e^{-\eta t}}
{(\l_+-\l_-)^2}\right]\!\bigg\}.
\ee
Next, we observe
that at large times the exponents $\e^{-\l_+t}$ and $\e^{-\eta t}$ die
out, whereas $\e^{-\lambda_- t}$ grows. Keeping at $t\gg (-\l_-+\eta)^{-1}=\l_+^{-1}$ only the terms with growing exponent, we arrive at Eq.~(\ref{PspecAsymp}) from the main text.

\section{Resumming re-crossings for Kramers turnover}
\label{app:WH}

Here we sum the series (\ref{Pescape}) describing the probabilities for a particle to re-cross the barrier after $n$ oscillations around the metastable minimum. The energy distribution $\P_1(E)$
is given by Eq.~(\ref{PE1}) and $\P_n(E)$ is defined inductively from $\P_{n-1}(E)$ through the convolution of the type (\ref{PE2}).
It turns out that the distribution after the $n$th oscillation
can be cast into the form,
\be
\label{PEn}
\P_n(E)=-\sum_{k=1}^{n-1} \P_k(E)+\frac{\e^{-E/T}}{T}\bigg[1-f_{n,\Delta}\bigg(\frac{E}{\sqrt{4T\ve}}\bigg)\bigg]\;,
\ee
where
\be
\label{DeltaKram}
\Delta=\sqrt{\frac{\ve}{4T}}
\ee
and the function $f_{n,\Delta}(z)$ is defined as an iterated 
convolution of Gaussians,
\be
\label{fnconvol}
f_{n,\Delta}(z)=\dint_{-\infty}^0\!\!\!\!\ldots\dint_{-\infty}^0\frac{\diff z_{n-1}\ldots
  \diff z_0}{\pi^{n/2}}
\exp\big[\!-(z\!-\!z_{n-1}\!-\!\Delta)^2-\ldots-(z_1\!-\!z_0\!-\!\Delta)^2\big]\;.
\ee
The expression (\ref{PEn}) is proved by induction.
Substituting it into (\ref{Pescape}), we find
\be
\label{Pmpinfty}
R^{(-)}=1-\dint_0^\infty \frac{\diff E}{T}\e^{-E/T} 
f_{\infty,\Delta}\bigg(\frac{E}{\sqrt{4T\ve}}\bigg)\;,
\ee
with $f_{\infty,\Delta}(z)=\lim\limits_{n\to\infty}f_{n,\Delta}(z)$. The latter
function satisfies an integral equation,
\be
\label{finteq}
f_{\infty,\Delta}(z)=\dint_{-\infty}^0\frac{\diff z_1}{\sqrt{\pi}}\,
\e^{-(z-z_1-\Delta)^2} f_{\infty,\Delta}(z_1)\;,
\ee
with the asymptotics
\be
\label{finfasymp}
\lim\limits_{z\to-\infty}f_{\infty,\Delta}(z)=1~,~~~~~~~~\lim\limits_{z\to\infty}f_{\infty,\Delta}(z)=0\;.
\ee
This integral equation is similar to 
the equation derived in \cite{Melnikov:1986}.\footnote{It would coincide
  exactly if instead of $f_{\infty,\Delta}(z)$ we consider
  $\e^{-4z\Delta}f_{\infty,\Delta}(z)$.}   
We solve it
using the 
Wiener--Hopf method.
To ease notations, we suppress the subscript $\Delta$ in what follows. 

We introduce one-sided Fourier transforms of the function $f_\infty$,
\be
\label{phipm}
\vf_+(y)=\dint_0^\infty \diff z\,f_{\infty}(z)\,\e^{izy}~,~~~~
\vf_-(y)=\dint_{-\infty}^0 \diff z\,f_{\infty}(z)\,\e^{izy}\;.
\ee
With the asymptotics (\ref{finfasymp}), the functions $\vf_+(y)$ and
$\vf_-(y)$ are regular in the upper and lower half-planes of $y$,
respectively. If we further assume that $f_\infty(z)$ asymptotes to $1$ at
negative $z$ exponentially, $|1-f_{\infty}(z)|<\e^{\l z}$ for some
$\l>0$, then $\vf_-(y)$ is regular in the half-plane $\Im y<\l$, apart
from a simple pole at $y=0$ where it behaves as $\vf_-(y)\approx
\frac{1}{iy}$. Note that the re-crossing rate
(\ref{Pmpinfty}) is expressed as
\be
\label{Pmpvf}
R^{(-)}=1-4\Delta\;\vf_+(4i\Delta)\;.
\ee

Equation (\ref{finteq}) implies
\be
\label{inteqFourier}
\vf_+(y)+\vf_-(y)=g(y)\vf_-(y)\;,
\ee
where
\be
\label{FourGauss}
g(y)=\e^{-y^2/4+iy\Delta}
\ee
is the Fourier transform of the Gaussian.
Note that the function $1-g(y)$ does not have zeros in the strip
$0<\Im y < 4\Delta$.
Let us introduce two
functions
\be
\label{Gpm}
G_\pm(y)=\exp\bigg[\pm\frac{1}{2\pi i}
\int\limits_{-\infty+i\epsilon_\pm}^{+\infty+i\epsilon_\pm}\frac{\diff y'\ln\big(1-g(y')\big)}{y'-y}\bigg]\;,
\ee
where the integration contours run parallel to the real axis and are chosen such that
$0<\epsilon_+<\epsilon_-<\min(\l,4\Delta)$.
The function $G_+(y)$ is regular and have no zeros in the half-plane $\Im y>\epsilon_+$, whereas $G_-(y)$ is
regular and have no zeros at $\Im y<\epsilon_-$. In the overlap of these half-planes we have
$G_+(y)G_-(y)=1-g(y)$ and eq.~(\ref{inteqFourier}) can be written as
\be
\label{spliteq}
\frac{\vf_+(y)}{G_+(y)}=-\vf_-(y)G_-(y)\;.
\ee
Here l.h.s. is a regular function at $\Im y>\epsilon_+$, and 
r.h.s. is analytic at $\Im y<\epsilon_-$, with a simple pole at
$y=0$. Moreover, since these two functions coincide in 
a
strip, they actually form a single analytic function $\Phi(y)$ defined on the
whole complex plane with a pole at $y=0$. The most general function
with these properties
has the form  $\Phi(y)=c_1+c_2/y$, and the asymptotics at infinity and at
$y=0$ fix it to be:
\be
\label{Phisol}
\Phi(y)=-\frac{G_-(0)}{iy}~~~\Longrightarrow~~~
\vf_+(y)=-\frac{G_-(0)G_+(y)}{iy}~,~~~\vf_-(y)=\frac{G_-(0)}{iyG_-(y)}\;.
\ee
The inverse Fourier transform of the sum $\vf_-+\vf_-$ then provides the
solution to Eq.~(\ref{finteq}).

The re-crossing rate is determined using Eq.~(\ref{Pmpvf}),
\be
\label{Pmpvf1}
R^{(-)}=1-G_-(0)G_+(4i\Delta)\;.
\ee
We deform the integration contours in the definitions of $G_\pm$ 
to run along the line ${\Im y'=2\Delta}$, on which the function
$g(y')$ is real. After some straightforward
algebra, we arrive at Eq.~(\ref{Pmpfin}) from the main text.

\section{Initial state preparation}
\label{app:instate}

Here we give details on how to prepare the surface-flux ensemble for the calculation of the dynamical rate in Secs.~\ref{sec:num} and \ref{sec:finite}.
This consists of choosing the dividing surface and specifying the initial field and momentum 
distributions on that surface.

\subsection{Quadratic approximation}
\label{app:instate1}

We choose the TS $\phi=\phi_\text{ts}(x;L/2)$ as discussed in Sec.~\ref{ssec:non-pert}. Next, we 
expand the field and momentum $\pi=\dot{\phi}$ in the basis of linear perturbations around the TS,
\begin{align}
& \phi(x) = \phi_{\rm ts}(x;L/2) + q_-\chi_-(x) + X_c \chi_0(x)+ \sum_{i=1}^{N-2} q_i \chi_i(x) \;,\label{Phi_expansion}\\
& \pi(x) = p_-\chi_-(x)+P_c \chi_0(x)+ \sum_{i=1}^{N-2}p_i\chi_i(x) \;.\label{Pi_expansion}
\end{align}
Here $\chi_\a$, $\alpha=\;$`$-$',$0,1,...,N-2$, are a complete set of $N$ orthonormal eigenmodes around $\phi_{\rm ts}$. 
They satisfy the equation
\be\label{LinPertEq} 
-\triangle\chi_\alpha(x) + V_4''\big(\phi_\text{ts}(x;L/2)\big)\chi_\alpha(x) = \mu_\alpha \chi_\alpha(x) \;,
\ee
where $\triangle$ is the (lattice) Laplacian and $\mu_\alpha$ are eigenvalues.
The mode $\chi_-$ is unstable (${\mu_-=-\omega_-^2}$) and corresponds to the transition coordinate. The mode $\chi_0(x)=\phi'_\text{ts}(x)/\sqrt{\Es}$ is the zero mode ($\mu_0=0$) corresponding to the space translations parameterized by the collective coordinate $X_c$. The modes $\chi_i$ with $i\geq 1$ are positive (stable) modes ($\mu_i>0$). We fix the dividing surface $\d\R$ by the condition $q_-=0$.

We prepare an ensemble of initial states 
lying on $\d\R$ as follows.
We fix $q_-=0$ and pick the momentum $p_-$ of the negative mode from the distribution
\be
\label{P_minus}
\rho_-(p_-)= \theta(-p_-)\frac{|p_-|}{T}\exp\left(-\frac{p_-^2}{2T}\right)\;,
\ee
which is consistent with Eq.~(\ref{Rho_init2}) in the quadratic approximation of the potential around the barrier.
The theta-function ensures that the TS is initially pushed back to the false vacuum.
Next, excitations along the zero mode correspond to 
the center-of-mass motion of the critical bubble. 
The center-of-mass velocity $v_{\rm CM}$ of the bubble follows the Maxwell distribution, $\rho_0(v_{\rm CM})\propto \exp\big(-v_{\rm CM}^2\Es/(2T)\big)$ \cite{Pirvu:2023plk}.
The momentum $P_c$ of the zero mode is normalized so that $P_c=v_{\rm CM}\sqrt{\Es}$. Consistently with the discussion in Sec.~\ref{ssec:prelim}, we fix $X_c=0$ and pick $P_c$ 
from the distribution
\be
\label{P_0}
\rho_0(P_c)=\frac{1}{\sqrt{2\pi T}} \exp\left(-\frac{P_c^2}{2T}\right)\;.
\ee
Finally, in the quadratic approximation the coefficients of the positive modes $(q_i,p_i)$, $i\geq 1$, follow the Gaussian distributions: 
\be \label{P_plus}
\rho_i(q_i,p_i) = \frac{\sqrt{\mu_i}}{2\pi T} \exp\left( -\frac{p_i^2}{2T} \right) \exp\left( -\frac{\mu_i q_i^2}{2T} \right) \;.
\ee

\subsection{Non-perturbative algorithm}
\label{app:instate2}

For the study of Sec.~\ref{ssec:NumTest}, we need non-perturbative definitions of the metastable region $\R$ and the surface-flux ensemble, which we construct as follows.
Consider a static field configuration~$\phi(x)$. We call a configuration unstable if the eigenmodes of linear perturbations around it have one or more negative eigenvalues.
The eigenmodes satisfy the equation
\be\label{LinPertEqGen} 
-\triangle\chi_\alpha(x) + V''\big(\phi(x)\big)\chi_\alpha(x) = \mu_\alpha \chi_\alpha(x) \;,~~~ \a=1,...,N \;.
\ee
We declare that $\phi(x)$ belongs to $\R$ if the Hamiltonian evolution along its most unstable direction $\a_*$ 
is pulled back towards the false vacuum $\phi=0$ by the Hamiltonian evolution.
This condition can be expressed as follows, 
\be  \label{eq:conditionForNonperturbativeR}
\bigl(\chi_{\a_*}(x), \phi(x)\bigr) \cdot \big( \chi_{\a_*}(x) , \dot{\pi}(x)\big) < 0 \;,
\ee
where we introduced the scalar product $\big(f(x),g(x)\big)\equiv\int\diff x f(x) g(x)$ and $\dot{\pi}(x)$ is found from the Hamilton equations, $\dot{\pi}(x)=-\left.\delta H/\delta\phi\right\vert_{\phi=\phi(x)}$.
The former dot product indicates in which direction the false vacuum is from the configuration $\phi$ along the negative eigenmode direction $\chi_{\a_*}$. The latter indicates the direction into which the Hamiltonian evolution pushes the configuration along the negative eigenmode.
Adding all stable (i.e.\ without negative eigenmodes) configurations to $\R$ completes its definition (we assume that there is no stable phase as is the case of the potentials studied in Secs.~\ref{sec:num}, \ref{sec:finite}). 
The dividing surface $\d\R$ is thus defined by the condition $\big( \chi_{\a_*}(x) , \dot{\pi}(x)\big)=0$.

In the MRT approach, the region $\R$ is defined in terms of the values of an order parameter which can be chosen rather freely. The order parameter is only required to map field configurations onto the real numbers in such a way that $\R$ corresponds to some identifiable range of values. Our definition of $\R$ can be interpreted through this lens, though the corresponding order parameter \eqref{eq:conditionForNonperturbativeR} is rather more complicated than e.g.\ the volume-averaged expectation value of the field.

To sample the surface-flux ensemble, we note that 
the probability flow through the infinitesimal area element $\Delta S\, \mathbf{n}$ of $\d\R^{(+)}$ at the point $\mathbf{z}\in \d\R^{(+)}$ during a finite time $\Delta t$ is given by
\be 
\Delta P(\bz)\propto \Delta t \Delta S \, \mathbf{n}  \cdot \dot{\mathbf{z}}(\bz) \, \e^{-H(\mathbf{z})/T} \;,
\ee
where $\mathbf{n}$ is a unit normal vector pointing outward and $\dot{\mathbf{z}}(\bz)$ is determined by the Hamiltonian flow. Comparing with Eq.~(\ref{outaverage}), we see that $\Delta P(\bz)$ samples $\bz\in\d\R^{(+)}$ with the same weight as the surface-flux ensemble.
This enables the following strategy. We prepare configurations, as described in Sec.~\ref{ssec:NumTest}, in the thin layer inside the dividing surface,
\be 
\big|\big(\chi_{\a_*}(x),\dot{\pi}(x)\big)\big| < C \;,  ~~~ \bigl(\chi_{\a_*}(x), \phi(x)\bigr) \cdot \big( \chi_{\a_*}(x) , \dot{\pi}(x)\big) < 0 \;,
\ee
where we choose $C=3.1\sqrt{\hat{T}}$, and the $\mathcal{O}\left(\sqrt{\hat{T}}\right)$ scaling matches the leading-order perturbation theory.
We then evolve the configurations for a short time $\Delta t=0.14/m$ and see if they cross $\d\R^{(+)}$.

Note that we do not check within the time of $\Delta t$, consisting of multiple time steps, whether the boundary was crossed -- only by the end points.
Hence, a trajectory can cross $\d\R$ back and forth several times during the time $\Delta t$ without registering these crossings. Luckily, we are interested in the rate, where each crossing only contributes with a weight divided by the number of crossings, \textit{cf.} e.g.\ Eq.~\eqref{GammaMRT0} or Sec.~\ref{sec:finite}. This leads to the result being insensitive to the missed crossings due to the finite $\Delta t$.

\bibliographystyle{JHEP}
\bibliography{Refs}

\end{document}